\documentclass[a4paper,fleqn,usenatbib]{mnras}

\usepackage[T1]{fontenc}
\usepackage{ae,aecompl}

\usepackage{graphicx,color}	% Including figure files
\usepackage{amsmath}	% Advanced maths commands
\usepackage{amssymb}	% Extra maths symbols
\usepackage{caption,subfig}

\newcommand\mearth{{\,{\rm M}_{\oplus}}}
\newcommand\mj{{\,{\rm M}_{\rm J}}}
\newcommand\msun{{\,{\rm M}_{\odot}}}

\newcommand\be{\begin{equation}}
\newcommand\ee{\end{equation}}

\title[Dust grain sedimentation in giant planets]{Planet formation inside proto-giants: First 3D simulations}
\author[]{Sergei Nayakshin
$^{1}$\thanks{E-mail}
\\
$^{1}$Department of Physics and Astronomy, University of
  Leicester, Leicester LE1 7RH, UK.
}

\date{Accepted XXX. Received YYY; in original form ZZZ}

\pubyear{2016}

\begin{document}
\label{firstpage}
\pagerange{\pageref{firstpage}--\pageref{lastpage}}
\maketitle

% Abstract of the paper
\begin{abstract}
Numerical simulations of pebble dynamics inside gas clumps formed by gravitational instability of protoplanetary discs are presented. We find that dust-mediated Rayleigh-Taylor instabilities  transport pebbles inward rapidly via dense metal-rich "fingers". This speeds up sedimentation of small pebbles by up to two orders of magnitude and yet does not impede grain growth because grains of all sizes sediment at the same collective speed as long as Stokes number is less than unity. In simulations with a fixed pebble size, solid planetary cores form  if pebble size exceeds a few cm. Pebble growth leads to core formation in some hundreds of years even when pebbles injected into clumps are of mm or smaller sizes. Properties of the gas clump dictate what kind of cores can be made. Low central temperature clumps allow formation of solid cores out of refractory materials, whereas in the highest temperature clumps pebbles of any composition are vaporised and make fuzzy cores only. These results confirm that gravitational instability of protoplanetary discs is a robust mechanism of hatching cores from sub-Earth to Neptune mass, as well as gas giants with massive cores, solid or fuzzy. This mode of planet formation is especially promising for environs too young and distant (such as the ALMA-observed HL Tau disc) or too violent (such as circum-binary planets), to form via the Core Accretion scenarios.
\end{abstract}

% Select between one and six entries from the list of approved keywords.
% Don't make up new ones.
%\begin{keywords}
%keyword1 -- keyword2 -- keyword3
%\end{keywords}

\section{Introduction}

Gas clumps formed by  gravitational instability of protoplanetary discs \citep[][]{KratterL16} present a viable environment in which grains can grow, sediment to the centre and form a solid core there \citep{Kuiper51,Kuiper51b,McCreaWilliams65,CameronEtal82,Boss98}. The central temperatures of these clumps vary, depending on their mass and evolutionary state, from $\sim 100$~K to $\sim 2000$~K \citep{Bodenheimer74}, in principle allowing grains of various compositions to reach the clump centre. 
%Smaller solids such as chondrules  and possibly planetesimals may also be formed or reprocessed inside the clumps \citep{BoleyEtal10,Vorobyov11,NayakshinCha12,IleeEtal17}.

\cite{Kuiper51b} believed that planets form on fixed orbits. We now know that massive self-gravitating gas discs hatch clumps by disc fragmentation at separations $\sim 100$~AU \citep{Gammie01,Rafikov05} but the clumps may migrate closer to the star in a matter of a few thousand years \citep{MayerEtal04,VB06,VB10,MachidaEtal10,BaruteauEtal11,MichaelEtal11,MachidaEtal11}. Those clumps that manage to contract and collapse into second cores \citep[also called post-collapse gas giants, or "hot start models" in different contexts][]{Larson69,BurrowsEtal00}, and survive the migration phase, become gas giant planets. Clumps that contract too slowly are tidally disrupted \citep{BoleyEtal10,Nayakshin10c}. 

If a solid core is synthesized inside the clump by the time it is disrupted, the core is released back into the world. However, the total condensible mass of metals inside a gas clump of mass $M$ is only $\sim 3 \mearth (M/1 M_J) (Z/0.01)$, where $Z$ is clump metallicity.  Grains also need to be as large as 1 cm in radius for an efficient grain sedimentation into the core, and this may not occur in time before the clump is disrupted. Finally, the internal regions of the clump may be too hot for grains to exist. 

Previous {\em isolated} clump studies found that these challenges are not easily overcome \citep{HelledEtal08,HS08,ForganRice13b}. However, 
the dust content of gas clumps may be far greater due to accretion of $\sim 1$ mm or larger grains from the disc \citep{HN18} via a process known as pebble accretion  \citep{OrmelKlahr10,JohansenLacerda10,LambrechtsJ12,LambrechtsEtal14}. Furthermore, spiral arms and gas clumps may be enriched with solids already at birth \citep{RiceEtal04,BoleyDurisen10,BoleyEtal11a,GibbonsEtal12,GibbonsEtal14}. 

\cite{NayakshinFletcher15,Nayakshin16a}  included the process of pebble accretion in their population synthesis, with results showing some promise in terms of core masses, compositions, orbital separations and host star metallicity correlations \citep[for a broad comparison of the model results with observations, see][]{Nayakshin_Review}.

%\cite{HN18} recently presented 3D simulations of how a gas clump, modelled as a sink particle, interacts with gas and dust grains of the parent protoplanetary disc. Pebbles  (dust grains of radius $\sim$ 0.1-10 cm) were found to accrete onto the clumps very efficiently.
%The thermal and the feedback challenges were studied via population synthesis by  

%In all of this previous work, however, the processes of grain sedimentation, core formation and possible planetesimal formations inside gas clumps were studied either analytically or via 1D spherically symmetric numerical codes \citep[e.g.,][]{HelledEtal08,HS08,Nayakshin10a,Nayakshin10b}. This stands in stark contrast to the extensive recent progress in 2D and 3D simulations of coupled gas-dust dynamics in large scale ($\sim 100$~au) protoplanetary discs within which the clumps form. For example, \cite{BoothClarke16} studied relative velocity distributions of grains in self-gravitating discs. \cite{LABate16} found an interesting toroidal instability which may help to limit inward migration speeds for dust. \cite{BateLA17} studies grain dynamics during protoplanetary disc formation. \cite{LP14,PL15} developed new smoothed particle hydrodynamics \citep[SPH;][]{Monaghan92} methods to model dust dynamics. \cite{DipierroEtal15,DipierroEtal16a} applied these methods to the observed deep dust gaps in the protoplanetary disc of HL TAU \citep{BroganEtal15}. 

However, the processes of grain sedimentation and core formation inside gas clumps were studied previously either analytically or via 1D spherically symmetric numerical codes only. In this paper we present first 3D numerical simulations of coupled gas and dust dynamics inside the gas clumps. We are in particular interested in the fate of the additional grains accreted by the clump from the parent disc because these grains may outnumber by total mass those native to the clump. To achieve higher numerical resolution, isolated gas clumps are studied here but the initial conditions are tailored to mimic clumps in their protoplanetary disc birth environment. We start with simulations in which grain size is fixed, the initial conditions are spherically symmetric, but later relax these assumptions. Table 1 (see \S \ref{sec:ic}) gives a summary of simulations presented here and main results learned from these. Animations of two simulations, Sp1Z1a01F and DarkCollapse, are available via online supplementary material, and at these links, respectively:

\noindent {\scriptsize \verb!https://www.dropbox.com/s/7e56pxlnhtkrqk6/SpZ1a01N8e5.mp4?dl=0!}
{\scriptsize \verb!https://www.dropbox.com/s/wvnailpusfg8x7t/DarkCollapse.mp4?dl=0!}

%\newpage

\section{Preliminaries}\label{sec:expect}

%\subsection{The gas clump and its environment}\label{sec:clump}

The protoplanet is introduced as a pebble-free polytropic sphere of mass $M_0 = 3\mj$. This is motivated by the fact that detailed gas clump contraction calculations show that the energy transfer within the clumps is strongly dominated by convection \citep[e.g.,][]{HelledEtal08}. The central temperature of our fiducial clump is $T_{\rm cen} = 300$~K, the central density is $\rho_{\rm cen} = 1.8\times 10^{-9}$ g/cm$^3$ and the clump radius is $R_{\rm p}\approx 1.5$~AU. Radiative cooling of the clump is neglected (see \S \ref{sec:ic}). This clump would be destroyed by tidal forces if it reaches the planet-star separation $D\sim 20$~AU when its radius is comparable to the Hill radius. 

%In realistic protoplanetary discs, the clump accretes pebbles and may be over-abundant in metals by a factor of several within $\sim$ a few thousand years, see \S 2.3 in \cite{Nayakshin15a} or fig. 10. in \cite{HN18}\footnote{and note that the latter paper assumed very low pebble abundance, $Z=0.002$ by mass, in the parent disc; in realistic discs this can be much higher, resulting in much more vigorous accretion of pebbles onto the clump}. Pebbles first arrive in the outer layers of the planet and it is our aim to understand their further dynamics inside the clump. Although we shall consider gradual "loading" of the planet with pebbles below, most tests will be ran under the assumption of an instantaneous loading of pebbles into the planet to save computational resources. 

\subsection{Drag laws and sticking grain growth}\label{sec:dynamics}

We use equations (7-9) from \cite{Weiden77} for the aerodynamical friction force between gas and a pebble particle of internal material density $\rho_a$, radius $a$, moving through the gas at a relative velocity $\Delta v$. In the Epstein regime, the magnitude of the friction force is given by
%\begin{equation}
$ F = (4\pi/3) \rho a^2 v_{\rm th} \Delta v$, 
%\label{Fep}
%\end{equation}
where $\rho$ and $v_{\rm th} = [8 k_{\rm b} T/\pi \mu]^{1/2}$ are density and mean thermal speed of gas with temperature $T$, $k_{\rm b}$ is the Boltzmann constant and $\mu = 2.45 m_{\rm p}$ is the mean molecular weight. The Epstein drag law is used for particles with size $a < a_{\rm tr} = (3/2) \lambda$, where $\lambda \approx 4 \rho_{-9}^{-1}$~cm is the mean free path of hydrogen molecules, with $\rho_{-9} = \rho/(10^{-9}$~g cm$^{-3}$). For larger particle sizes, the friction force depends on the Reynolds number, $Re = 2 a \Delta v/\nu_{\rm visc}$, where $\nu_{\rm visc}$ is the viscosity coefficient. We use the ideal gas viscosity law, $\nu_{\rm visc} = (1/3) \lambda v_{\rm th}$. 

The dependence of the drag coefficient $C_{\rm d}$ on the Reynolds number iss specified in \cite{Weiden77} and implemented in our numerical code, but for the analytical understanding of the problem, it suffices to use the Stokes law for particles $a > (3/2) \lambda$ because $\Delta v$ is usually much smaller than $v_{\rm th}$, in which case 
%\begin{equation}
$F = 2\pi a (\mu/\sigma_{\rm H}) v_{\rm th} \Delta v $, 
%\label{Fst}
%\end{equation}
with $\sigma_H$ is the H2 molecule collision cross section.%, taken to be $10^{-15}$~cm$^2$. 
The equation of motion for a dust particle with velocity $\mathbf{v}$ is
\begin{equation}
\frac{d \bf v}{dt} = \mathbf{g} + \mathbf{F} =
\mathbf{g} - \frac{\mathbf{v - v_{\rm g}}}{t_{\rm st}}\;,
\label{dvdt0}
\end{equation}
where $\mathbf{g}$ is the gravitational acceleration, $\mathbf{v_{\rm g}}$ is the surrounding gas velocity, and we defined the stopping time of the particle by
\begin{equation}
t_{\rm st} = \frac{m_a \Delta v}{F}\frac{\rho + \rho_{\rm p}}{\rho}\;,
\label{tstop}
\end{equation}
where $m_a = (4\pi/3)\rho_a a^3$ is the particle mass. 
%\newpage

%\subsection{Grain dynamics inside a clump}\label{sec:inside}

The terminal sedimentation velocity is found by setting $F = m_a g$, where $g(R) = G M(R)/R^2$ is the inward directed gravitational acceleration, and $M(R)$ is the mass of the clump interior to radius $R$,
\begin{eqnarray}
v_{\rm sed} =
\frac{2}{3} \frac{\rho_a a^2 \sigma_{\rm H}}{\mu v_{\rm th}} g(R) \quad \text{Stokes drag} \;,\\
v_{\rm sed} =
\frac{\rho_a a}{\rho v_{\rm th}} g(R) \quad \text{Epstein drag}\;.
\end{eqnarray}
In the centre of the clump, the gas density is constant to a good approximation, $\rho = \rho_{\rm cen}$, so $g(R) \approx (4\pi/3) G\rho_{\rm cen} R$, so
\begin{eqnarray}
v_{\rm sed} =
4.3 \;\text{m/s} \; \frac{\rho_a a_1^2}{T_{300}^{1/2}} \rho_{-9} R_{\rm au}\quad \text{Stokes drag } (a\gtrsim 10\; \text{cm})\;,\\
v_{\rm sed} =0.26 \;\text{m/s} \; \frac{\rho_a a_0}{T_{300}^{1/2}} R_{\rm au}
 \quad \text{Epstein drag } (a\lesssim 10\; \text{cm})\;,
 \label{vsed_num}
\end{eqnarray}
where $a_1 = a/(10$~cm), $a_0 = a/(1$~cm), $T_{300} = T/(300$~K). The corresponding sedimentation times are,
\begin{eqnarray}
t_{\rm sed} =
1100 \;\text{yr} \; (\rho_{-9} \rho_a a_1^2)^{-1} T_{300}^{1/2} \quad \text{Stokes drag} \;,\\
t_{\rm sed} =1.8\times 10^4 \;\text{yr} \; (\rho_a a_0)^{-1} T_{300}^{1/2}
 \quad \text{Epstein drag}\;.
 \label{tsed_num}
\end{eqnarray}
Eq. \ref{tsed_num} is formally correct only in the centre of the clump. However, sedimentation is slowest in the clump centre, where gravity is weak, so that eq. \ref{tsed_num} is actually fairly accurate.

%\subsection{Grain growth}\label{sec:gg_anal}

For a grain of radius $a$ moving through a background of much smaller grains with volume density $\rho_{\rm bg}$ at a relative velocity $\Delta v_{\rm bg}$, the rate of grain mass ($m_{\rm a}$) increase by perfectly sticking collisions is $ d m_{\rm a}/dt = \pi a^2 \rho_{\rm bg} \Delta v_{\rm bg} = 3 m_{\rm a}/t_{\rm gr}$,
where we defined the growth time scale as
\begin{equation}
t_{\rm gr} = \frac{4 \rho_a a}{\rho_{\rm bg} \Delta v_{\rm bg}}\;.
\label{tcoll1}
\end{equation}
With this definition, the grain size grows with time as
\begin{equation}
\frac{d a}{dt} = \frac{a}{t_{\rm gr}}\;.
\label{tcoll2}
\end{equation}
We shall consider effects of high speed collisions in which grains fragment below.

\begin{figure*}
\includegraphics[width=1.\textwidth]{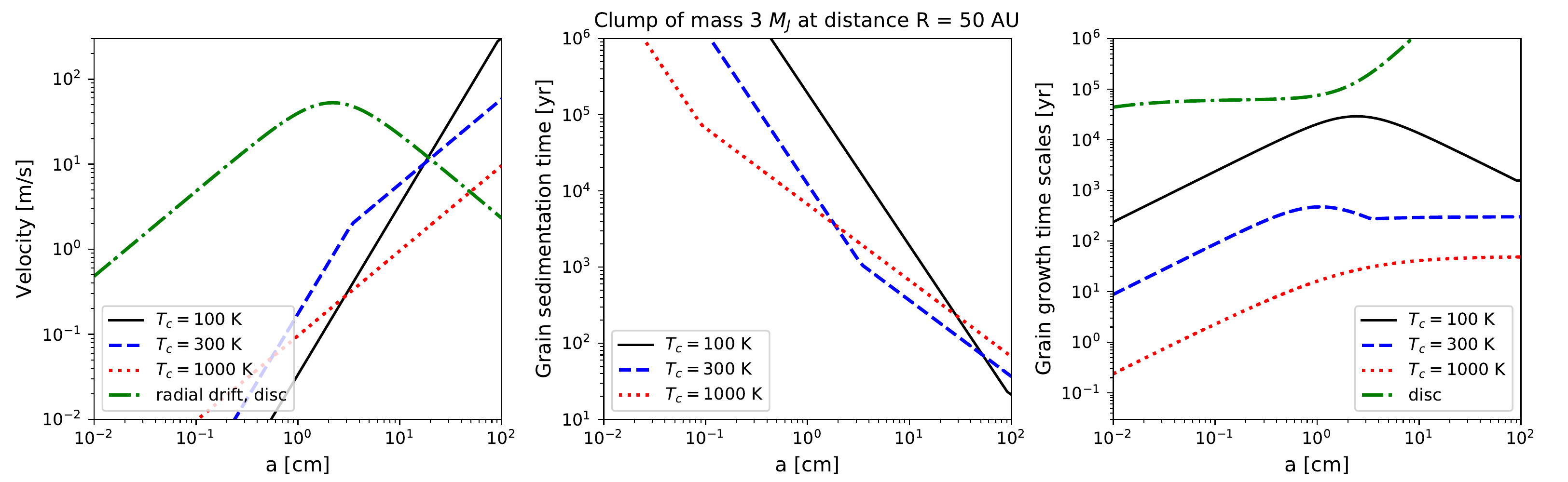}
\caption{{\bf Left:} Sedimentation velocities for grains versus grain size $a$ inside gas clumps. The curves are computed for three different polytropic gas clumps, all of mass $3 \mj$. The green dash-dot curve shows the radial velocity of the grains in a massive disc at separation of 50 AU. {\bf Middle:} Grain sedimentation time scale as a function of $a$ for the same three clumps. {\bf Right:} Grain growth or fragmentation time scales. The green dash-dot curve shows grain growth time inside the disc. See \S \ref{sec:ggrowth} for more detail.}
 \label{fig:grain1}
 \end{figure*}

\subsection{Clumps: safe heavens for grain growth}\label{sec:ggrowth}

The left panel of Fig. \ref{fig:grain1} shows grain sedimentation velocity as a function of grain size calculated for clumps with $T_{\rm cen}= 100$, 300, and 1,000~K. Grain material density is set to $\rho_a = 5$~g~s$^{-1}$.  The panel also shows the radial drift velocity of grains in the protoplanetary disc (green dash-dot curve) calculated following \cite{Weiden77} at radial distance of 50 AU. The disc is assumed to have radial profile $\Sigma \propto 1/R$, temperature profile $T(R) \propto 1/R^{1/2}$, $H/R = 0.1$ at 50 AU, and the mass $M_{\rm disc} = 0.2\msun$. 

The maximum radial drift velocity of pebbles in the disc occurs for the Stokes number $St = 1$ particles, which for our disc model corresponds to size $a \sim 5$~cm. \cite{BoothClarke16} concluded that velocity dispersion of pebbles in self-gravitating gas discs is too high to allow grain growth to proceed beyond Stokes number $\sim 0.01 - 0.1$. 
%This result can be approximately understood from the radial velocity curve for pebbles in the disc. At this grain size the radial drift velocity is a few m~s$^{-1}$ (cf. fig. \ref{fig:grain1}). Any realistic disc must contain a wide range of grain sizes at any given radius, and therefore the drift velocity is of the order of the typical velocity with which grains of the given size may collide with grains of a similar size. For most materials, grains fragment when colliding at velocities exceeding  a few m~s$^{-1}$ \citep[e.g.,][]{BlumWurm08}. 
Therefore, we should expect that grains entering the gas clump will be a few mm in size. The middle panel of fig. \ref{fig:grain1} shows the sedimentation time scales. For grains of a few mm size, sedimentation time is very long, $t_{\rm sed} \sim $ a few $\times 10^4$ to $\sim 10^6$~years. By the time the grains could sediment, the clump is likely to either collapse or be tidally disrupted, none of which is promising for solid core formation.

The right panel of fig. \ref{fig:grain1} depicts the grain growth or fragmentation time scale (eq. \ref{tcoll1}), assuming that the background grain density is $\rho_{\rm bg} = 0.02 \rho_{\rm cen}$. When using eq. \ref{tcoll1}, we added to $\Delta v$ a Brownian motion velocity of $20$~cm/s \citep[see, e.g.,][]{DD05}. For small grains, the grain growth time scale is generally much shorter than the grain sedimentation time implying that grains may increase in size rapidly. If they grow to the size of $\sim 10$ cm, then they will sediment into the clump centre onto the sedimentation time for such grains, which is relatively short. On the other hand, large grains, $a\ge 10$~cm, sediment inward very rapidly (cf. the left panel of fig. \ref{fig:grain1}), and are likely to be affected by grain fragmentation. 

%One caveat to this is to note that grain growth is actually slower at the outer regions of the clump than the right panel of fig. \ref{fig:grain1} shows because grain density at the outskirts of the clump should be expected to be much lower than in its centre. Grain dynamics and growth are therefore complex coupled time dependent processes even in 1D models of gas clumps. 

The dash-dot green curve in the right panel of fig. \ref{fig:grain1} shows the grain growth time for a grain in the disc at radial separation of 50 AU. Here we use the radial drift velocity plotted in the left panel as the estimate for $\Delta v$ in eq. \ref{tcoll1}, and we also added the Brownian motion component to it. Comparison of the green dash-dotted curves in the left and right panels of fig. \ref{fig:grain1} with the respective curves for the clumps show that clumps are a safe heaven for grain growth. For definitiveness, consider materials with fragmentation velocities of 3 m~s$^{-1}$. In the disc, such grains will only grow to a few mm size, when the collisions start to shatter them. In contrast, grains can grow to sizes of a few cm to almost half a metre inside the clumps before collisions become fragmenting. Further, this growth happens quickly, in tens of years to perhaps $10^4$ years. 

The physical reason why gas clumps provide much more promising environs for grain growth compared to discs is their much higher density (many orders of magnitude, typically). Due to this, grain-grain collisions in the clumps are much milder, occurring at smaller relative velocities, and yet they are much more frequent than collisions in the disc.

\subsection{Numercal method and initial conditions}\label{sec:ic}

The numerical method employed in this paper is presented in \cite{HN18}. In brief, Gadget 3, a widely used Smoothed Particle Hydrodynamics (SPH) with N-body code \citep[][]{Springel05} is employed to model the coupled dynamics of gas (SPH particles) and dust grains (N-body particles). Gravitational forces on all components are calculated. Dust particles interact with the SPH neighbouring particles also via the aero-dynamical friction force (\S \ref{sec:dynamics}). The SPH equations of motion for gas contains the aerodynamical friction term with the minus sign, guaranteeing momentum conservation in the interaction between the two species. The heat generated by the dust particles as they move through the surrounding gas is also included in the energy equation for the gas. An ideal equation of state with the adiabatic index $\gamma = 7/5$ is used. We neglect radiative cooling of the gas. This is reasonable since the clump cooling time is a few $\times 10^4$~years \citep{HelledEtal08,Nayakshin15a}, and is much longer than the duration of the simulations.

\begin{table*}
	\centering
	\caption{Simulations presented in the paper}
	\label{tab:sim_table}
	\begin{tabular}{lcccr} 
		\hline
		Section & Notes & Simulation names& Figures & Main conclusions \\
        \hline
        \hline
		\ref{sec:coll} & Dynamics of fixed size grains  & & \\
        \hline
        \ref{sec:spherical} & Collective versus test particle & SpZ1a01F, SpZ1a01TP & \ref{fig:3Dsim},\ref{fig:2} & Pebble-rich finger sedimentation \\
        \ref{sec:pert} & Non-spherical initial condition & NonSpZ1a01 & \ref{fig:pert6} & Rayleigh-Taylor mushroom heads \\
        \ref{sec:psize} & Various grain sizes & SpZ1a01 --SpZ1a100 & \ref{fig:tsed2} & Finger sedimen. is rapid yet gentle \\
        \ref{sec:Zdep} & Dependence on metallicity & SpZ05a01 -- SpZ4a01 & \ref{fig:tsedZ} & Instability grows faster at larger $Z$\\
        \ref{sec:geometry} & Various geometries & BulletZ1a01, SlabZ1a01 & \ref{fig:bullet}, \ref{fig:Slab} & Non-spherical IC speed up sedimen.\\
        \ref{sec:origin} & Sinusoidal perturbations & SinZ1a01N5e4 -- SinZ1a01N32e5 & \ref{fig:pert_t0}--\ref{fig:pert_growth} & The instability is Rayleigh-Taylor\\
        \ref{sec:inst_decay} & Instability decay for large $a$ & SinZ1a01N8e5--SinZ1a26N8e5 & \ref{fig:pert_decay},\ref{fig:pert_vs_a} & Instability is suppressed at large $a$\\
		\ref{sec:core} & Core collapse & SpZ1a01N8e5--SpZ1a100N8e5 & \ref{fig:Density_structure}, \ref{fig:Core_dens} & Core collapse occurs at $t=t_{\rm sed}$\\
        \hline
        \ref{sec:growth} & Variable grain size \\
        \hline
		\ref{sec:exp_growth} & Grain growth and fragmentation & WedgeZ2a01V1, ... & \ref{fig:Frag_a100}, \ref{fig:Frag_a01} & Grain size evolution is rapid  \\
        \ref{sec:vap} & Grain vaporisation in hot clumps & WedgeZ2a10Tc100, ... & \ref{fig:evap}& Solid vs fuzzy core formation \\
        \hline
       \ref{sec:loading} & Effects of pebble weight on the clumps \\
       \hline
  		\ref{sec:loading} & Uniform idealised metal loading & UniZload & \ref{fig:test1} & Agreement with theory\\
       \ref{sec:dark} & Dark Collapse          & DarkCollapse & \ref{fig:Dark_Collapse_Map}, \ref{fig:Dark_Collapse}&  
       	Clump collapse due to pebble weight \\
       %6.2 & Solid core accretion feedback & & & Massive cores destroy host clumps\\
       %6.3 & Dependence on pebble abundance & & & Pebbles may dictate gas clump fate\\
	\end{tabular}
\end{table*}

%We start our analysis with simulations in which all the grains are of equal fixed size (\S \ref{sec:coll}), but will relax this assumption in \S \ref{sec:growth}. The material density of grains is set to $\rho_a = 5$~g cm$^{-3}$, appropriate for a mixture of silicates and iron. The parameters of the gas clump are the same as the 'fiducial' clump. 

%These choices are not important for the fixed grain runs, but will be important in actual astrophysical applications. In \S \ref{sec:growth} and \S \ref{sec:loading} we shall consider clumps of different central temperatures and icy grains as an important contrast to rocky grains. %For rocks  and Fe the vaporisation temperature is $\sim 1300$~K, while for water ice it is only $\sim 150$~K for typical gas densities in the clump \citep[e.g.,][]{HelledEtal08}. Further, grain vaporisation should also include the latent heat of grain vaporisation term in the gas energy balance equation \citep[e.g.,][]{Nayakshin15c}. We leave these complications for a future paper. 

Prior to pebble immersion into the clump,  the clump is relaxed for many dynamical times, keeping the polytropic constant $K$ in $P = K\rho^{\gamma}$ fixed as a global constant. 
%During this relaxation procedure SPH particle velocities become much smaller than the local sound speed, ensuring that the clump is very close to a polytropic sphere in hydrostatic balance. 
After grains are introduced inside the cloud, the polytropic constant $K$ is no longer kept a global constant, allowing it to evolve independently for each SPH particle, e.g., increase due to gas-dust frictional heating or gas shocks via artificial viscosity prescription (although the latter does not really occur in the tests presented below as gas motions are subsonic for parameters choices made).

The number of SPH particles used is $N_{\rm sph}=0.8$ Million for most of the simulations below, but is varied in some tests  from $N_{\rm sph}=5\times 10^4$ to 3.2 Million.  The total number of pebble particles used for most of the tests in this paper is set to 40\% the SPH particle number, unless specified differently. At the beginning of the simulations ($t=0$), pebbles of a specified size and of total mass $Z M_0$, where $Z \ll 1$, are deposited in the outer regions of the clump.  

%\subsection{Simulations table}\label{sec:table}

Table 1 lists for convenience all of the simulations presented in the paper. Each row shows a corresponding section, the main effects being investigated, simulation names, figures and main conclusions arising from the simulations.  
%In \S \ref{sec:coll} dynamics of fixed grains at a relatively low pebble abundance is investigated; in \S \ref{sec:growth} pebble growth and vaporisation are investigated; in \S \ref{sec:loading} results are extended on the situations in which pebble abundance is significant enough to cause significant clump contraction and eventual collapse due to H$_2$ molecule dissociation. The simulation names bear a hint on the parameters used in the corresponding runs. For example, SpZ05a01 is a simulation with a spherically symmetric initial condition, with pebble abundance Z = 0.5\% and fixed pebble size of $a=0.1$~cm.

\section{Collective effects}\label{sec:coll}

\subsection{Spherically symmetric initial conditions}\label{sec:spherical}

\begin{figure*}
\includegraphics[width=0.32\textwidth]{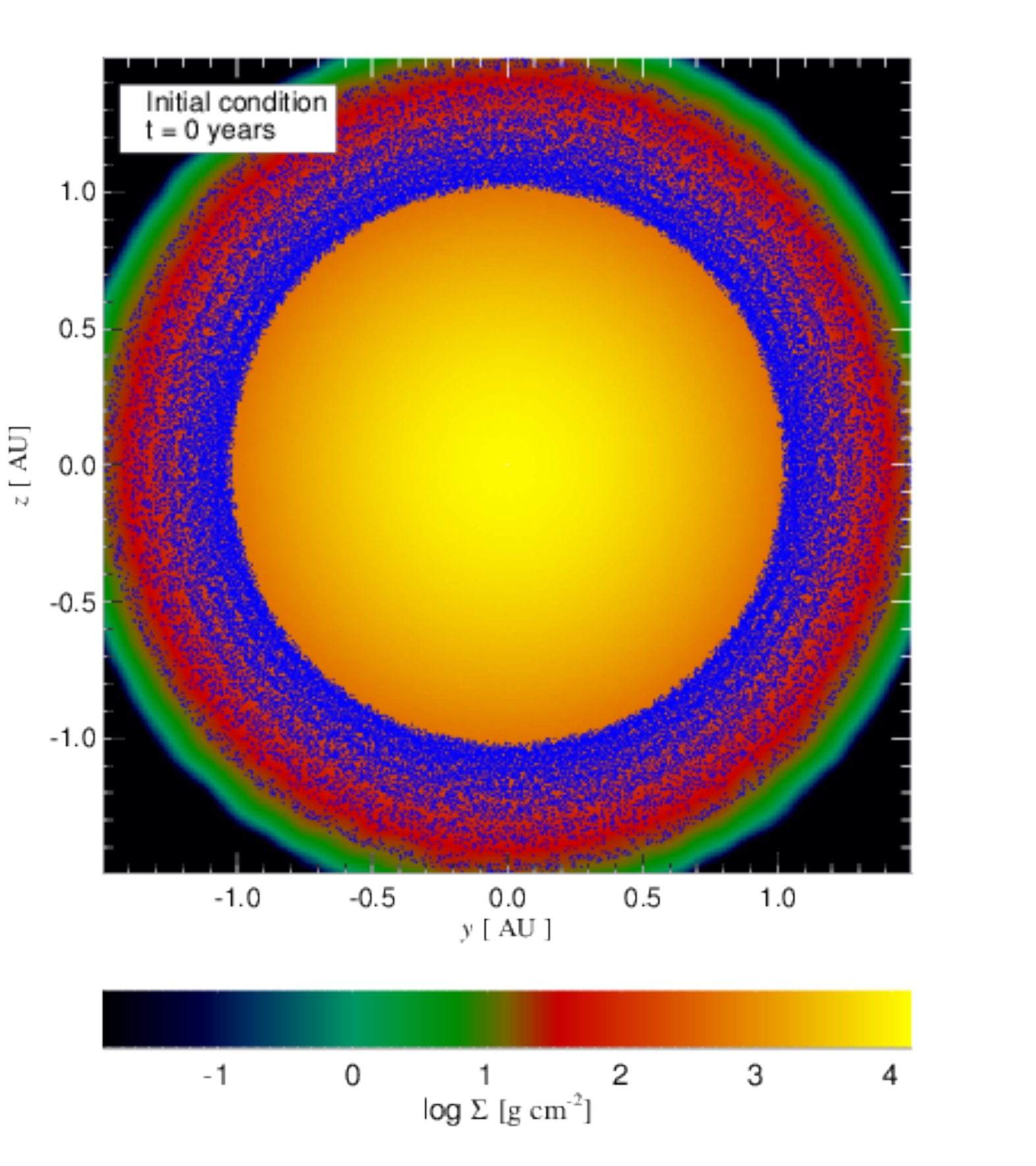}
\includegraphics[width=0.32\textwidth]{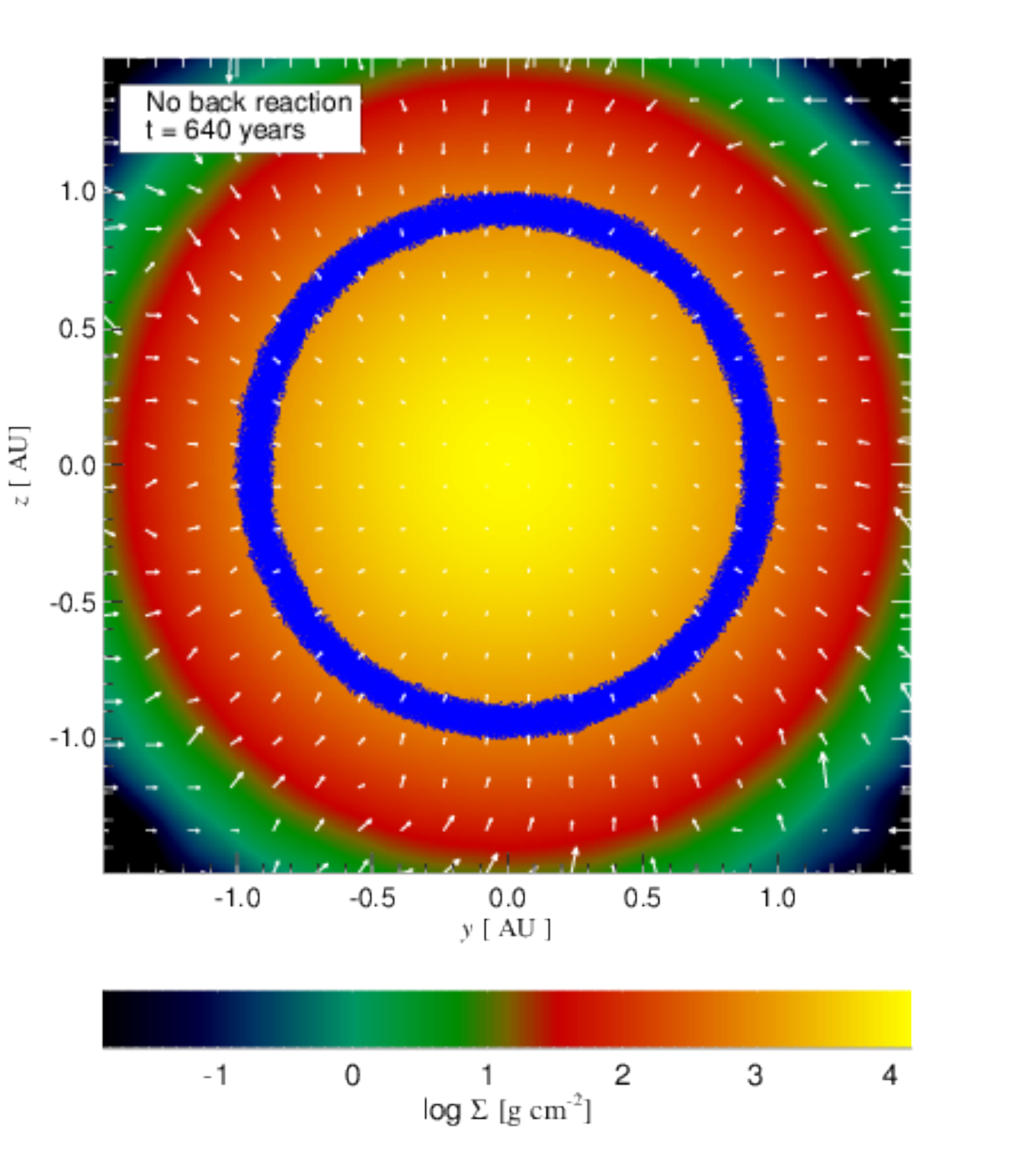}
\includegraphics[width=0.32\textwidth]{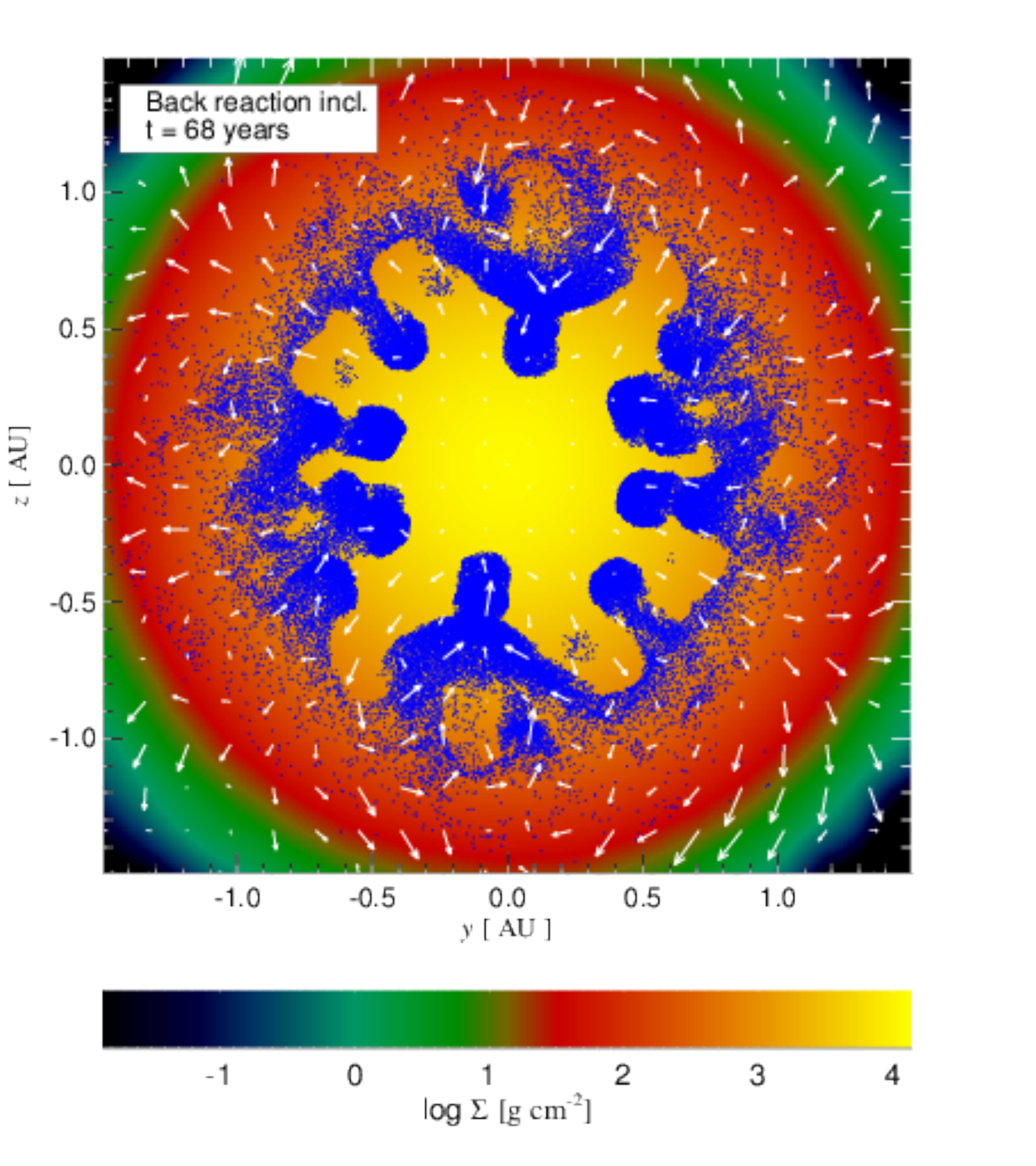}
\caption{Left: The initial condition for spherically symmetric simulations presented in \S \ref{sec:coll}. A gas clump is loaded in the outer regions with $Z=0.01$ of pebbles with size $a=0.1$~cm. Middle and Right: Same projections but at later times for the simulations that include back reaction force onto gas (right, SpZ1a01F) and not (middle, SpZ1a01TP).}
 \label{fig:3Dsim}
 \end{figure*}

In this section pebbles mirror the gas density profile for the outermost 10\% of the SPH particles. Pebble size is fixed at $a=0.1$ cm, $Z=0.01$, and the SPH particle number is $N_{\rm sph} = 0.8$ Million.

\begin{figure}
\includegraphics[width=0.4\textwidth]{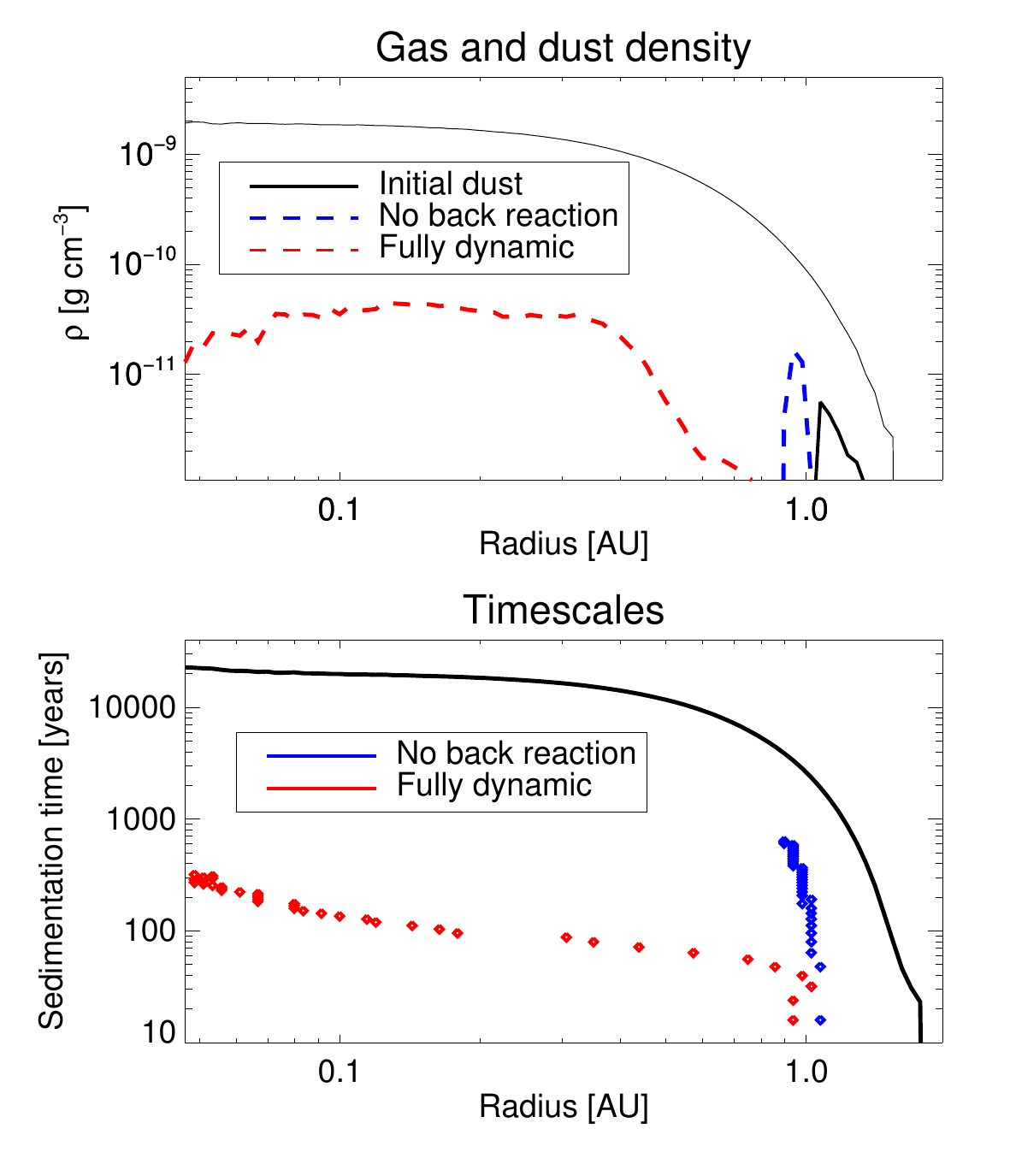}
\caption{{\bf Top}: Density profiles versus distance from the clump centre for two simulations presented in \S \ref{sec:spherical}. Black curves show initial condition for gas (thin line) and dust (thick line). The colored dashed curves are for dust profiles at later times. {\bf Bottom:} Sedimentation time as a function of distance from the centre (black curve), and the time actually taken for the sedimentation front to reach given radius (red and blue symbols). See text for more detail.}
 \label{fig:2}
 \end{figure}

The left panel of Fig. \ref{fig:3Dsim} shows the resulting initial configuration of the pebble-loaded gas clump.  The background image shows a slice of the gas projected density map between -0.25~au~$\le x \le$~0.25 au. The blue coloured dots show the positions of individual pebble particles. The white arrows show the map of the gas velocity field (set to zero identically at $t=0$). The initial radial density profiles of dust and gas can be also seen in Fig. \ref{fig:2}.
 
Simulation SpZ1a01F (an animation of the simulation is available in the online supplementary material) is ran as described in \S \ref{sec:ic}, whereas SpZ1a01TP is identical to it in every aspect except the frictional back reaction force on the gas is turned off.  Pebble particles are hence treated in the test particle approximation  in simulation SpZ1a01TP (at $Z=0.01$ the gravitational force on the gas from the pebbles is quite weak).

The middle and the right panels of Fig. \ref{fig:3Dsim} show the projected gas densities, the velocity field and pebble particle locations. The two projections are made at different times, $t= 640$ and $t=64$ years for the middle and the right panels, respectively.
This is done because the pebble distribution evolves much faster in the run SpZ1a01F. We observe that pebbles sediment in a purely radial, spherically symmetric and laminar fashion in the middle panel. In contrast, both pebbles and gas show strongly non-radial flows in the right panel. Grains arrive in the inner part of the clump much sooner in the fully self-consistent simulation.

Fig. \ref{fig:2} which presents the gas and dust density profiles averaged on concentric shells in the top panel. The solid curve shows the initial gas density profile. The dashed black curve shows the same for pebbles at time $t=0$ (corresponding to the left panel of Fig. \ref{fig:3Dsim}), whereas the blue and the red dashed curves show pebble profiles at times $t=640$ and $t=320$ (when the simulations were terminated) for the test particle SpZ1a01TP and the fully dynamic SpZ1a01F simulations, respectively. We observe that in the no back reaction case pebbles sediment so slowly that they essentially stall at radius of just smaller than 1 AU, at least on the time scales of these simulations. In contrast, in the fully dynamic simulation most of the pebbles are in the inner $\sim 0.4$~AU part of the clump.

Grain sedimentation time, defined as $R/v_{\rm sed}$ and calculated as described in \S \ref{sec:expect}, is plotted with the solid curve in the bottom panel of Fig. \ref{fig:2}.  
Sedimentation time scale is shortest at large $R$ because $\rho(R)$ is rapidly decreasing with increasing $R$ (cf. the top panel of Fig. \ref{fig:2}). This also explains why the grain shell becomes narrower with time in the no-back-reaction simulation: pebbles further from the centre tend to catch up with those deeper in.

The coloured diamonds in the bottom panel of Fig. \ref{fig:2} show the time on the $y$-axis and the corresponding radial position, $r_{\rm f}$,  of the sedimentation front as a function of time. In the spherical geometry, sedimentation front is defined simply as the innermost radius to which pebbles sedimented at a given time. For the non-spherical geometry we define the sedimentation front as following. At a given time, the average density of pebbles and gas in the planet is calculated on concentric spherical shells. The smallest radius of the shell where the ratio $\rho_{\rm peb}/\rho_{\rm gas}$ exceeds $Z_f = 0.005$ is then defined as the sedimentation front. The resulting function $r_{\rm f}(t)$ does not depend sensitively on the exact value of $Z_f$ provided $Z_f$ is not too large. 

The blue diamonds in the bottom panel of Fig. \ref{fig:2} show evolution of the sedimentation front with time in the test particle simulation SpZ1a01TP. These results are consistent with eq. \ref{tsed_num} for sedimentation of particles in the Epstein regime. If the simulations were ran for longer, the blue diamond sequence would converge onto the solid black curve. The sequence of red diamonds for $r_{\rm f}(t)$ shows clearly that pebbles penetrate into the inner regions of the planet much faster when their effects onto gas dynamics are properly included. 

%The steps in the blue diamonds curve corresponds to the finite binning of the concentric shells on which the position of the sedimentation front is calculated.

%The slight outward motion of the sedimentation front in the beginning of the simulation is due to a moderate radial oscillation of the gaseous clump which starts soon after the beginning of the simulation. The oscillation is excited by the instantaneous deposition of the extra $Z=0.01$ weight of pebbles onto the planet at $t=0$.

\subsection{A non-spherical initial perturbation test}\label{sec:pert}

\begin{figure}
\includegraphics[width=0.225\textwidth]{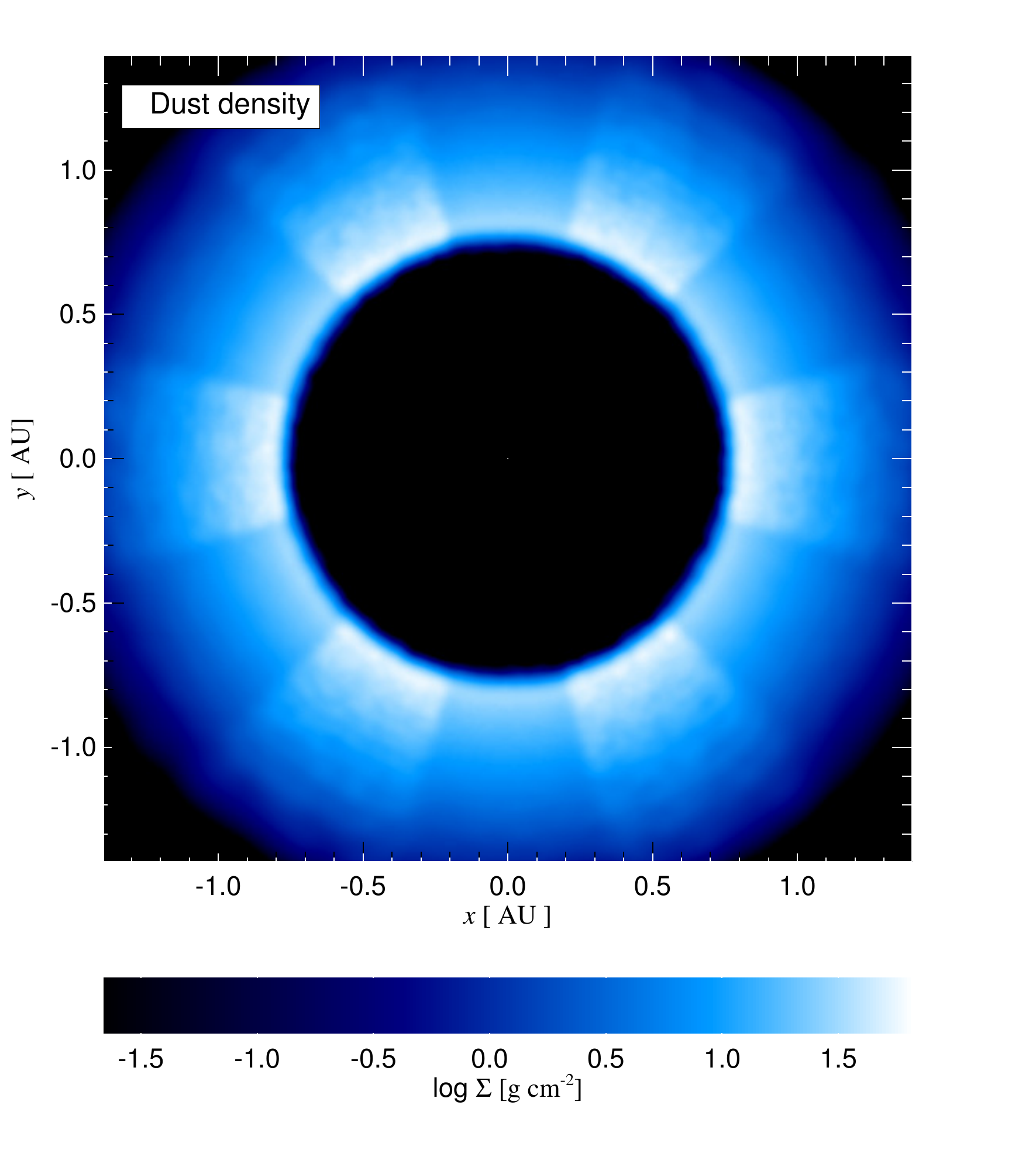}
\includegraphics[width=0.225\textwidth]{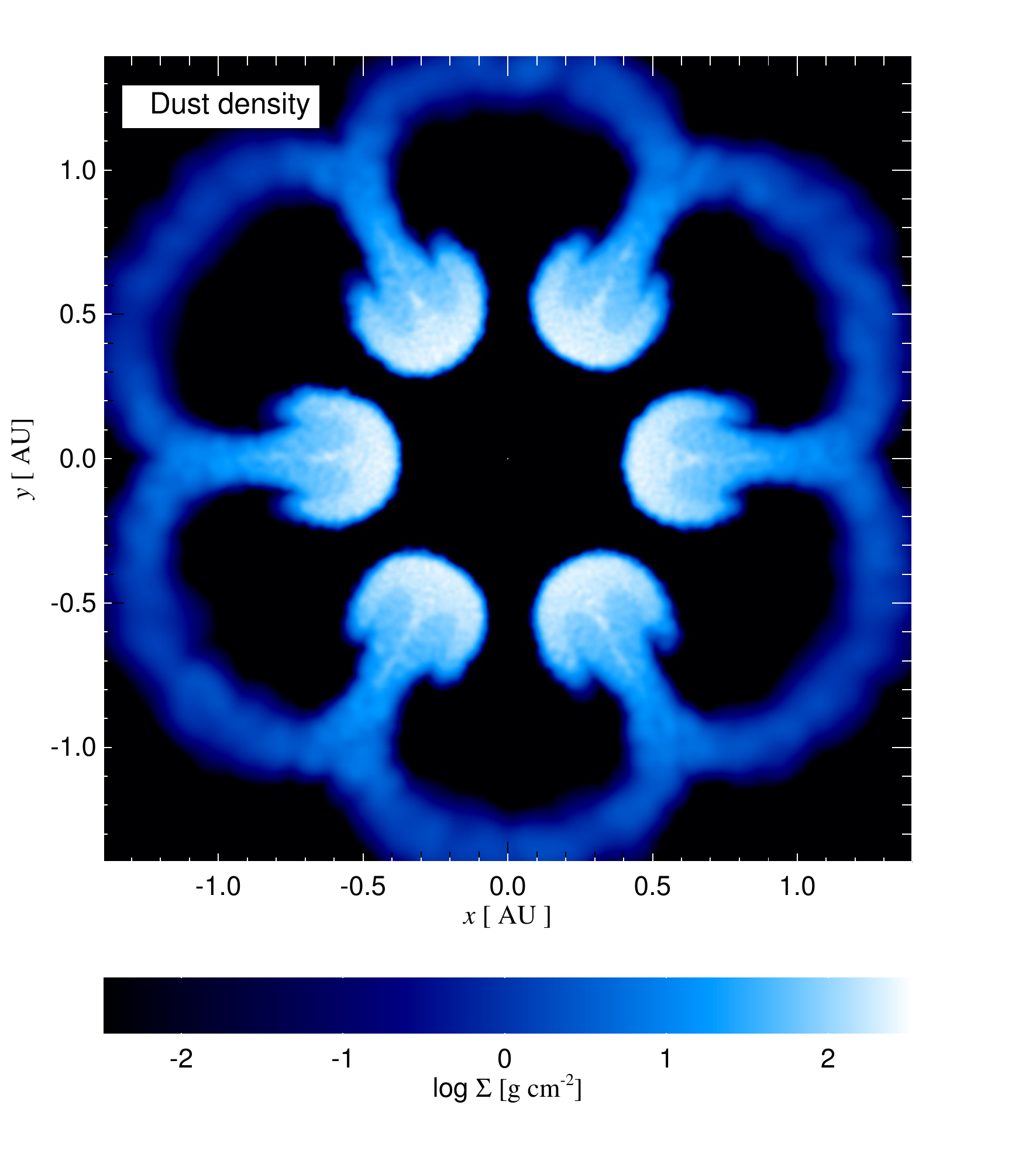}
\includegraphics[width=0.23\textwidth]{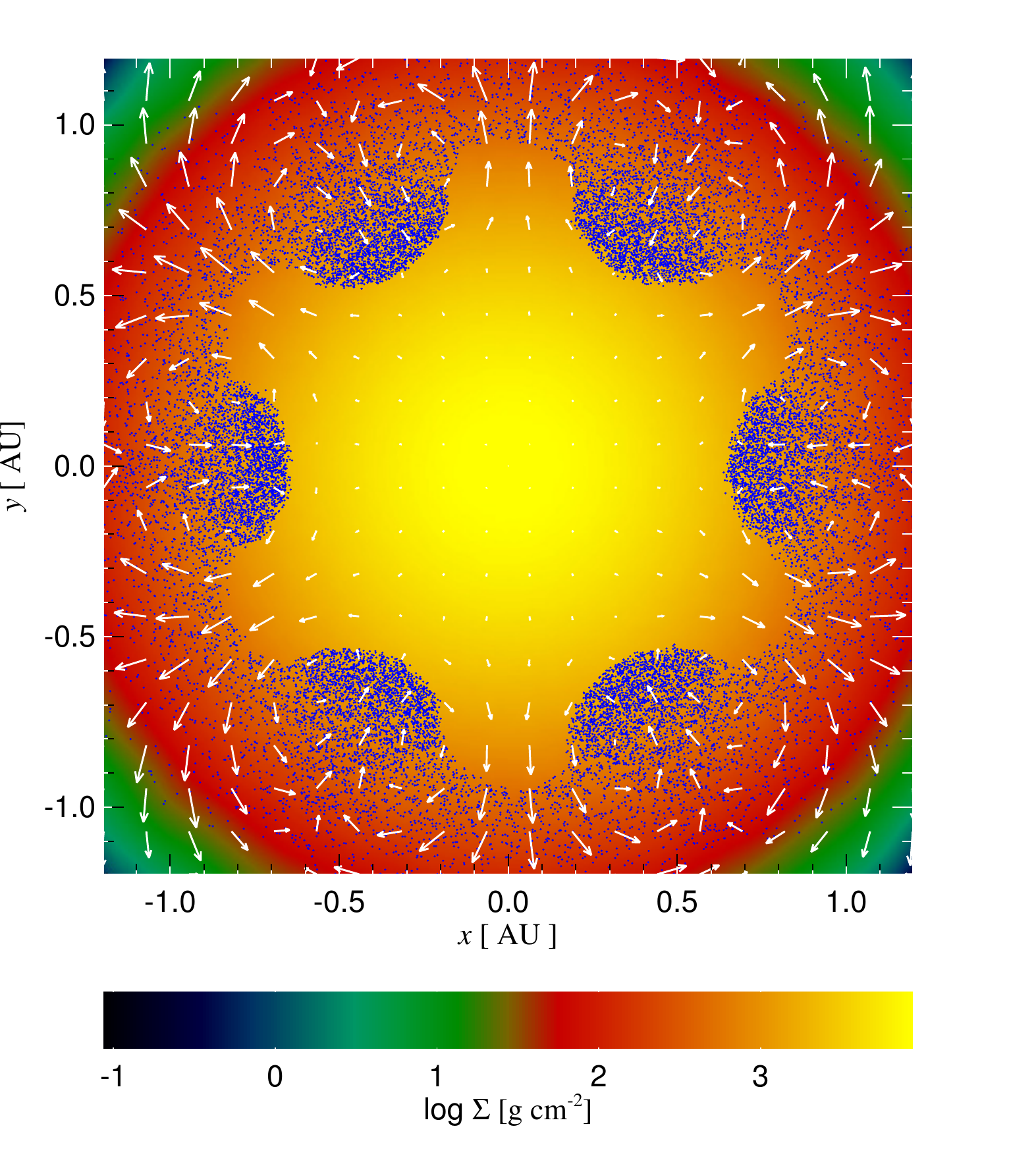}
\includegraphics[width=0.23\textwidth]{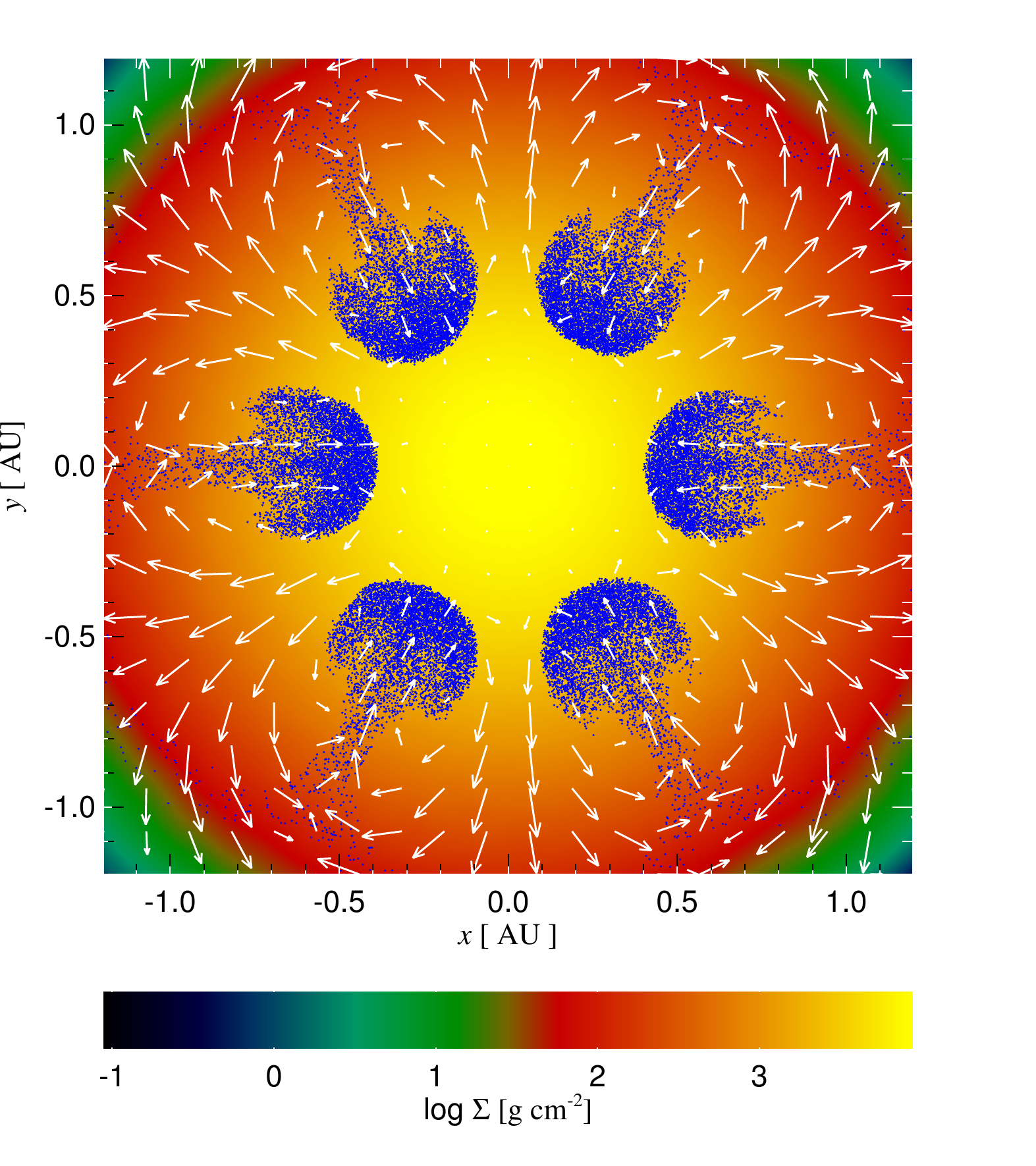}
\caption{Non-spherical initial perturbation test NonSpZ1a01 described in \S \ref{sec:pert}. {\bf Top panels:} A $z$-coordinate slice for the pebble particles at $t=0$ (left panel) and at $t=64$~years (right panel). {\bf Bottom panels:} similar projections for the gas (colors) and pebbles (blue points) at an early time (left) and at $t=64$~years (right).}
 \label{fig:pert6}
 \end{figure}

In simulation NonSpZ1a01 the initial radial distribution of pebble particles is the same as that described in \S \ref{sec:spherical}, but the angular distribution differs. Consider spherical coordinates in which $z = R \cos \theta$, $x= R\sin\theta \cos\phi$ and $y = R\sin\theta \sin\phi$.  First, the $3/4$ of the pebble particles are distributed isotropically.
%, so that $-1 \le \cos\theta\le 1$ and $0 \le \phi \le 2 \pi$ are uniform random variables. 
The reminder $1/4$ of pebble particles are then distributed uniformly in $\cos\theta$ only. In the azimuthal angle $\phi$, the distribution is uniform within 6 sectors out of 12 equal size sectors on which the full $0 \le \phi \le 2\pi$ circle is divided. The top left panel of Fig. \ref{fig:pert6} shows the $|z| \le 0.21$~au slice of the initial pebble density field, rendered with 40 dust particle neighbours. The pebble density at $t=0$ in the denser regions is $5/3$ times higher than in the unperturbed regions.

The density and velocity maps of gas and pebbles are presented at times $t= 32$ and $64$ years, respectively, in the bottom panels of Fig. \ref{fig:pert6}. The top right panel shows pebble density field at time $t=64$ years. The dynamics of pebbles and gas is highly correlated. In the regions where pebbles are abundant, gas "settles" together with the pebbles. Pebble-free inner clump regions rise buoyantly up. Note also the mushroom-like heads of the infalling fingers.

\subsection{Dependence on pebble size: rapid, non fragmenting sedimentation}\label{sec:psize}

For this section, the initial conditions and all other settings are the same as those for SpZ1a01F except for pebble size $a$, 
which is varied from $a=0.1$ cm to $a=100$~cm. The respective runs are labeled SpZ1a01 for $a=0.1$~cm through to SpZ1a100 for $a=100$~cm. For simplicity, pebble sedimentation time is now defined as the time it takes the sedimentation front to propagate to radius $R= 0.2$~au. 

\begin{figure}
\includegraphics[width=0.45\textwidth]{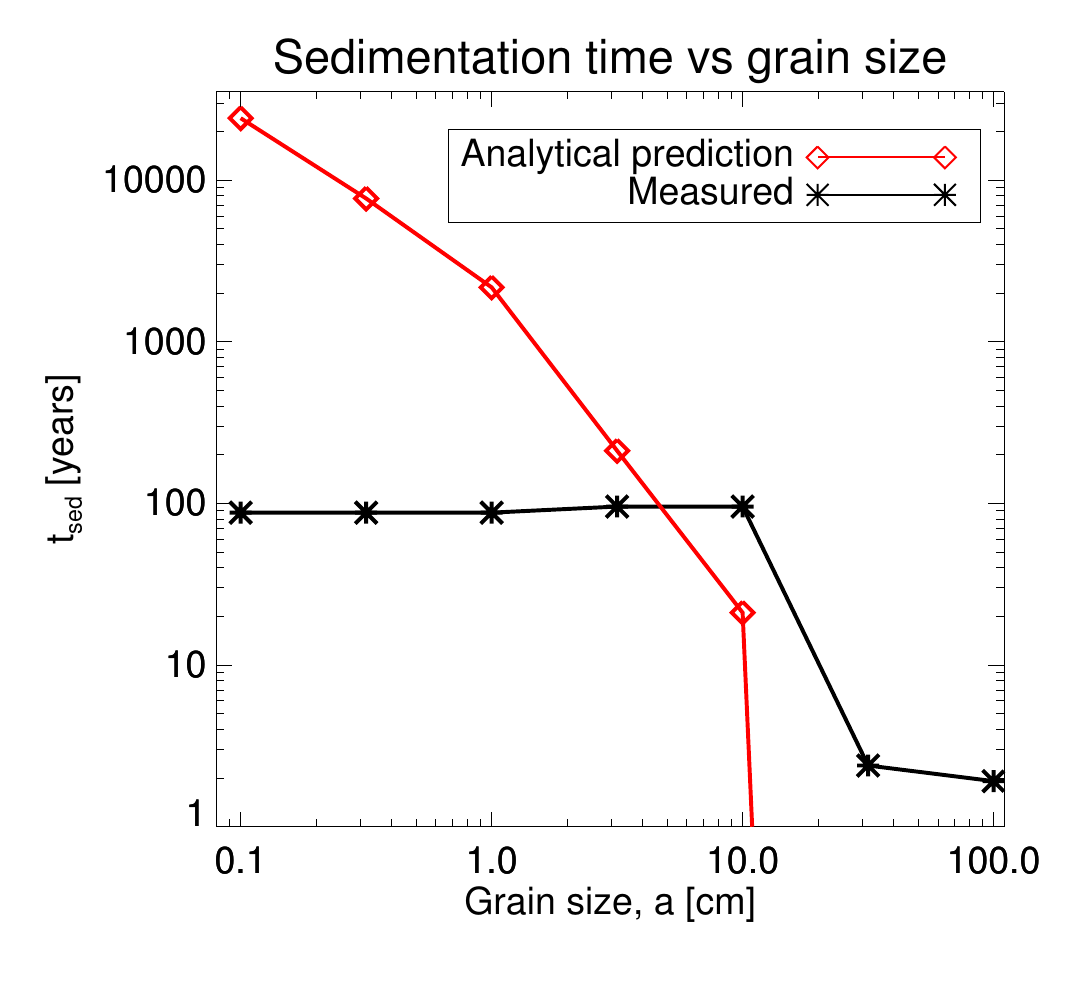}
\caption{Dependence of pebble sedimentation time on the pebble size $a$ for simulations SpZ1a01 -- SpZ1a100. The red curve gives the test particle analytic prediction which is clearly at odds with what is actually happening. Instead, small grains sediment into the clump centre much faster, at a speed independent of grain size since their motion is controlled by the collective effects. Large grains, on the other hand, sediment even faster in the test particle regime, and fall to the centre of the planet almost in free fall.}
 \label{fig:tsed2}
 \end{figure}

The black asterisks in Fig. \ref{fig:tsed2} show sedimentation time versus grain size from the simulations, whereas the red diamonds show the analytical test particle result. According to the later, small grains should take a very long time to sediment. Due to collective effects, these particles sediment much more rapidly. The measured sedimentation time is independent of $a$ for small grains because small grains are well coupled to the gas and co-move with it as it sinks to the centre. This is contrary to the test particle prediction, where sedimentation time should be inversely proportional to $a$ or $a^2$.

The speed with which small pebbles sediment due to collective effects is rather large,
\begin{equation}
v_{\rm sed} \sim \frac{\rm 1 AU}{\rm 100 yr} \approx 50 \; {\rm m s}^{-1}\;.
\label{vsed_coll}
\end{equation}
This is much larger than the break-up speed for most materials. However, since grains of all sizes up to some maximum size (which will be quantified in \S \ref{sec:inst_decay}) move together with the surrounding gas, their relative velocity dispersion is bound to be much smaller than the "macroscopic" sedimentation velocity (eq. \ref{vsed_coll}). Therefore, collective sedimentation speeds up small grain sedimentation inside the clumps while at the same time keeping the process gentle, avoiding violent collisions otherwise expected to fragment pebbles. This means that clumps are an even better environment for making solid cores than found in analytical 1D analysis in \S \ref{sec:ggrowth}.

For grains greater than a few cm, collective effects seem to slow down their sedimentation somewhat. The effect however exists over a small range in grain size only ($a=10$~cm in the figure), and is probably due to random motions caused by the instabilities which make such grains deviate from their otherwise purely radial inward motion. At the largest grain sizes in the figure, $a\ge 30$~cm, pebbles are so large that they sediment at nearly the free fall velocity. The sedimentation time of the largest pebbles in Fig. \ref{fig:tsed2}, $a=30$ and 100 cm, is just a little longer than dynamical time for the clump, $t_{\rm dyn} \sim (R^3/G M_0)^{3/2}\sim 2$ years. The red analytical curve does not take into account this physical limit.

\subsection{Dependence on metallicity of added pebbles}\label{sec:Zdep}

\begin{figure}
\includegraphics[width=0.45\textwidth]{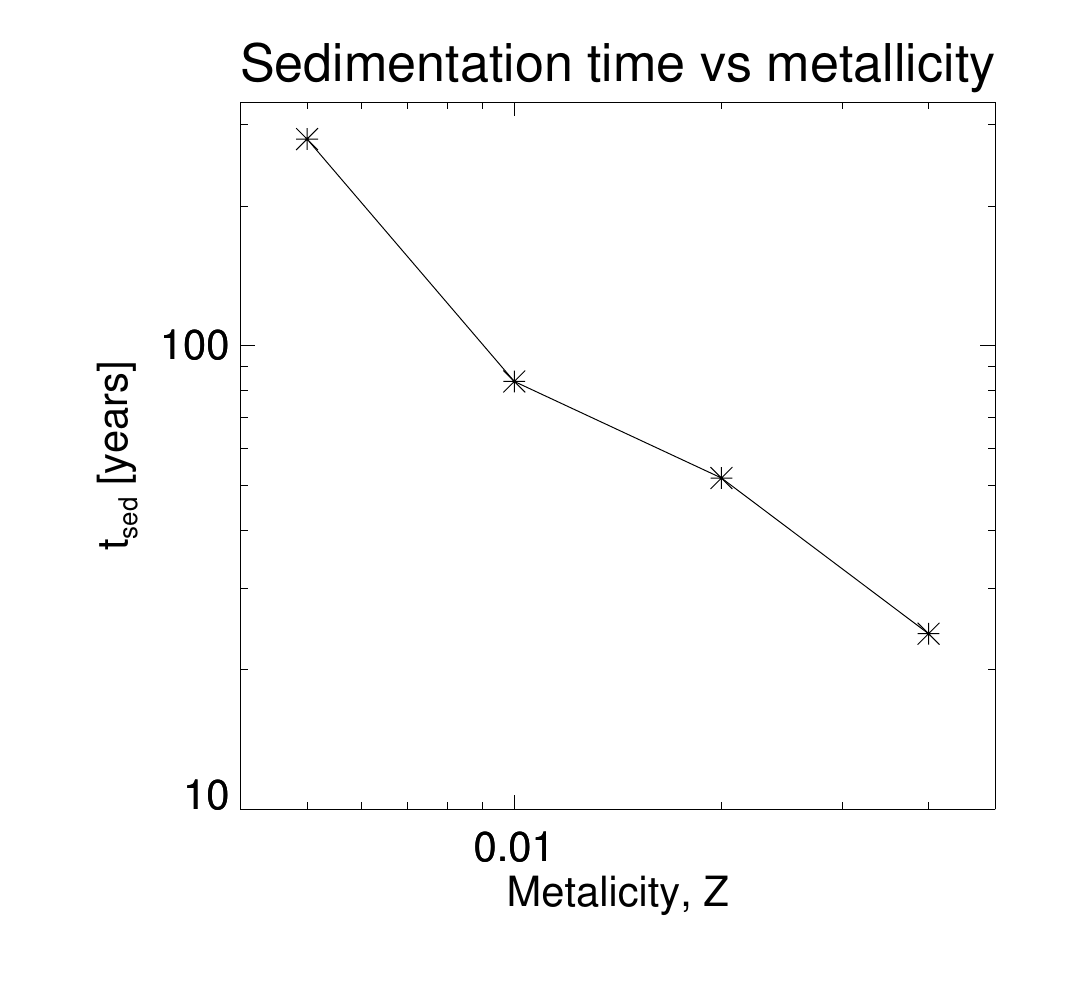}
\caption{Dependence of pebble sedimentation time on the total mass of added pebbles, defined via pebble metallicity $Z$, for spherically-symmetric initial dust configuration and grain size $a=0.1$~cm. Based on runs SpZ05a01 to SpZ4a01}
\label{fig:tsedZ}
\end{figure}

\begin{figure*}
\includegraphics[width=0.3\textwidth]{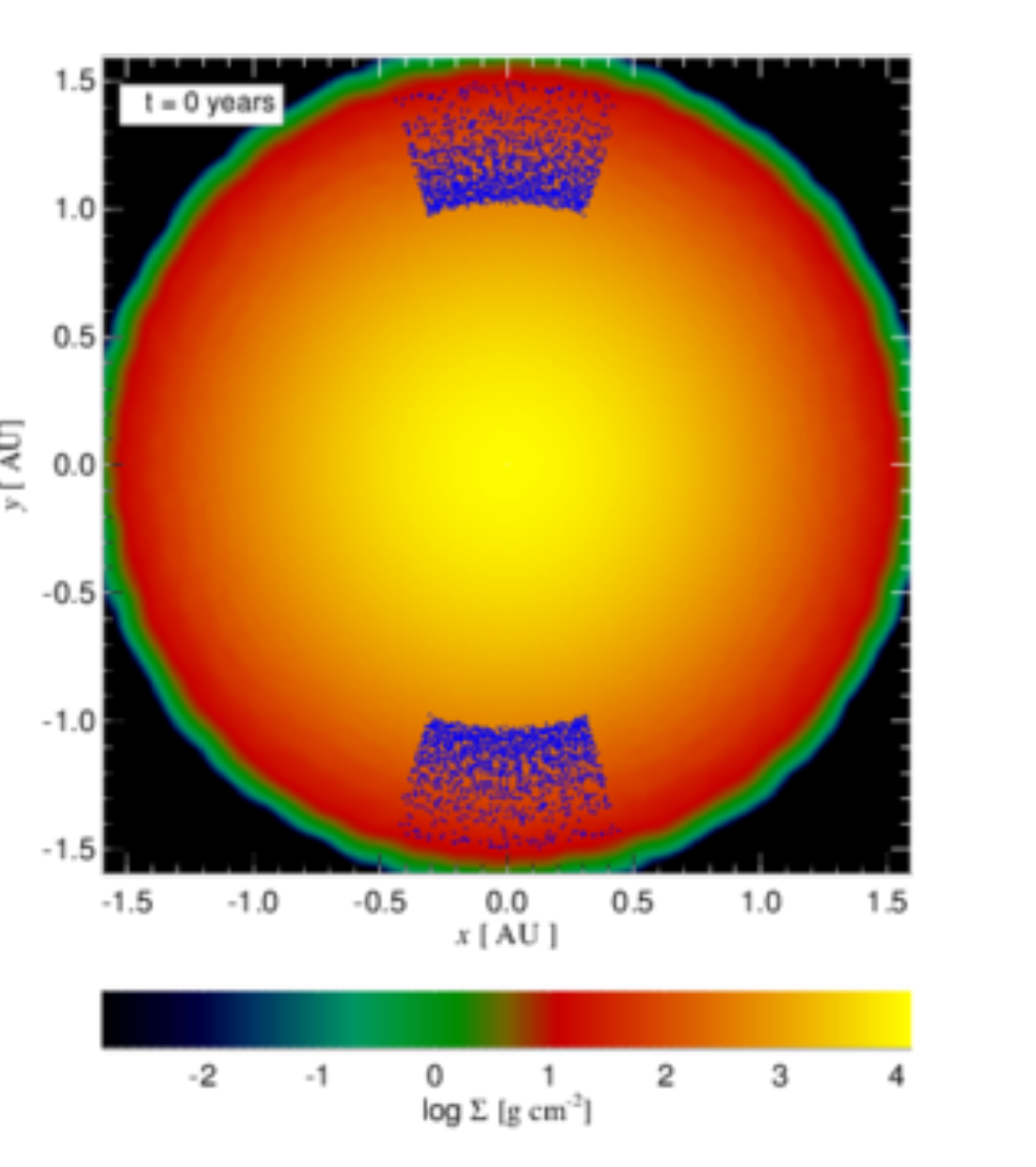}
\includegraphics[width=0.3\textwidth]{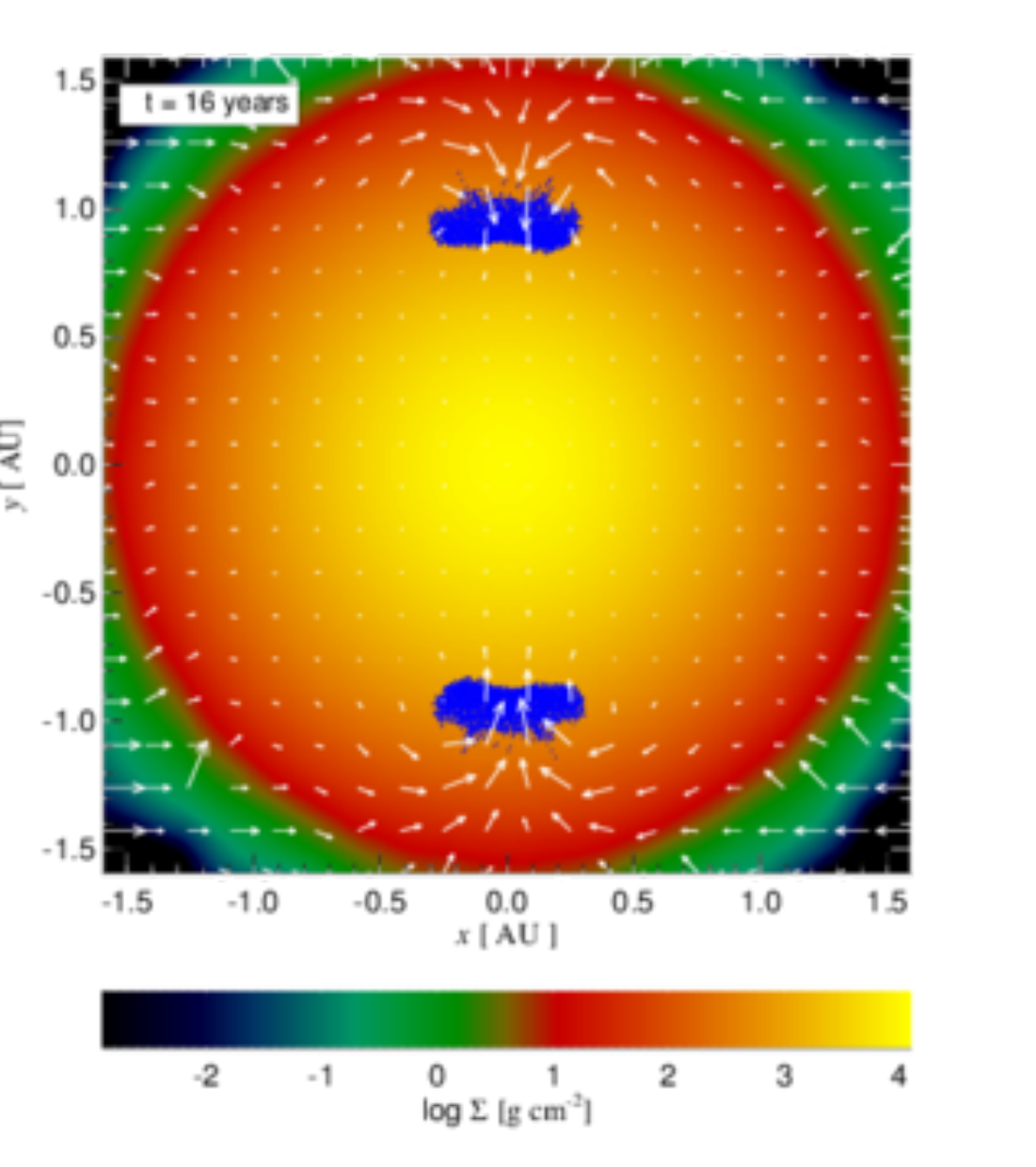}
\includegraphics[width=0.3\textwidth]{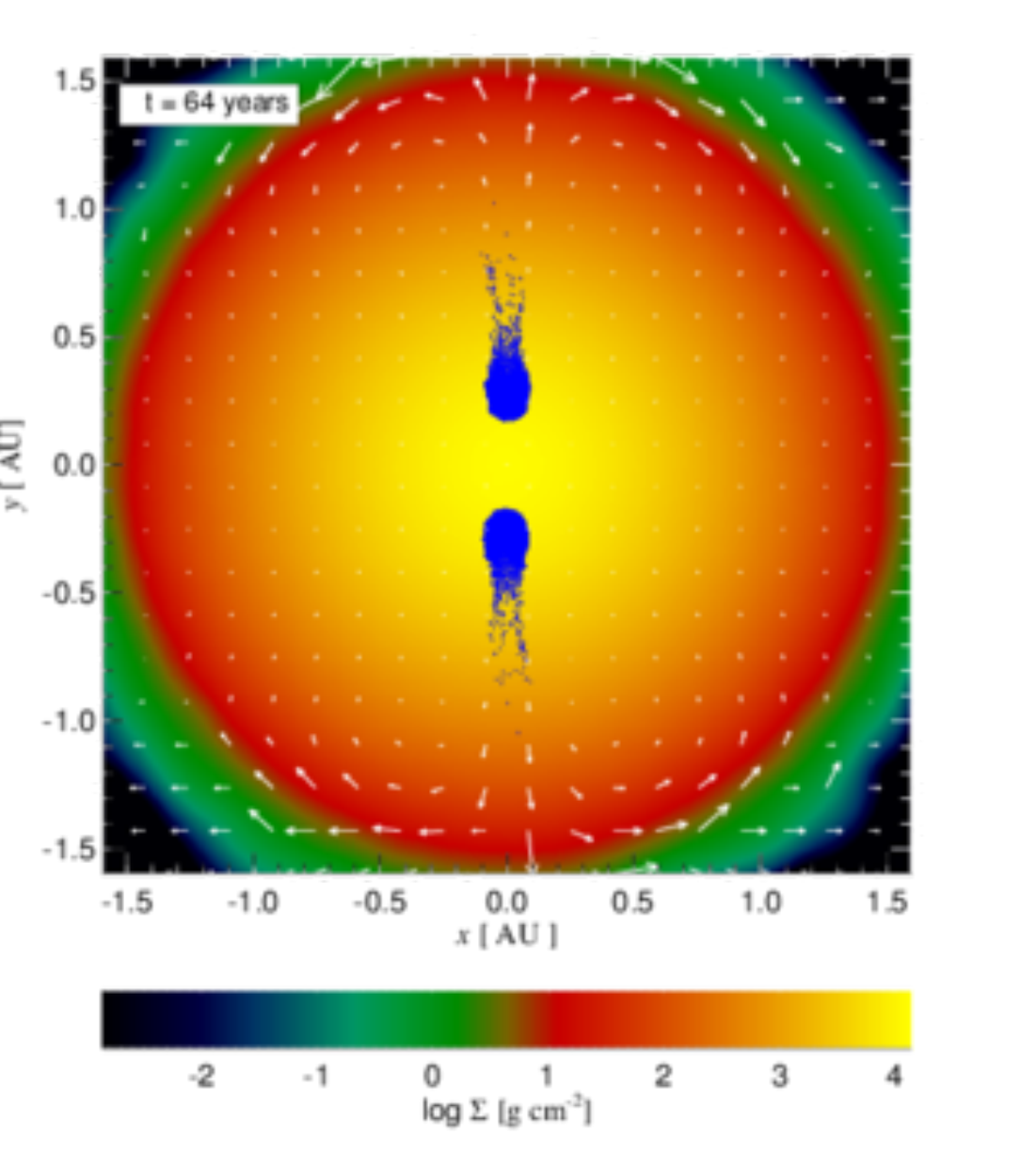}
\caption{Snapshots from a "bullet" simulation BulletZ1a01 with setup identical to SpZ1a01 (see \S \ref{sec:psize}) but with pebbles loaded only in a fraction of the full solid angle. See text in \S \ref{sec:geometry} for more detail.}
 \label{fig:bullet}
 \end{figure*}
 
In this section we use the same setup as in \S \ref{sec:psize}, keeping $a=0.1$~cm, and instead vary the mass of pebbles, covering pebble metallicity from $Z=0.005$ to $Z=0.04$ in steps by a factor of two. The runs are labeled SpZ05a01 to SpZ4a01. Fig. \ref{fig:tsedZ} shows how sedimentation time depends on $Z$. The higher the metallicity, the faster the pebbles sediment down. The scaling is approximately 
$ t_{\rm sed} \propto Z^{-1}$. However, in these tests geometry of the problem was the same for all $Z$. We shall see later on that it is the {\em local} gas metallicity, defined as $Z_{\rm loc} = \rho_{\rm peb}/\rho$, 
%\label{eq:zloc0}
where $\rho_{\rm peb}$ and $\rho$ are the pebble and gas local densities, respectively, rather than the global metallicity $Z$ that determines the sedimentation time scale. 

\begin{figure}
\includegraphics[width=0.45\textwidth]{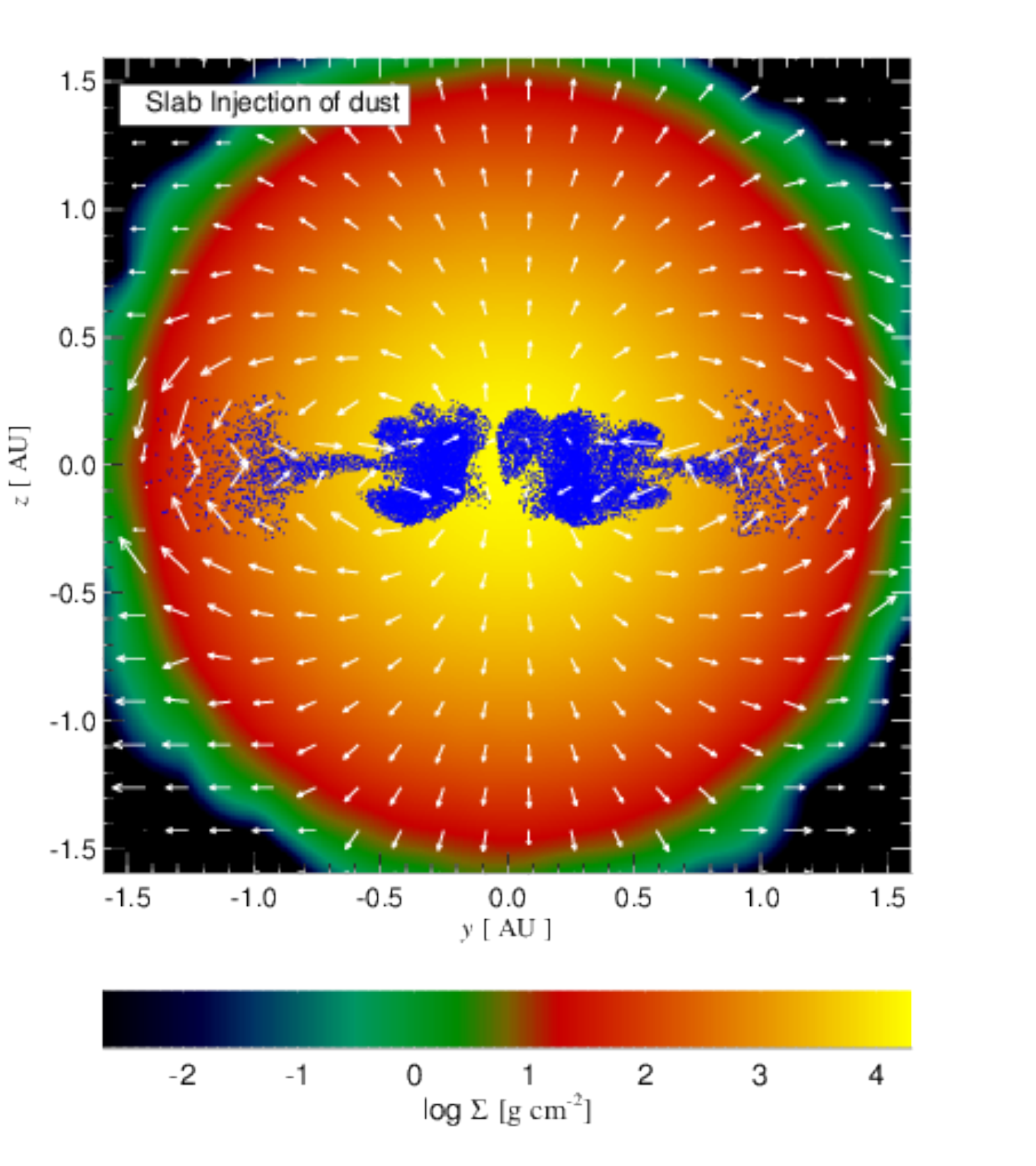}
\caption{Simulation SlabZ1a01 in which pebbles are injected in a slab rather than in a sphere, and at regions of gas density below $10^{-10}$~g/cm$^3$. The snapshot corresponds to time $t=35$~years. See text for more detail.}
\label{fig:Slab}
\end{figure}

\subsection{Different geometries}\label{sec:geometry}

Deposition of pebbles into a gas clump from the parent protoplanetary disc is far from a spherically symmetric process. Pebbles may be entering the clump mainly along the disc midplane, and could be further constrained to streams by the spiral density waves \citep{BoleyDurisen10}. For the following "bullet" simulation, BulletZ1a01, the numerical parameters and setup are all the same as in simulation SpZ1a01, except that pebbles are deposited only in the two fragments of the full solid angle, cut out by the conditions $|z|/R \le 0.3$ and $|x|/R \le 0.3$. Pebble particle mass in this simulation is the same as in SpZ1a01, so that the local pebble metallicity $Z_{\rm loc}$ is also the same, but only in the regions loaded with pebbles. 
%The global pebble metallicity of simulation BulletZ1a01 is obviously much smaller, $Z \approx 6 \times 10^{-4}$ compared to $Z=0.01$ in SpZ1a01.

Initially (see fig. \ref{fig:bullet}), pebbles fall in radially, as in SpZ1a01. When test particle sedimentation stall at a higher gas density, a pancake like patch of pebbles develops. Collective effects then develop and re-shape the patch into a bullet-like formation. Qualitatively, the dynamics of the bullet is fairly similar to the dynamics of one of the dense fingers from simulation SpZ1a01. The sedimentation time for the bullet is a little shorter than in simulation SpZ1a01, around $70$ years versus 100 years. This proves that it is the local pebble metallicity, $Z_{\rm loc}$, rather than the global clump metallicity enhancement, $Z$, that controls the rate at which pebble-rich material settles into the centre.

%It is also interesting to ask whether the instabilities would develop if pebbles are deposited in a continuous fashion, as expected for a realistic clump embedded into a parent gas disc \citep[e.g., see][]{HN18}, rather than loaded instantaneously. 

In simulation SlabZ1a01 pebbles are injected at a constant rate in a disc-like configuration. To enable that, only SPH particles that satisfied the following conditions were allowed to hatch new dust particles: (a) gas density at the particle location is below $10^{-10}$~g/cm$^3$, and (b) the $z$ coordinate of the particle satisfies $|z|< 0.3$~AU. The rate of pebble particle creation by any SPH particle satisfying this condition is set to $t_{\rm birth}^{-1}$, where  $t_{\rm birth}= 8$ years. In practice, at every SPH particle time step $\Delta t$, a uniform random variable is $0 \le \xi\le 1$ is drawn, and a new pebble particle created if $\xi < \Delta t/t_{\rm birth}$. Newly born Pebbles have mass of 0.1 SPH particle mass, and they are injected with position and velocity equal to that of the SPH parent.

The results of the test are fairly similar to previously presented ones except here the instability develops only when a significant amount of pebbles is injected into the clump. Fig. \ref{fig:Slab} shows one snapshot for this simulation at time $t=35$~years. The total pebble metallicity of the gas clump at this time is $Z\approx 0.015$. The inner edge of the dust disc develops instabilities  and rains down via finger-like structures. Pebble-free gas tends to stream away from the dust disc upwards, perturbing the fingers away from the symmetry plane somewhat. There is also a back flow of gas to replenish pebble-enriched gas in the midplane that sank in together with pebbles. 

%One cursory but potentially significant note arising from the assymetric initial condition simulations presented in this section relates to radiative cooling, -- this can be a subject of a future paper.

\subsection{On the nature of the instability}\label{sec:origin}

Rayleigh-Taylor (RT) instability occurs when a denser liquid is on top of a less dense one in a gravity field with a downward acceleration $g$. For 1D sinusoidal perturbations, the RT instability grows exponentially, with the growth time scale
\begin{equation}
t_{\rm RT} = \left(2 \pi {\lambda \over g} {\rho_2 + \rho_1 \over \rho_2 - \rho_1}\right)^{1/2}\;,
\label{trt00}
\end{equation}
where $\rho_2$ and $\rho_1$ are the gas densities of the heavy and the light liquids, respectively, and $\lambda$ is the wavelength of the perturbation \citep[e.g.,][]{Drazin02}. Eq. \ref{trt00} can be used to verify performance of numerical codes \citep[e.g.,][]{CalderEtal02}.

\begin{figure}
\includegraphics[width=0.47\textwidth]{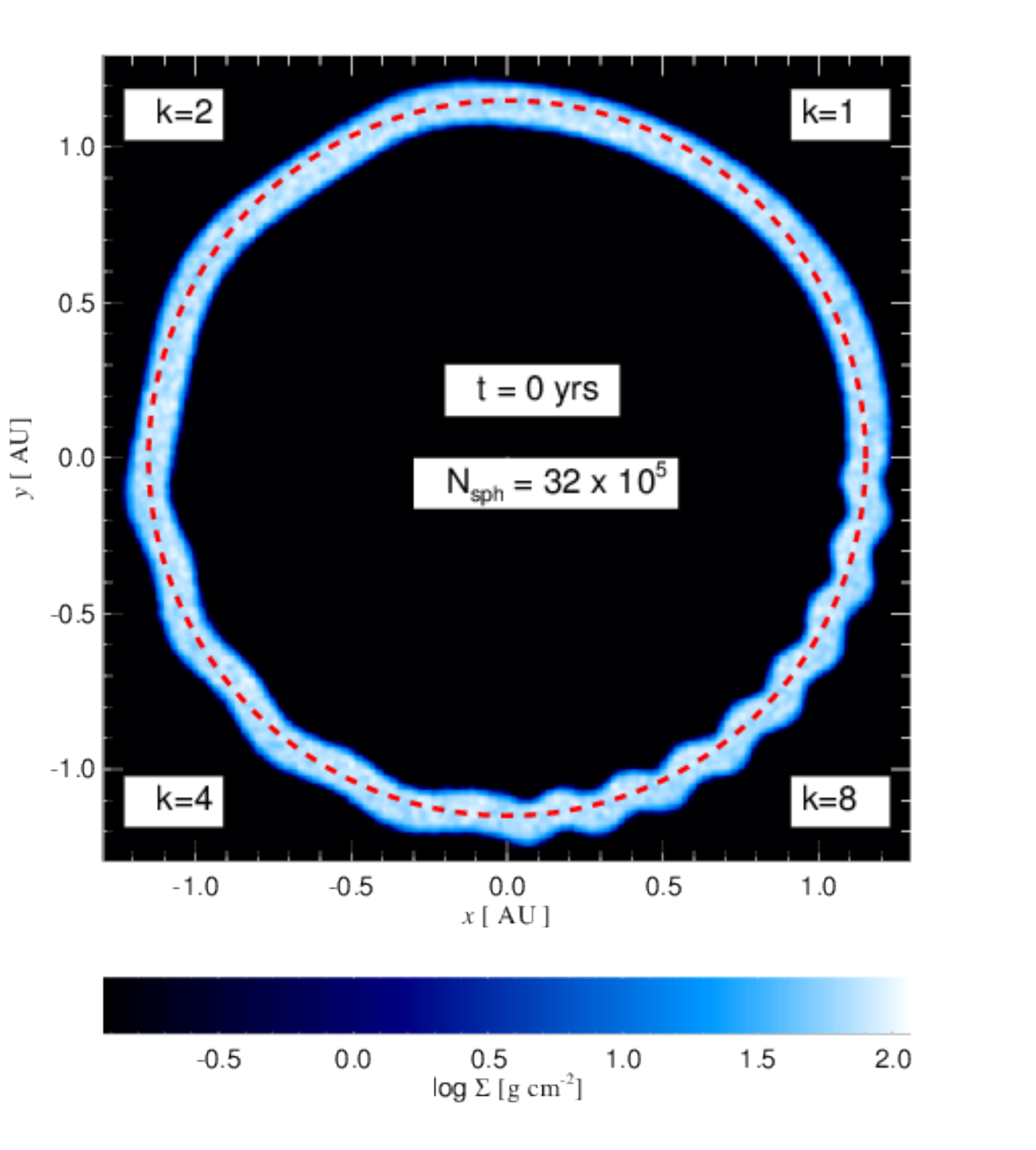}
 \caption{Pebble particle density for the initial condition in the sinusoidal perturbation runs SinZ1a01N5e4 to SinZ1a01N32e5 discussed in \S \ref{sec:origin}. The case with the highest SPH resolution is shown here. The red curve is a circle with radius 1.1 AU.}
 \label{fig:pert_t0}
 \end{figure}

In this section we start with a pebble-free gas clump and then add pebbles in a shell with a specific sinusoidal perturbation to then compare the numerical results to eq. \ref{trt00}. First, we create a spherical shell of pebbles uniformly filling a concentric shell between radii $R_{\rm pert} -\Delta R_{\rm pert}$ and 
$R_{\rm pert} +\Delta R_{\rm pert}$, where $R_{\rm pert} = 1.1$ AU and $\Delta R_{\rm pert} = 0.1$~AU. 
We discard from the shell the regions outside $-1/4 \le \cos \theta \le 1/4$ since we focus our analysis near $z=0$ to make the problem approximately 2D as eq. \ref{trt00} assumes. Then a radial position shift in the pebble location is made, $R \rightarrow R + \delta R(\phi)$, where $\delta R(\phi)$ is a function of the azimuthal angle $\phi$:
\begin{equation}
\delta R(\phi) =  \delta R_0 \sin \left[\frac{2(\phi-\phi_n)}{k \pi} \right]\;,
\label{phi-pert}
\end{equation}
where $\delta R_0 =0.02$~AU, $\phi_n = (\pi/2) (n-1)$ and $k = 2^{n-1}$. $n$ here is the quadrant number in the azimuthal angle $\phi$, running from 1 to 4. This way we test a range of wavelengths in a single simulation, to save numerical costs. The first quadrant of azimuthal angle $\phi$, $0 \le \phi \le \pi/2$, $n=1$, contains exactly one full phase (oscillation) of the sine-wave perturbation. 
The next one, $n=2$, $\pi/2 \le \phi \le \pi$, contains exactly two; the third contains four and the last, $(3/2) \pi \le \phi \le 2\pi$, contains eight. Fig. \ref{fig:pert_t0} shows the $xy$-plane projection of a thin slice ($ |z| \le 0.05$ AU) of the pebble density distribution at $t=0$ created in this way. The red dashed curve shows the circle $R=R_{\rm pert}$. The wavelength of the perturbation in the four quadrants is 
\begin{equation}
\lambda_k =  \frac{\pi R_{\rm pert}}{2 k}\quad \text{ for }\;  k = 1, 2, 4, 8
\label{lambda_k}
\end{equation}

We run this initial condition with varying SPH particle numbers in steps of a multiplicative factor of $4$, from the minimum of $N_{\rm sph}= 0.05$ Million to the maximum of $N_{\rm sph} = 3.2$ million. The number of pebble particles for these runs is always set at 10\% of the SPH particle number, the total clump metallicity of the added pebbles is $Z=0.01$, and the grain size is $a=0.1$ cm. The minimum resolvable length scale of SPH simulations is roughly equal to the SPH smoothing length, $h_{\rm sml}$, over which all particle quantities are averaged \citep{Lucy77,Monaghan92}. For GADGET in particular \citep{Springel05},
\begin{equation}
h_{\rm sml} = \left[\frac{3 n_{\rm nb} m_{\rm sph}}{4\pi \rho}\right]^{1/3} \propto \; N_{\rm sph}^{-1/3}\;,
\label{hsml}
\end{equation}
where $n_{\rm nb} = 40$ is the number of SPH neighbor particles used here; the last relation comes from the fact that mass of an SPH particle is $m_{\rm sph} = M/N_{\rm sph}$, where $M$ is the total gas mass of the clump. 

%A simple prediction is then that at a given $N_{\rm sph}$, the instability should be growing the fastest in the quadrant in which $\lambda_k \approx h_{\rm sph}$, because larger scales grow slower (cf. eq. \ref{trt00}), whereas smaller scales are insufficiently resolved. 

\begin{figure*}
\includegraphics[width=0.47\textwidth]{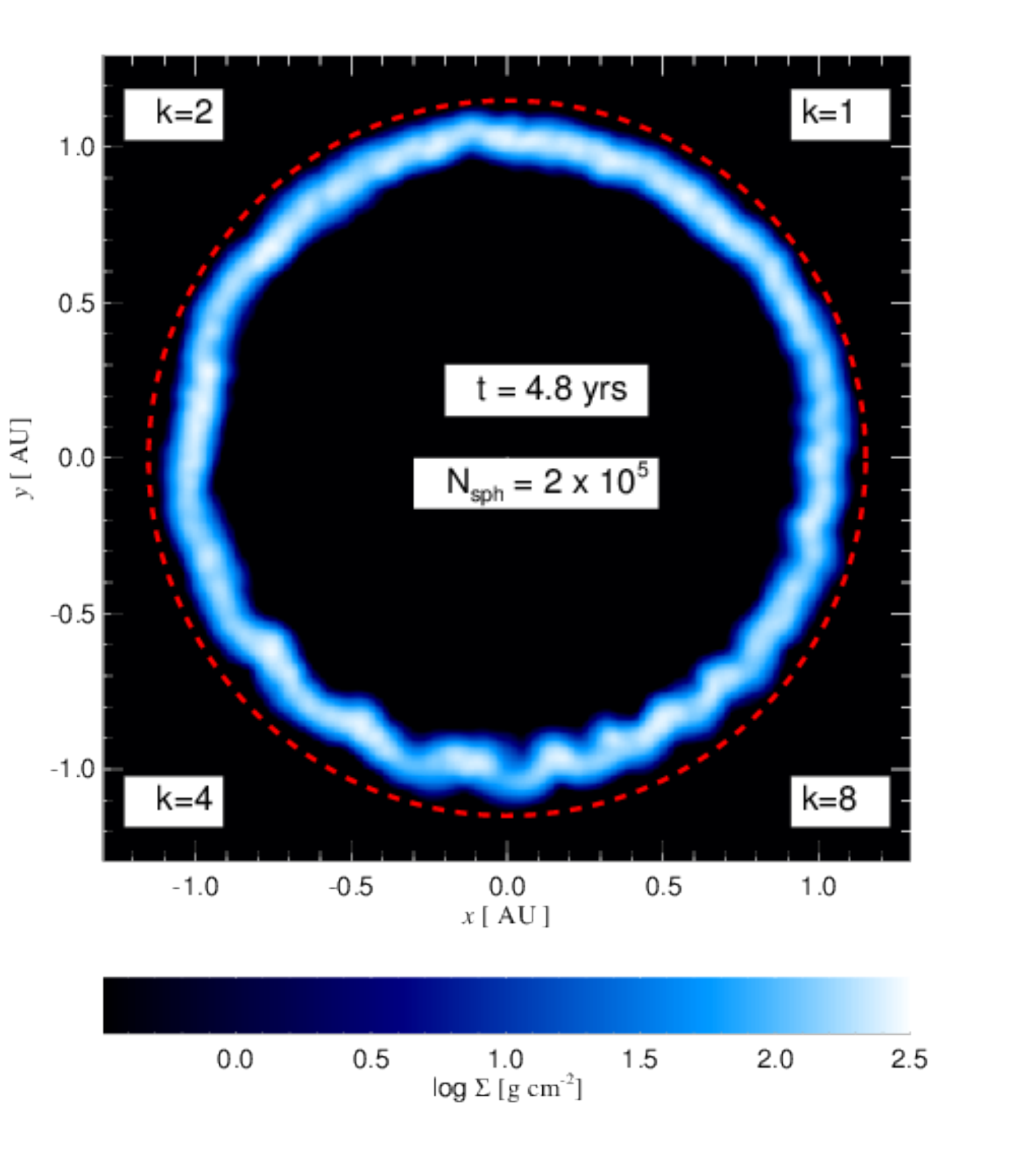}
\includegraphics[width=0.47\textwidth]{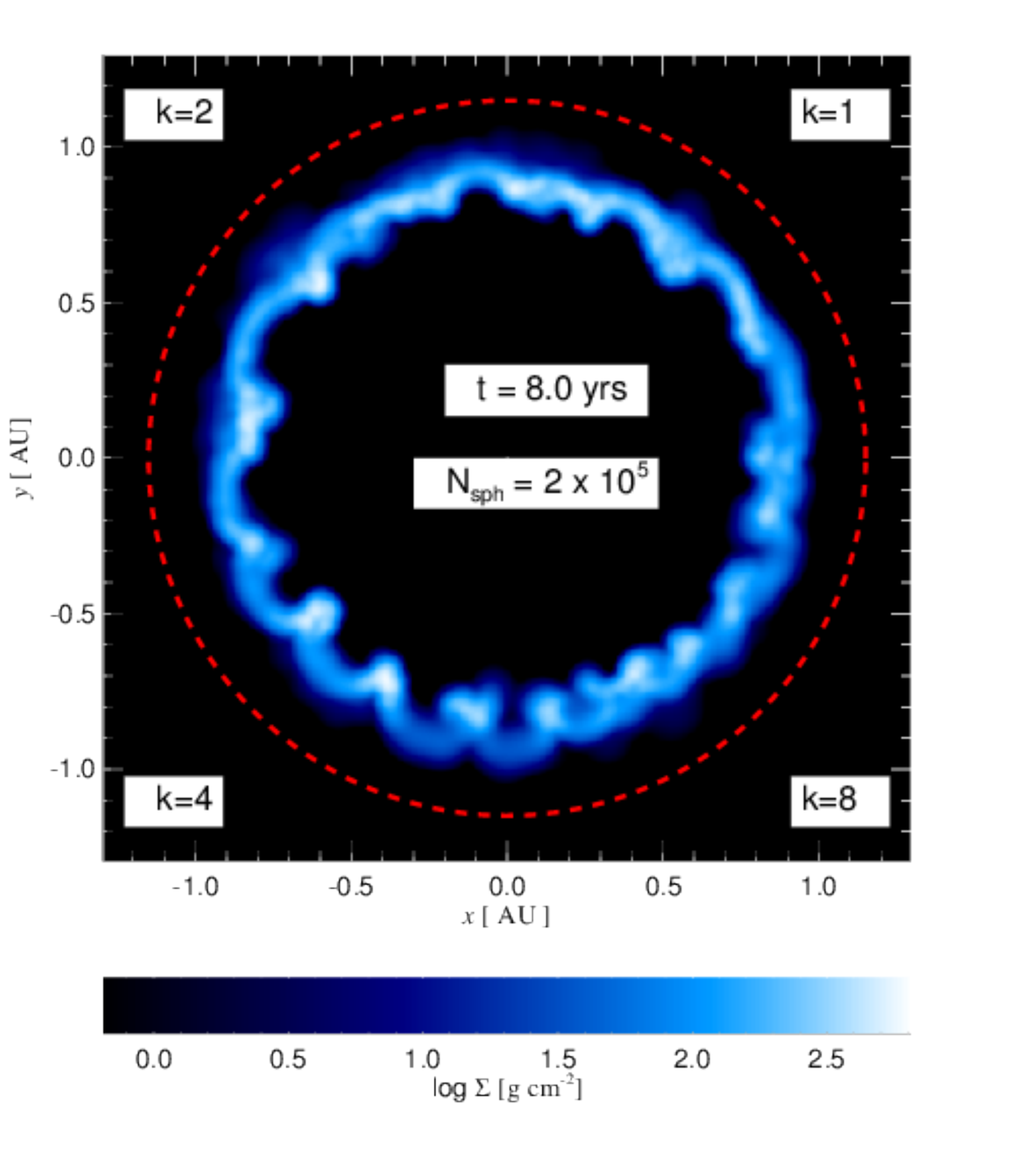}
\includegraphics[width=0.47\textwidth]{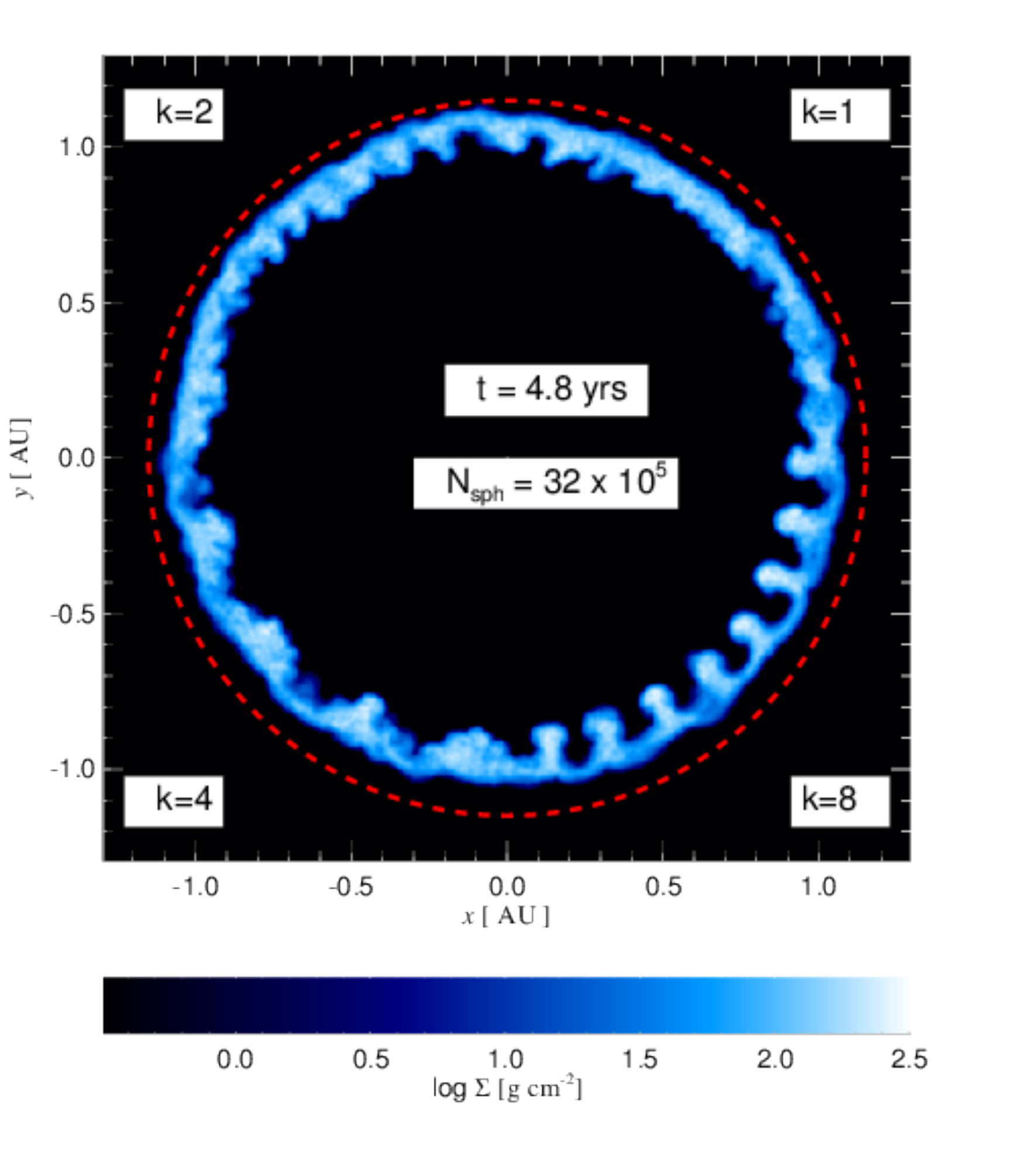}
\includegraphics[width=0.47\textwidth]{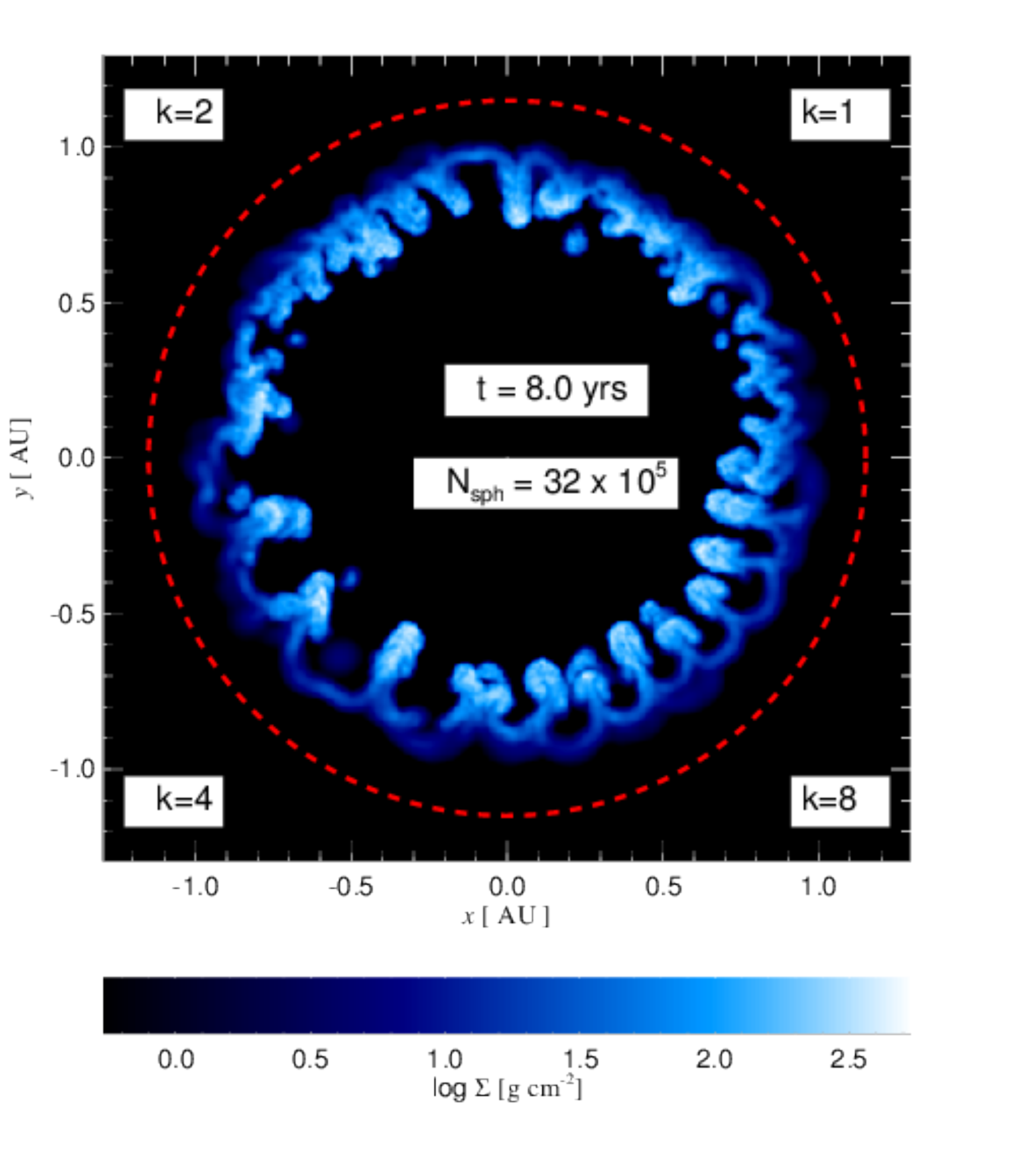}
\caption{Growth of perturbations in the sinusoidal perturbation test described in \S \ref{sec:origin}. {\bf Top panels:}  Snapshot from simulation with $N_{\rm sph} = 2\times 10^5$ SPH particles. {\bf Bottom panels:} same but for $N_{\rm sph} = 3.2 \times 10^6$ particles. Note that different wavelengths grow differently at different numerical resolutions.}
 \label{fig:res_st}
 \end{figure*}

The top and the bottom panels of fig. \ref{fig:res_st} show the development of the instability for $N_{\rm sph} = 0.2$ Million and $N_{\rm sph} = 3.2$ million, respectively. In the two earliest snapshots at $t=4.8$ years on the left of fig. \ref{fig:res_st} we can identify the fastest growing modes visually. For $N_{\rm sph} = 0.2$ Million, the growth of the perturbation is the largest in the third quadrant, $k=3$, whereas for SinZ1a01N32e5 the peaks of the perturbation is in the $k=4$ quadrant grow the fastest. At lower resolution some of the peaks in the highest $k$ quadrant merge as there is not enough numerical resolution to follow their growth properly. In SinZ1a01N32e5 the low-$k$ perturbations ($k=1$ and $k=2$) evolve to display features on scales smaller than originally present at $t=0$ in these quadrants.

\begin{figure}
\begin{center}
\subfloat{\includegraphics[width=0.5\textwidth]{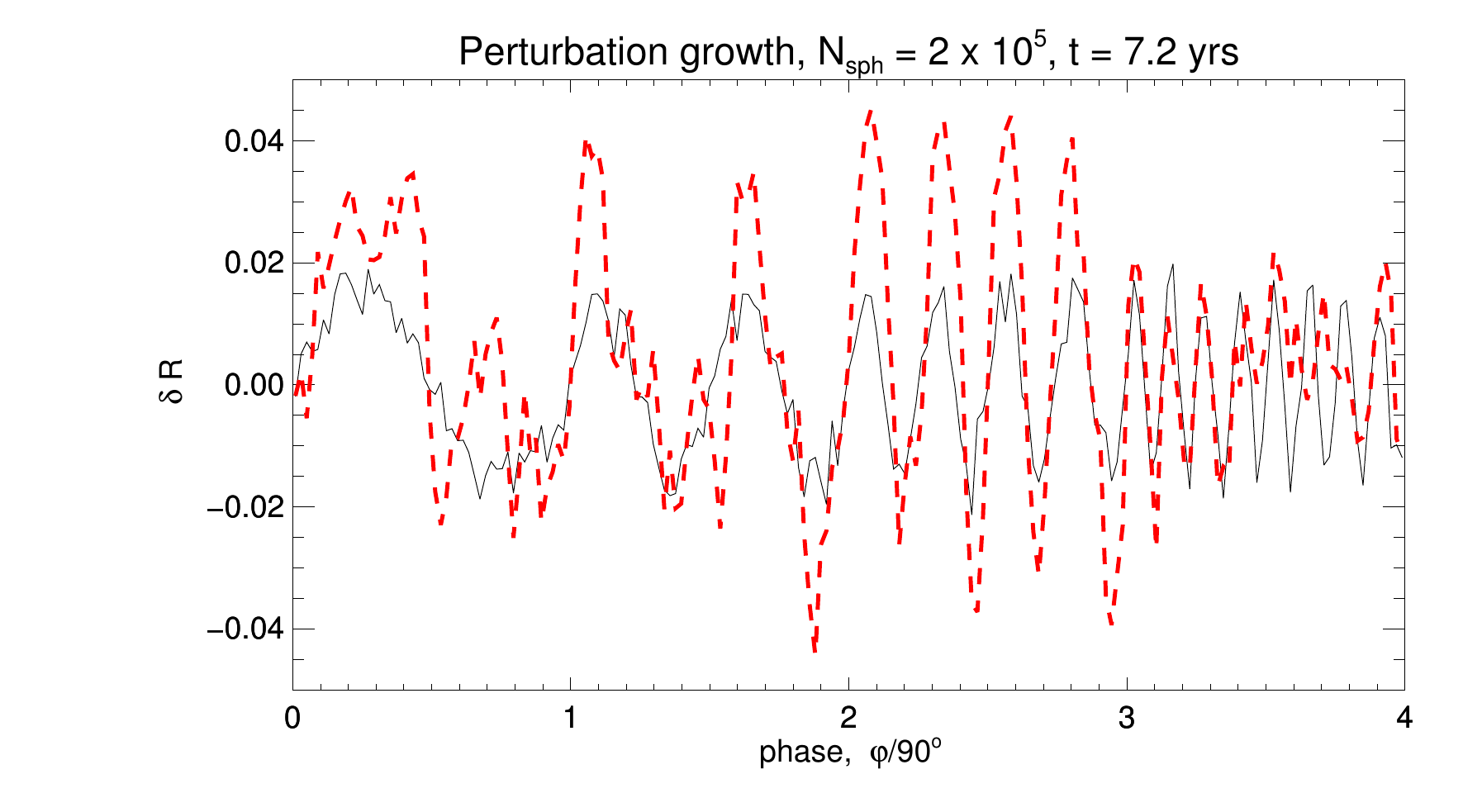}} \\
\subfloat{\includegraphics[width=0.5\textwidth]{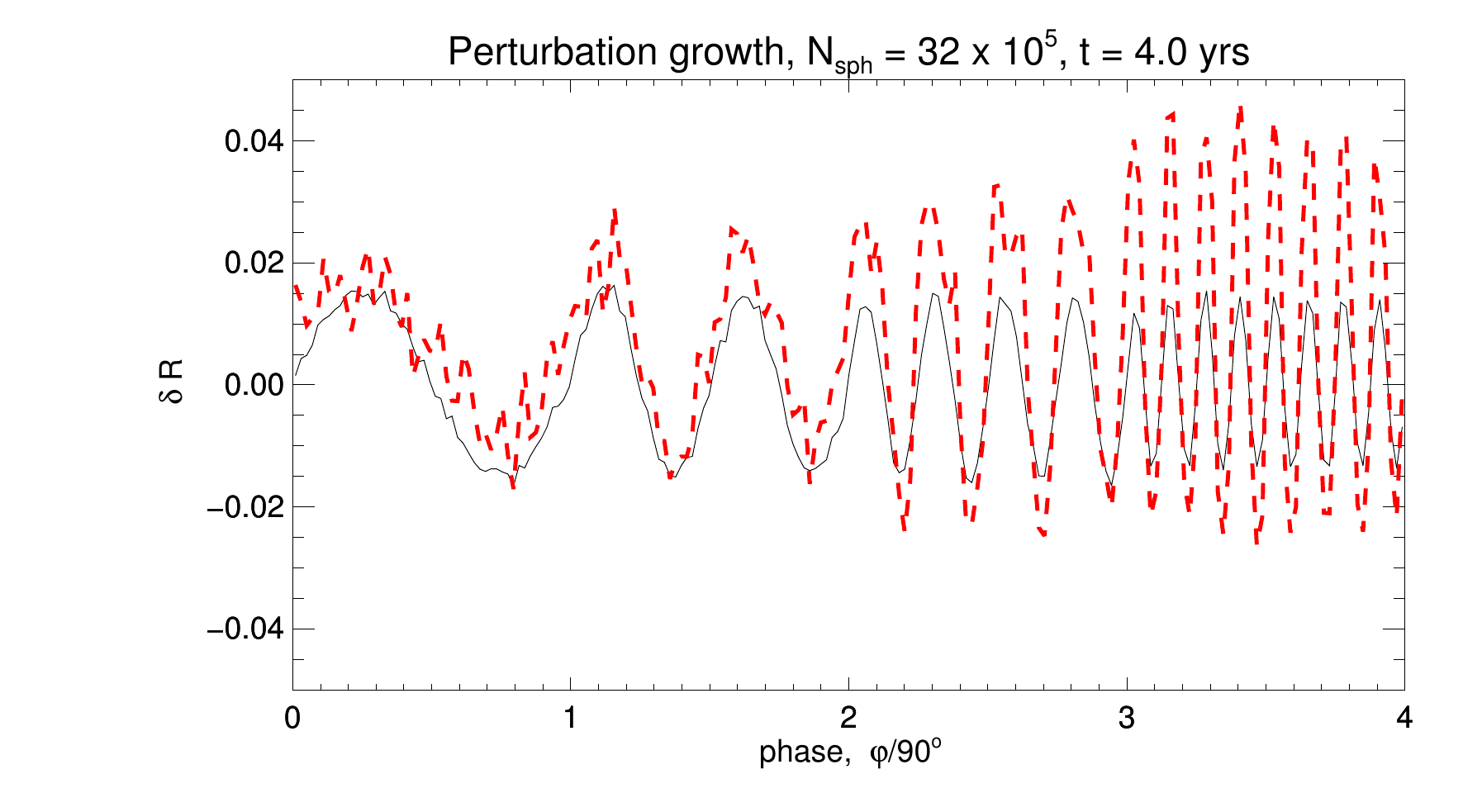}}\\
\caption{Azimuthal analysis of perturbation growth at two different numerical resolutions, $N_{\rm sph}$, for simulations SinZ1a01N2e5 and SinZ1a01N32e5. The black solid and the red dashed curves show the initial perturbation and that at later time as indicated above the respective panel. Note that the higher the resolution, the smaller the length scales that dominate perturbation growth. Also note that a factor of $\sim 2$ growth is achieved sooner at higher resolution.}
 \label{fig:phi_sine}
\end{center}
\end{figure}

For a more quantitative analysis, we calculate the deviation of the mean $R$ for pebbles as a function of the azimuthal angle $\phi$. Fig. \ref{fig:phi_sine} shows the resulting curves for the $N_{\rm sph} = 0.2$ Million and $N_{\rm sph} = 3.2$ million cases at $t=0$ (black curves) and later times. As in fig. \ref{fig:res_st}, we find that perturbations grow the fastest in the third quadrant for SinZ1a01N2e5 and ain the fourth quadrant for SinZ1a01N32e5. We define and auto-correlation function
\begin{equation}
A_n(t) = \frac{4}{\pi}\;\int_{\phi_{n-1}}^{\phi_{n}} d\phi \; \delta R(\phi, 0) \delta R(\phi, t)\;,
\label{An0}
\end{equation}
where the integration limits are the limits of the quadrants, $\phi_{n} = (\pi/2)(n-1)$. Here $\delta R(\phi, t)$ is defined as a local deviation of the mean radius of the shell at this $\phi$ from the $2\pi$ average of the pebble shell radius, $R_0(t)$. In particular, for each azimuthal $\phi_i$ bin, we find all the particles within the bin and then calculate the mean radial distance of the pebbles from the centre of the clump, $R(\phi_i, t)$. The deviation in that bin is then $\Delta R(\phi_i, t) = R(\phi_i, t)-R_0(t)$. With this definition, $A_n(0) = 1$ for all $n$, and the theoretically expected scaling with time is
\begin{equation}
A_n^{\rm exp}(t) = \exp\left[ \frac{t}{t_n}\right]\;,
\end{equation}
where $t_n$ is the growth time of the RT instability. We then use two definitions of the minimum resolvable wavelength, $\lambda_{\rm min}$. In the first we assume that $\lambda_{\rm min} = 4 h_{\rm sml}$ (to resolve the four quadrants of a sine wave). In the other definition, we consider the $A_n(t)$ plots such as fig. \ref{fig:phi_sine} and find visually the quadrant that grows the most at a given SPH resolution. We take the perturbation wavelength of that quadrant as $\lambda_{\rm min}$. In cases when two adjacent quadrants grow equally rapidly we take the mean $\lambda$ of the two quadrants. With $\lambda_{\rm min}$ defined, we use eq. \ref{trt00} to find the expected fastest growth rate at a given SPH particle number. 

\begin{figure}
\includegraphics[width=0.47\textwidth]{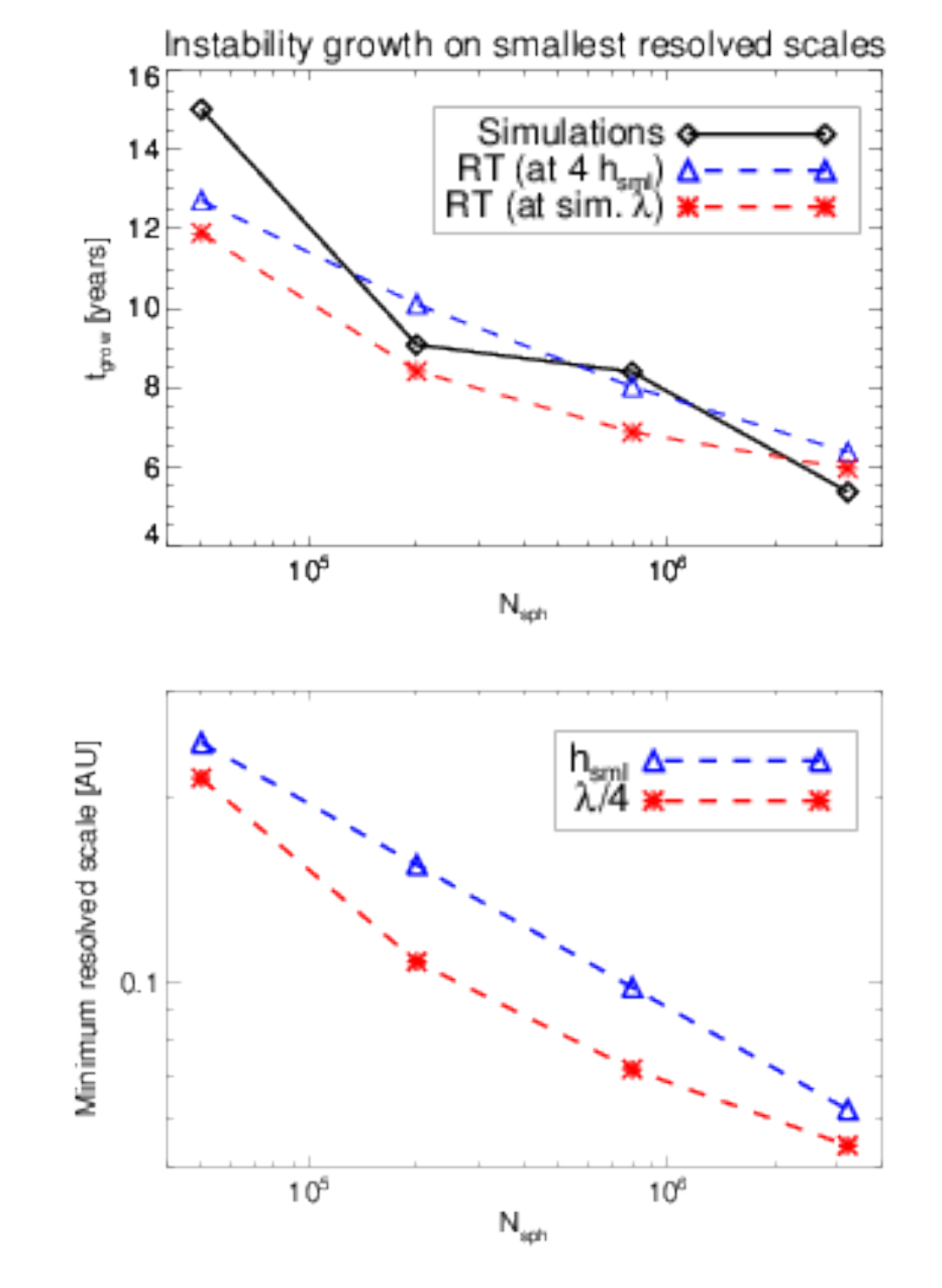}
 \caption{Perturbation growth rates and the corresponding length scales, measured from the simulations, compared to those theoretically expected. See text in \S \ref{sec:origin} for more detail.}
 \label{fig:pert_growth}
 \end{figure}

The top panel of fig. \ref{fig:pert_growth} shows the fastest growing instability time scales $t_{\rm inst}$ versus $N_{\rm sph}$, whereas the bottom panel shows the two respective definition of the minimum resolvable scale. As expected, the instability grows faster at higher resolution. The blue dashed curves assume eq. \ref{4h} scaling, and provide the closest match to the instability growth rates measured from the observations. Although the agreement is not perfect,  it is good enough for us to accept that Rayleigh-Taylor instability is the culprit for the collective effects in grain sedimentation found here.

%\subsection{Collective versus dispersive dynamics}\label{sec:coll_vs_disp}

%In this section we study grain dynamics for a bimodal grain distribution to try and understand what determines whether grains of very different sizes will move together, with similar velocities, or separately as predicted by the test particle approximation. 

\subsection{Instability decay for large grain sizes}\label{sec:inst_decay}

\begin{figure}
\includegraphics[width=0.5\textwidth]{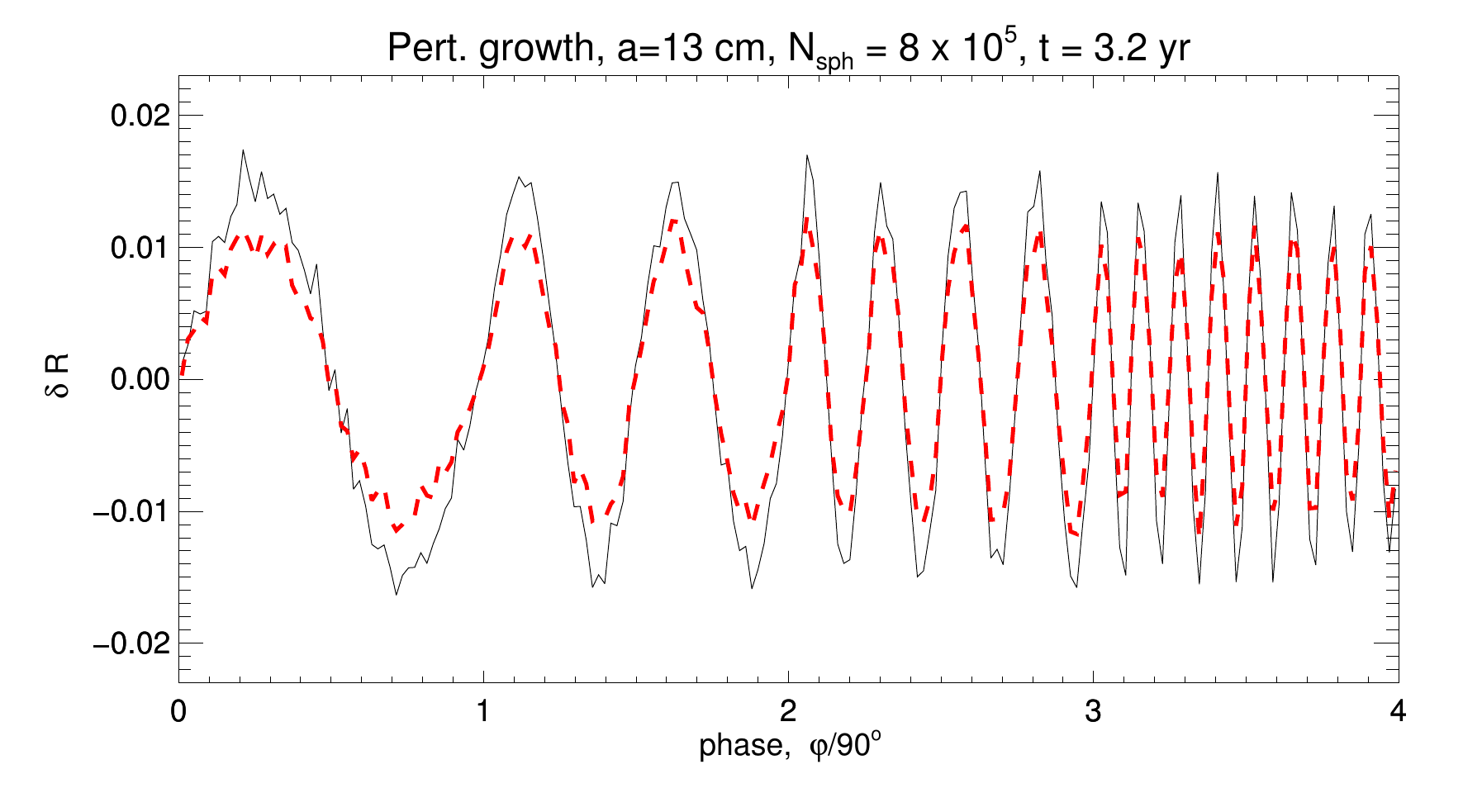}
 \caption{Similar to fig. \ref{fig:phi_sine} but for simulation SinZ1a13N8e5, i.e., for a much larger grain size, $a=13$~cm. Perturbations now decay rather than grow, and the decay rate is independent of perturbation wavelength.}
 \label{fig:pert_decay}
 \end{figure}

Small grains and the gas are tightly coupled and so move together as one heavy fluid. However, large grains are able to separate themselves from the surrounding gas, which must affect dust-RT instability growth. Simulations SinZ1a02N8e5, etc., are exactly analogous to SinZ1a01N8e5 but for grain sizes increasing by a factor of two from $a=0.2$~cm (SinZ1a02N8e5) to $a=26$~cm (SinZ1a26N8e5). In brief, for large grains the perturbation amplitude in fact decreases with time rather than increases. Fig. \ref{fig:pert_decay} shows this for $a=13$~cm, at one particular time. Compare $A_{\rm n}(t)$ for this simulation with those obtained earlier for $a=0.1$~cm (fig. \ref{fig:phi_sine}). We see that $A_{\rm n}(t)$ decreases for SinZ1a13N8e5 with time for all wavelengths considered.

\begin{figure}
\includegraphics[width=0.5\textwidth]{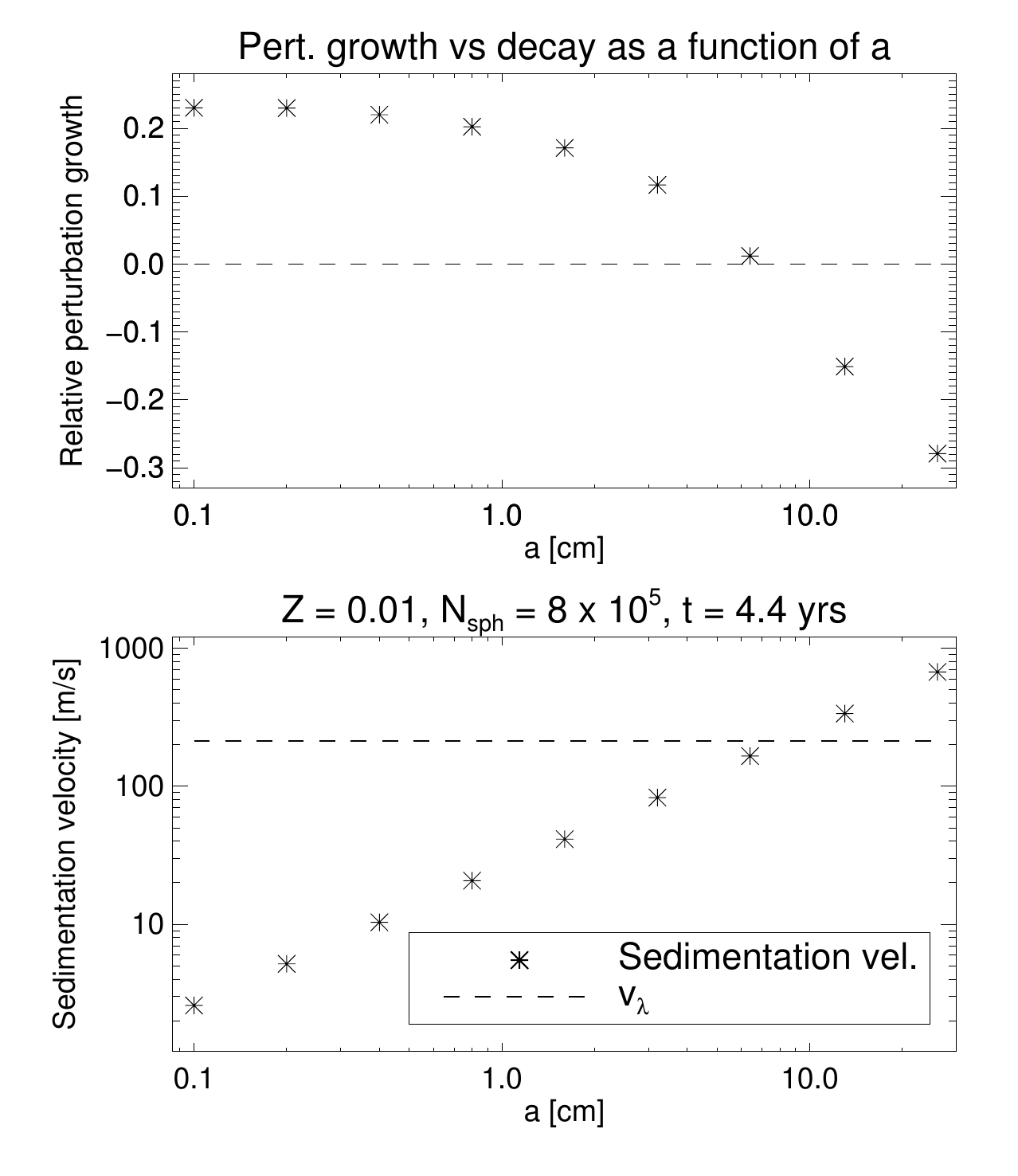}
 \caption{{\bf Top panel:} Perturbation growth or decay versus grain size for simulations SinZ1a01N8e5 to SinZ1a26N8e5. {\bf Bottom panel:} Grain sedimentation velocity compared to the perturbation growth velocity $v_{\lambda}$. Note that perturbations decay when sedimentation velocity exceeds $v_{\lambda}$.}
\label{fig:pert_vs_a}
\end{figure}

To understand this, define a characteristic perturbation velocity, $v_{\rm per}$, with which sinusoidal perturbations grow, as
$v_{\rm per} \sim \delta R_0/t_{\rm RT}$, 
%\label{vpert0}
where $t_{\rm RT}$ is the RT instability growth time scale defined for the shortest resolvable wavelength, set to $\lambda = 4 h_{\rm sml}$. For runs presented here, $v_{\rm pert} \approx 210$~m~s$^{-1}$. If dust particles are tightly coupled to gas then they move with velocities comparable to $v_{\rm pert}$. However, dust particles also move radially inward through the gas with the sedimentation velocity derived in \S \ref{sec:dynamics}. The bottom panel of Fig. \ref{fig:pert_vs_a} shows the dust particle sedimentation velocity as a function of $a$ at the location of the sinusoidal perturbation. The dashed horisontal line shows the perturbation velocity $v_{\rm pert}$.

The top panel of fig. \ref{fig:pert_vs_a} shows relative perturbation growth as a function of $a$ at time $t=4.4$~yr, defined as $\delta A_n = (A_n(t)-A_n(0))/A_n(0) = A_n(t) - 1$.
%For the figure, we use the average $\delta A = (\delta A_3 + \delta A_4)/2$ as for $N=8\times 10^5$ particles the fastest growing mode of the perturbation has wavelength in the middle between $n=3$ and $n=4$. 
For small grains dust sedimentation velocity is smaller than $v_{\rm pert}$, and perturbations grow; for large $a$ sedimentation velocity is larger than $v_{\rm per}$, and perturbations decay. In the latter case the dust particles free themselves up from the surrounding gas faster than the instability could grow; they are approximately in the test particle regime. 
%The perturbation amplitude then decays due to the sedimentation velocity {\em decreasing} into the clump. Just as in the middle panel of fig. \ref{fig:3Dsim}, grains further out from the centre fall into the clump faster than those closer in, therefore leading to the gas shell becoming narrower with time and hence $A_n(t)$ decreasing.

The inviscid one-fluid RT instability grows the fastest the smaller the perturbation length scale (eq. \ref{trt00}. In contrast, the dust-RT instability has a minimum length scale below which it does not grow. Define $v_\lambda$ as the maximum perturbation velocity corresponding to a wavelength $\lambda$ growing in the {\em linear} regime. In the linear regime, the perturbation amplitude is by definition smaller than $ \max(\delta R) \sim \lambda/2\pi$, so 
\begin{equation}
v_{\lambda} \sim \frac{\max(\delta R)}{t_{\rm RT}} = \frac{\lambda}{2\pi t_{\rm RT}(\lambda)}\;.
\label{vlambda0}
\end{equation}
Requiring this velocity to be larger than the grain sedimentation velocity, we derive the minimum wavelength of a perturbation that can grow,
\begin{equation}
\frac{\lambda_{\rm min}}{R} \sim (2\pi)^3 A \frac{g(R)}{R} t_{\rm stop}^2\;,
\label{lambda_min0}
\end{equation}
where we used $v_{\rm sed} = g(R) t_{\rm stop}$, and introduced the Atwood number $A = (\rho_2+\rho_1)/(\rho_2 - \rho_1)$. Since this analysis is approximate, it is reasonable to replace $g(R)/R \approx \Omega_{\rm p}^2 = G M_0/R_{\rm p}^3$, and then re-write this equation in a more transparent way,
\begin{equation}
\frac{\lambda_{\rm min}}{R} \sim (2\pi)^3 A \; \textrm{St}^{2}\;, 
\label{lambda_min1}
\end{equation}
where we must remember that the Stokes number $St = t_{\rm stop} \Omega_{\rm p}$ is a function of not only grain size but also radius inside the clump at $R < R_{\rm p}$ since the grain stopping time changes within the clump.

Equation \ref{lambda_min1} shows that large particles (large Stokes numbers) are unlikely to cause dust-RT instability because for them $\lambda_{\rm min}/R > 1$. Small Stokes number particles, St$\ll 1$, on the other hand, may cause the instability for wavelengths provided $\lambda_{\rm min} \le \lambda < R$. A fixed size particle will often be in the intermediate regime. At the gas clump outer edge, the corresponding $\lambda_{\rm min}$  is too large (larger than a good fraction of $R$), so the particles will tend to sediment in the test particle regime. However, when they reach dense enough layers in the clump, $\lambda_{\rm min}$ may drop sufficiently to allow the dust-RT instability to grow. Sedimentation of grains from that point inward proceeds via the dust-RT heavy fingers. Such behavior of the sedimenting particle -- first test-particle like, laminar, and then via the collective effects in the deeper regions -- is indeed observed in our simulations.

%This behavior is to be expected. We shall now see that $a=13$~cm grains are able to decouple from the surrounding gas. 

\subsection{Core formation}\label{sec:core}

\cite{NayakshinEtal14a} found that massive solid cores forming in the very centre of the pre-collapse planets could enforce gravitational collapse {\em of the whole planet}. For this to be the case one needs radiative cooling to be rapid to transport the heat outside of the gas clump.  This is not the case for our simulations. On the other hand, accretion heat released by the core could puff the clump up, making it more prone to tidal disruption \citep{Nayakshin16a}. We leave both of these important issues for future work, terminating simulations if/when gravitational collapse of the dust component occurs, before the collapsed core could becomes massive enough to affect the gas clump.

\begin{figure*}
\includegraphics[width=0.45\textwidth]{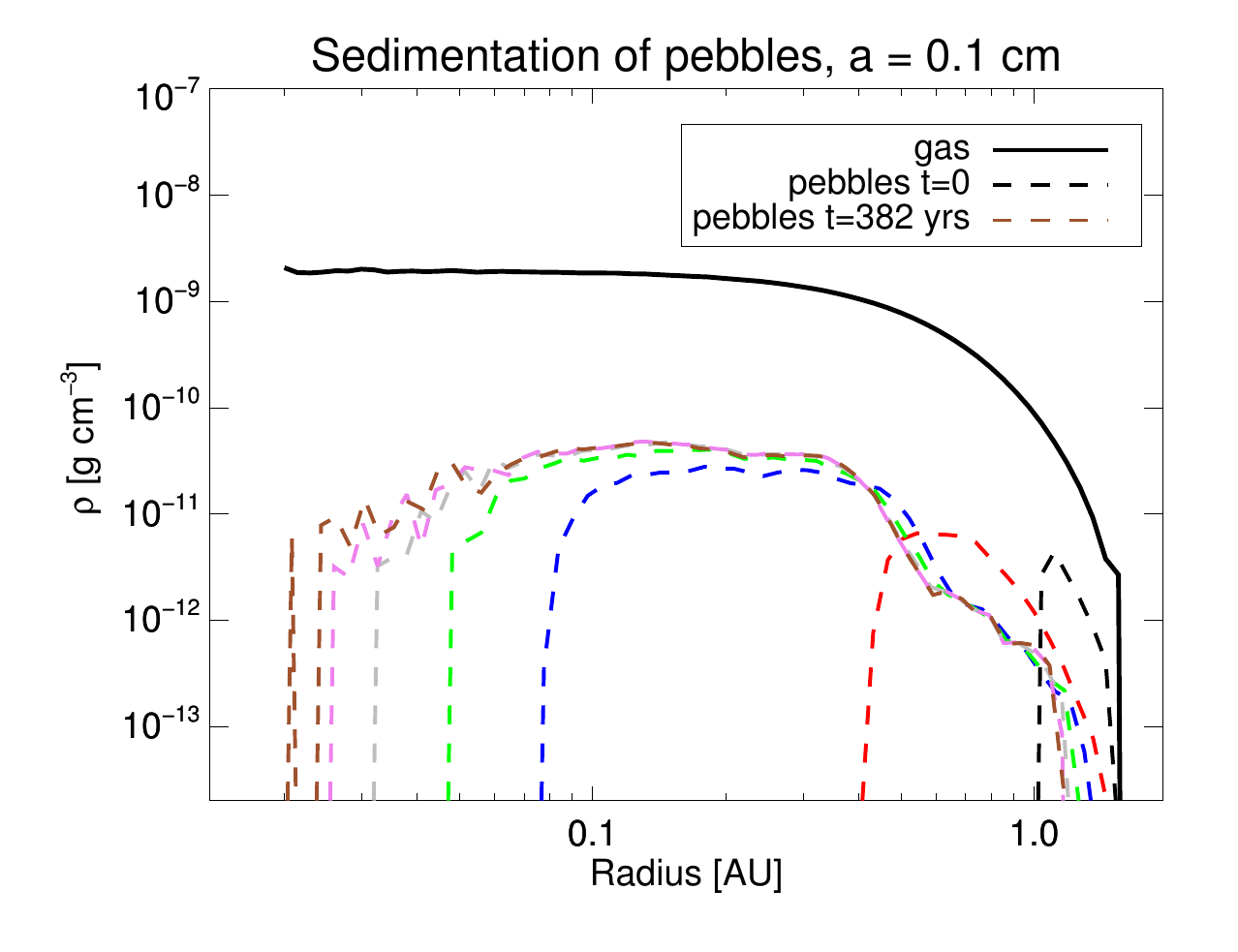}
\includegraphics[width=0.45\textwidth]{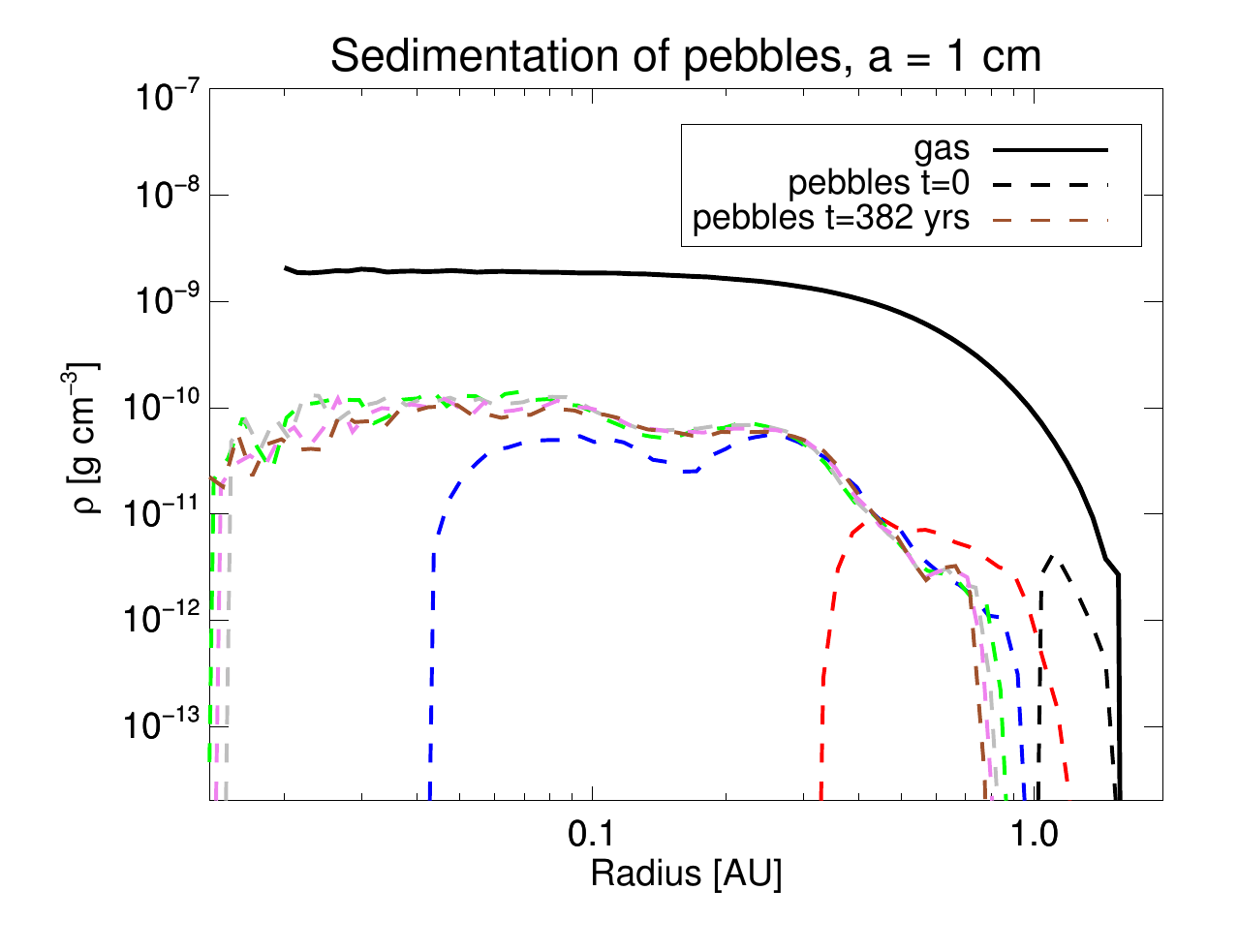}
\includegraphics[width=0.45\textwidth]{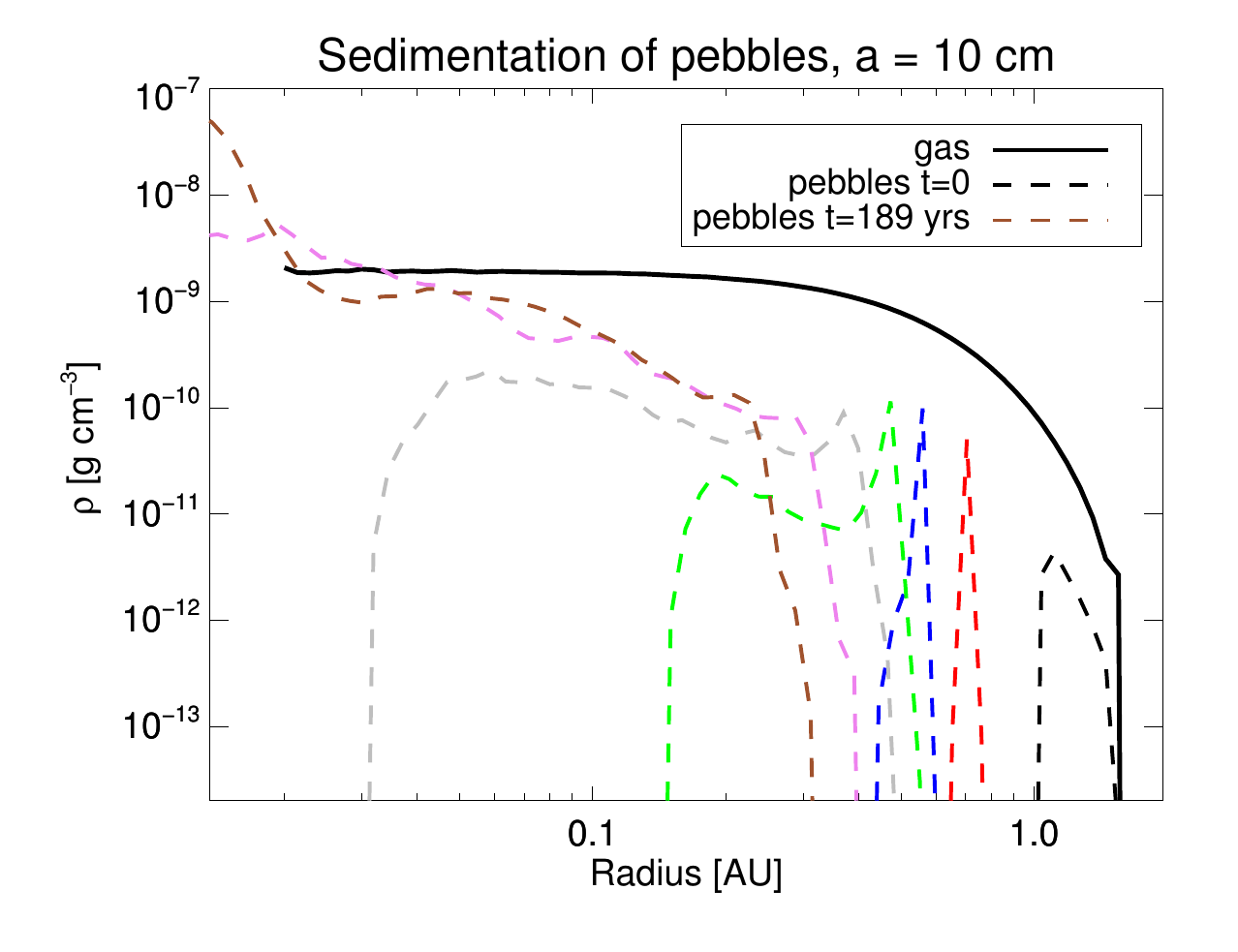}
\includegraphics[width=0.45\textwidth]{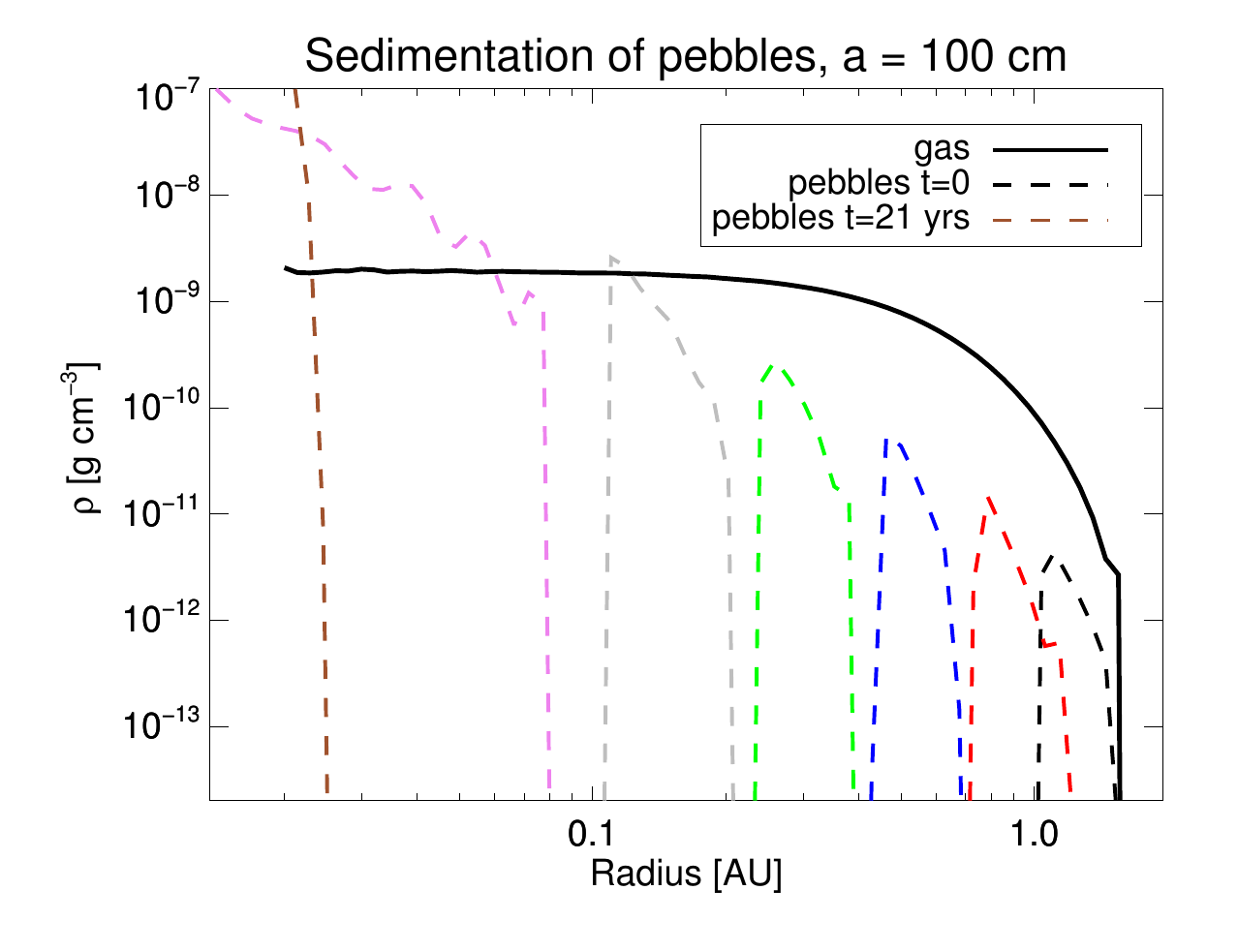}
 \caption{Gas (solid black curve) and pebble density (dashed coloured) for different pebble sizes $a$. Pebble densities are shown at 6 equally spaced time intervals, from $t=0$ to a maximum shown in the legend in each panel. Note that smaller pebbles ($a = $ 0.1 and 1 cm) do not separate into a core, whereas larger pebbles do.}
 \label{fig:Density_structure}
 \end{figure*}

Fig. \ref{fig:Density_structure} presents the gas (solid black) and the pebble density (colors) profiles averaged on concentric spherical shells as a function of  $R$ for simulations SpZ1a01N8e5, SpZ1a1N8e5, SpZ1a10N8e5 and SpZ1a100N8e5. The gas density profile evolves very little in these runs, so only the $t=0$  profile is presented for clarity. The maximum time shown in the panels corresponds to the last snapshot from the respective simulation, and may be different for different runs. We see that for $a=0.1$~cm pebbles remain sparse in the very centre of the clump up to the end of the simulation. For simulation SpZ1a1N8e5 pebbles sink into the very centre of the clump quickly, but after that there is very little evolution for the next few hundred years. In contrast, $a=10$ and $a=100$~cm grains sediment into the clump centre more rapidly and eventually form a very high density dust core. Both of these simulations stall soon thereafter.

Fig. \ref{fig:Density_structure} shows that sedimentation of small grain particles into the very centre of the gas clump, $R\rightarrow 0$, is inefficient. Firstly, the gravitational force drops with  $R$ in the clump centre, $g(R) \approx (4\pi/3) G\rho_{\rm cen} R$. Secondly, when most of the pebble-free gas is displaced from the centre by the falling pebble-rich fingers, the negative buoyancy approaches zero since there are no longer strong density contrasts. The final step in core formation has to be done via physical separation of pebbles and gas. When that occurs, pebbles start to dominate the density in the innermost part of the clump, and a gravitational collapse of the grain component is expected. \cite{Nayakshin10b} found analytically that the radial size of the grain "cluster" needs to be around 10\% of the radius of the gas clump for gravitational collapse. This is roughly borne out by fig. \ref{fig:Density_structure}. 

\begin{figure*}
\includegraphics[width=0.99\textwidth]{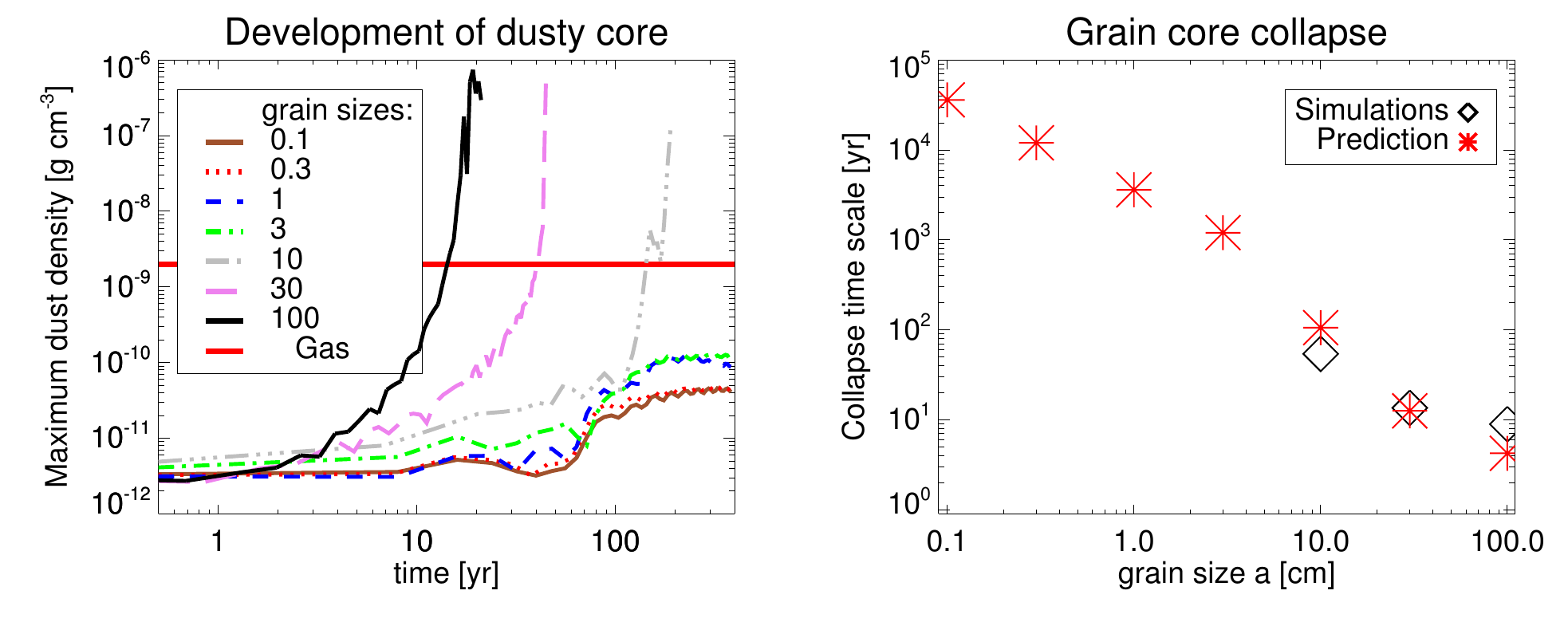}
\caption{{\bf Left:} Development of a dusty core, tracked by the maximum dust density in concentric shells, versus time, for simulations with different grain sizes. Grains with sizes smaller than 10 cm do not form a dense core during the $\sim 400$ years of the simulations. {\bf Right:} Theoretically expected grain core collapse timescale (eq. \ref{tcol_exp}), shown with the red asterisks, and the core collapse time scale measured from the simulations (black diamonds), versus grain size.}
 \label{fig:Core_dens}
 \end{figure*}

The expected dust core formation timescale is hence the time scale on which pebbles sediment in the test particle regime, eq. \ref{tsed_num}. This however needs to be corrected for very large grain sizes, $a \gtrsim 30$~cm. As nothing can collapse faster than the gravitational collapse time, $\sim 1/(G \rho_{\rm cen})^{1/2}$, we write
\begin{equation}
t_{\rm col} = t_{\rm sed} + \left(G\rho_{\rm cen}\right)^{-1/2}\;.
\label{tcol_exp}
\end{equation}
Fig. \ref{fig:Core_dens} analyses when and how gravitational collapse of pebble component occurs. The left panel of the figure shows the maximum density of the dust distribution, calculated on concentric shells, versus time. 
%These also include simulations SpZ1a03N8e5, SpZ1a3N8e5, SpZ1a30N8e5 not presented in fig. \ref{fig:Density_structure}. 
Only the pebbles with size $a\ge 10$~cm go through core collapse. The cores formed by the collapsing pebbles reach densities higher than $\rho_{\rm cen}$ by almost two orders of magnitude before the code stalls.

%At the core collapse point the code stalls, and the results cannot be trusted in any event due to the imposed softening of pebble gravitational potential ($h_{\rm soft} = 0.003$~AU) on small scales. Presumably gravitational collapse in real systems continues until pebble density of the order of the material density of the solids is reached.

The right panel of fig. \ref{fig:Core_dens} compares the expected gravitational collapse time scale (eq. \ref{tcol_exp}) with that actually measured from the simulation. To measure the latter, we first find the time $t_1$ when the pebble core density exceeds $4\times 10^{-10}$~g~cm$^{-3}$. The core collapse time scale is defined as $t_{\rm core} = t_{\rm end} - t_1$, where $t_{\rm end}$ is the end time of the simulation. While the density threshold chosen to mark $t_1$ is somewhat arbitrary, due to a very fast increase of the core density at that point (cf. the left panel of fig. \ref{fig:Core_dens}), the resulted measurement of $t_{\rm core}$ is fairly robust. The theoretically expected collapse time (red asterisks in the right panel of fig. \ref{fig:Core_dens}) has a noticeable break between $a=3$~cm and $a=10$~cm due to the switch from the Stokes to the Epstein drag at these scales. The fact that no core collapse occurs for pebbles smaller than $a=10$~cm is expected. For grains of size $a=3$~cm the collapse is only expected well after 1000 years, whereas our simulations were ran for just short of 400 years.

%While the agreement between the expected and the actually measured collapse time scales is only good to within a factor of two, this is encouraging since our simple theory does not take into account dynamics of gas, which may be affected by the gravitational collapse of the grain component and could therefore alter the dynamics of the pebbles in turn.

\section{Grain growth and fragmentation}\label{sec:growth}

%\subsection{Modeling methodology}\label{sec:method_gr}

%In \S \ref{sec:coll} we studied dynamics of fixed grain size, $a$. 
%Here we consider effects of grain-grain collisions and grain vaporisation when gas temperature is sufficiently high. 
%Analytical arguments in \S \ref{sec:expect} and previous 1D studies show that these processes are very important and can modify grain sizes and dynamics significantly \citep{McCreaWilliams65,Boss97,HelledEtal08,Nayakshin10a}. 

To model effects of grain-grain collisions, we use equations \ref{tcoll1}-\ref{tcoll2}. At high collision velocities grains fragment \citep{BlumWurm08}. We  introduce  grain fragmentation velocity $v_{\rm fr}$, such that for $\Delta v_{\rm bg} \ge v_{\rm fr}$ grain growth turns into grain fragmentation. The most frequently quoted values for $v_{\rm fr}$ from experiments and calculations are in the range of $\sim 1-10 $~m~s$^{-1}$ \citep{BlumMunch93,WyattDent02,SetohEtal07,BeitzEtal11}.

Resolving velocity difference of colliding pebble pairs directly in 3D simulations is currently numerically not attainable with SPH \citep{BoothClarke16}; we hence opt for an approximate treatment. Dust particles weakly coupled to gas have stopping times $t_{\rm st} \ll t_{\rm dyn}$ and therefore sediment at velocities of order the free fall velocity, which is in hundreds of m/s up to a few km/s. $\Delta v_{\rm bg}$ for these particles exceeds the fragmentation velocity for most materials by one-two orders of magnitude. Therefore, particles dominating grain growth are likely to $t_{\rm st} \lesssim 0.01 \times t_{\rm dyn}$. For such short stopping times it is reasonable to assume that dust particles move at terminate velocity with respect to gas, e.g., we can set $d\mathbf{v}/dt \approx 0$ in eq. \ref{dvdt0}, and thus gas-dust relative velocity is $\Delta v = |\mathbf{g}| t_{\rm st}$.  Assuming that collisions between particles of size $a$ and size $\sim a/2$ are most crucial at driving grain growth and fragmentation, we approximate
\begin{equation}
\Delta v_{\rm bg} = |\mathbf{g}| \frac{t_{\rm st}}{2}\;,
\label{tstop_ap0}
\end{equation}
where $\mathbf{g}$ is the gravitational acceleration at the location of the dust particle. 
%We call this a terminal velocity approximation to modeling grain growth. It is not immediately obvious that there is a good physical or numerical reason in trying to calculate the dust velocity dispersion to a better degree than equation \ref{tstop_ap0} while we do not resolve numerically the full particle distribution function at each point in the gas. The latter approach would effectively make the problem of dust dynamics and growth a 4-Dimensional problem and is numerically unassailable at the moment.
To delineate grain growth for $\Delta v_{\rm bg}\leq v_{\rm fr}$ from grain fragmentation for $\Delta v_{\rm bg}\geq v_{\rm fr}$, we further write 
\begin{equation}
\frac{d a}{dt} = \frac{a}{t_{\rm coll}}\; \frac{v_{\rm fr}^2 - \Delta v_{\rm bg}^2}{v_{\rm fr}^2 + \Delta v_{\rm bg}^2}\;.
\label{dadt_switch}
\end{equation}
Here we renamed the grain growth time scale $t_{\rm gr}$ introduced in \S \ref{sec:ggrowth} as the collision time $t_{\rm coll}$. The smallest grain particles are affected by Brownian motion \citep[see, e.g.,][]{DD05}. Accordingly, we add to equation \ref{tstop_ap0} a constant Brownian motion velocity $v_{\rm Br}$, set to $20$~cm/s.

\subsection{Grain growth or fragmentation}\label{sec:exp_growth}

Eight numerical experiments that include grain growth have same gas clump setup as SpZ1a01N8e5 before, with $N_{\rm sph} = 8\times 10^5$. Pebbles are added in the outer regions of the clump, but now within a wedge $|\cos\theta| \le 1/8$ and the total pebble abundance is $Z=0.02$. These choices do not compromise  the generality of our conclusions here. We test four initial grain sizes, from $a=0.1$~cm to $a=100$~cm. These tests are ran with two values of fragmentation velocity, $v_{\rm fr} = 1$ and 10~m~s$^{-1}$. The runs are labeled WedgeZ2a10V1, etc., reflecting pebble metallicity $Z=0.02$, grain size $a=10$~cm and $v_{\rm fr} = 1$~m/s in this example. 
\begin{figure*}
\includegraphics[width=0.45\textwidth]{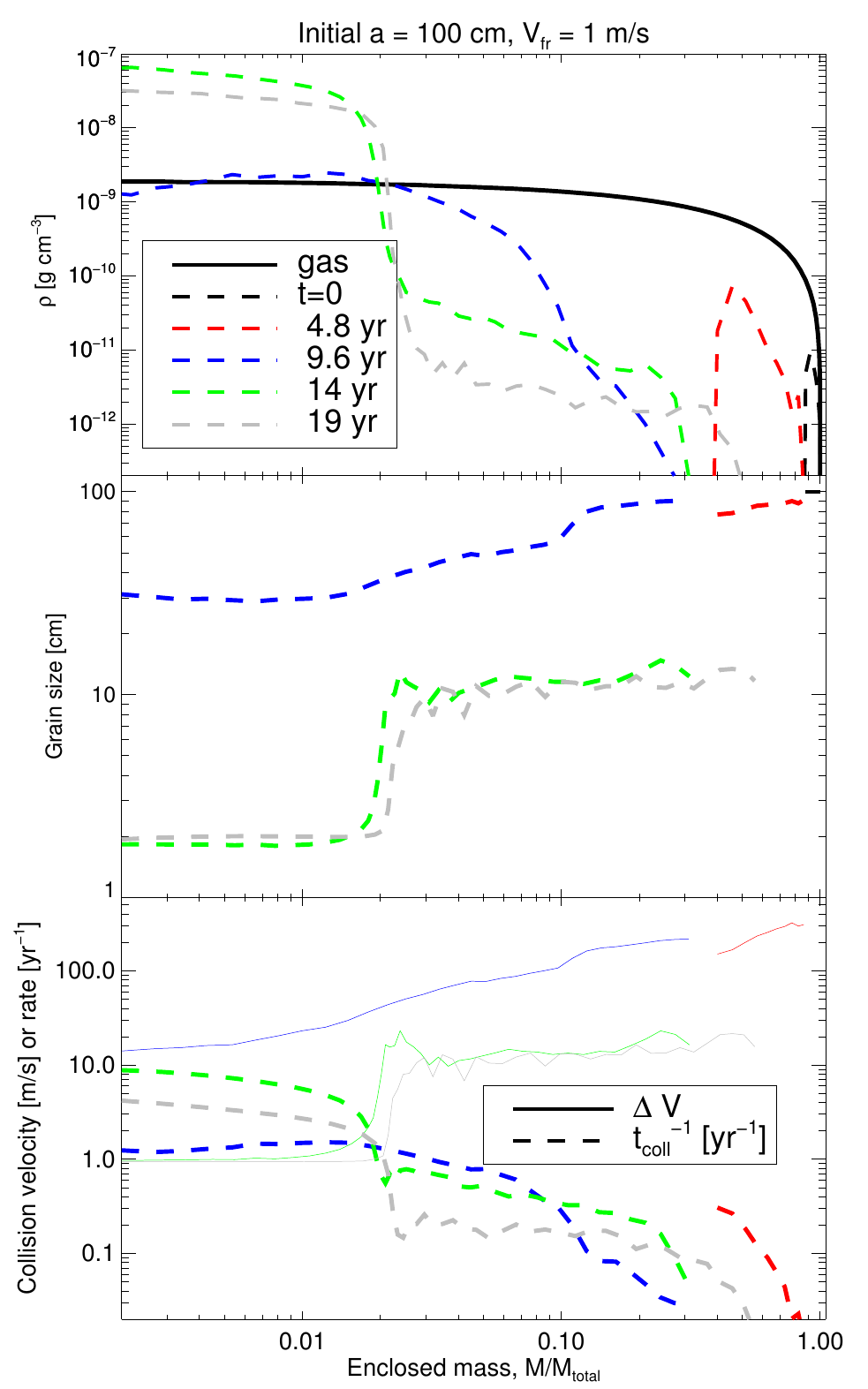}
\includegraphics[width=0.45\textwidth]{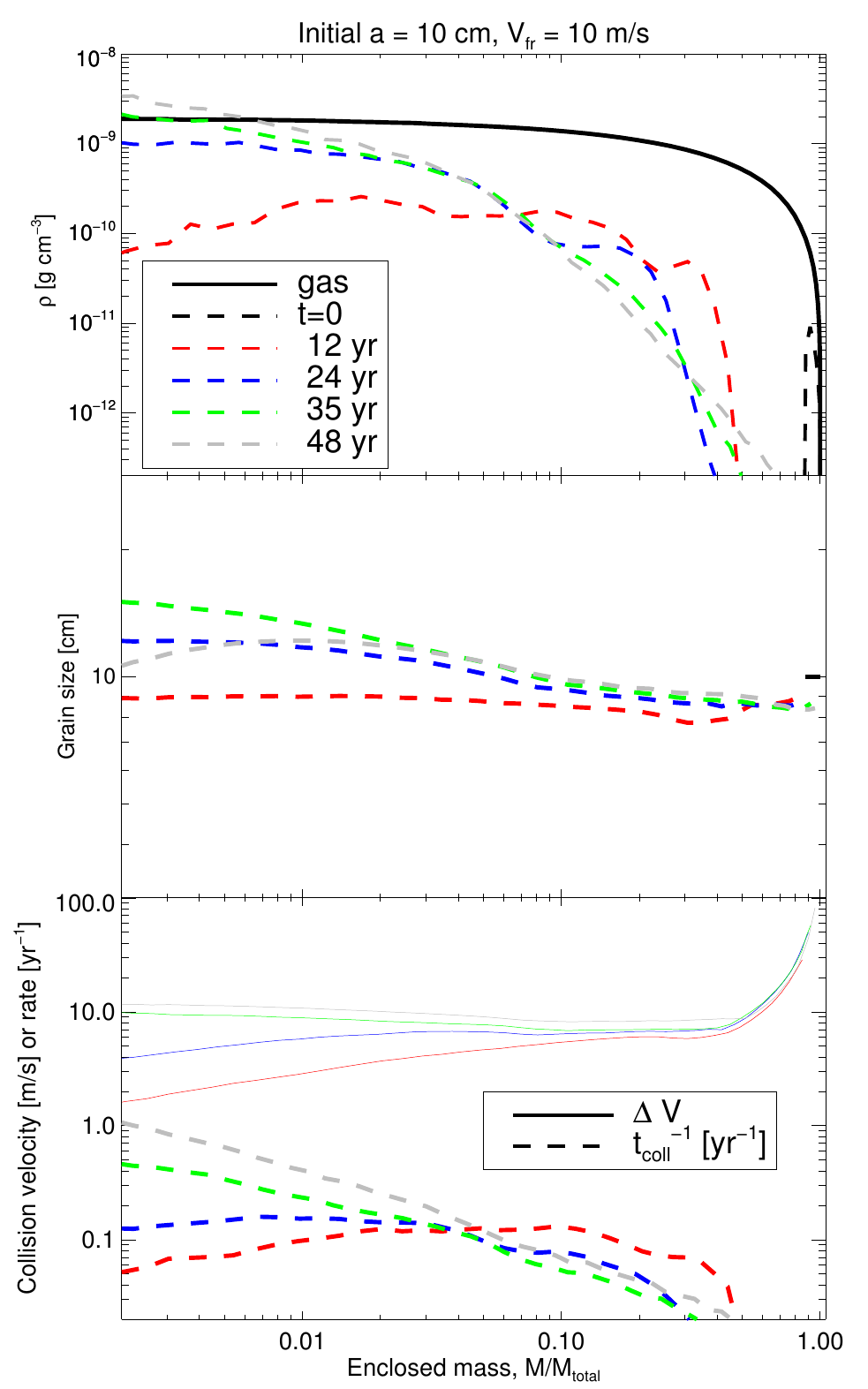}
\caption{Simulations WedgeZ2a100V1 and WedgeZ2a10V10, left and right panels respectively, exemplifying the effects of grain growth and fragmentation. {\bf The top} sub-panels show the gas and pebbles density averaged on concentric shells. {\bf The middle} sub-panels plot pebble sized averaged in the same way, and {\bf the bottom} sub-panels present collision velocity and the collision time scale. All of these quantities are shown as functions of the enclosed mass $M(R)$.}
\label{fig:Frag_a100}
\end{figure*}

%\subsubsection{Initially large grains}\label{sec:vlarge}
 
Fig. \ref{fig:Frag_a100} shows runs WedgeZ2a100V1 and WedgeZ2a10V10. The top, middle and lower panels show dust density, grain size, and collision velocity $\Delta v_{\rm bg}$ and the grain-grain collision rate ($t_{\rm coll}^{-1}$), respectively. These are plotted with differently coloured curves as a function of the enclosed total mass. The left panel in Fig. \ref{fig:Frag_a100} shows pebble properties at five different times. Focusing on the middle panel first, we notice that pebbles continue to be very large, close to their initial size $a=100$~cm, for the two earlier times shown. This is despite the fragmentation velocity set at just 1 m/s. The bottom left panel shows that this is because the collision time is long, $t_{\rm coll}\sim 10$ years and so most of the pebbles would not yet have had time to collide and fragment.

Just as in the fixed grain size simulation SpZ1a100N8e5, a dense self-bound grain core forms in the centre of the clump very rapidly. However, further evolution of the grains in the centre of the clump and core's growth is slower. Once pebbles reach central regions of the clump, their density becomes very large there, and collisions become frequent. Grain fragmentation starts to dictate their further evolution. This we can see from the fact that pebble-pebble collision velocity, $\Delta v_{\rm bg}$, is very close to the imposed fragmentation velocity $v_{\rm fr} = 1$~m/s in the central regions. Collisions thus self-regulate the  average pebble size to a size that has $da/dt \approx 0$. For this run, this corresponds to  $a\approx 2$~cm in the centre (cf. the middle left panel in fig. \ref{fig:Frag_a100}).

The right panel of fig. \ref{fig:Frag_a100} shows simulation WedgeZ2a10V10. In this case pebbles fall inward not as rapidly as in the left panel, and hence collisions start to affect pebble sizes sooner. This is again very clear from the fact that the collision velocity $\Delta v_{\rm bg}$ is quite close to $v_{\rm fr}$ over as much as $\sim$~70\% of the clump by mass. In the outer part of the cloud, this grain size regulation leads to a reduction in the grain size, whereas in the inner clump collisions increase the grain size. There is no core collapse in this simulation, but this is simply because WedgeZ2a10V10 was terminated at $t\approx 50$~yrs, before collapse could occur.

%\subsubsection{Grain growth for small grains}\label{sec:small_gr}

\begin{figure*}
\includegraphics[width=0.45\textwidth]{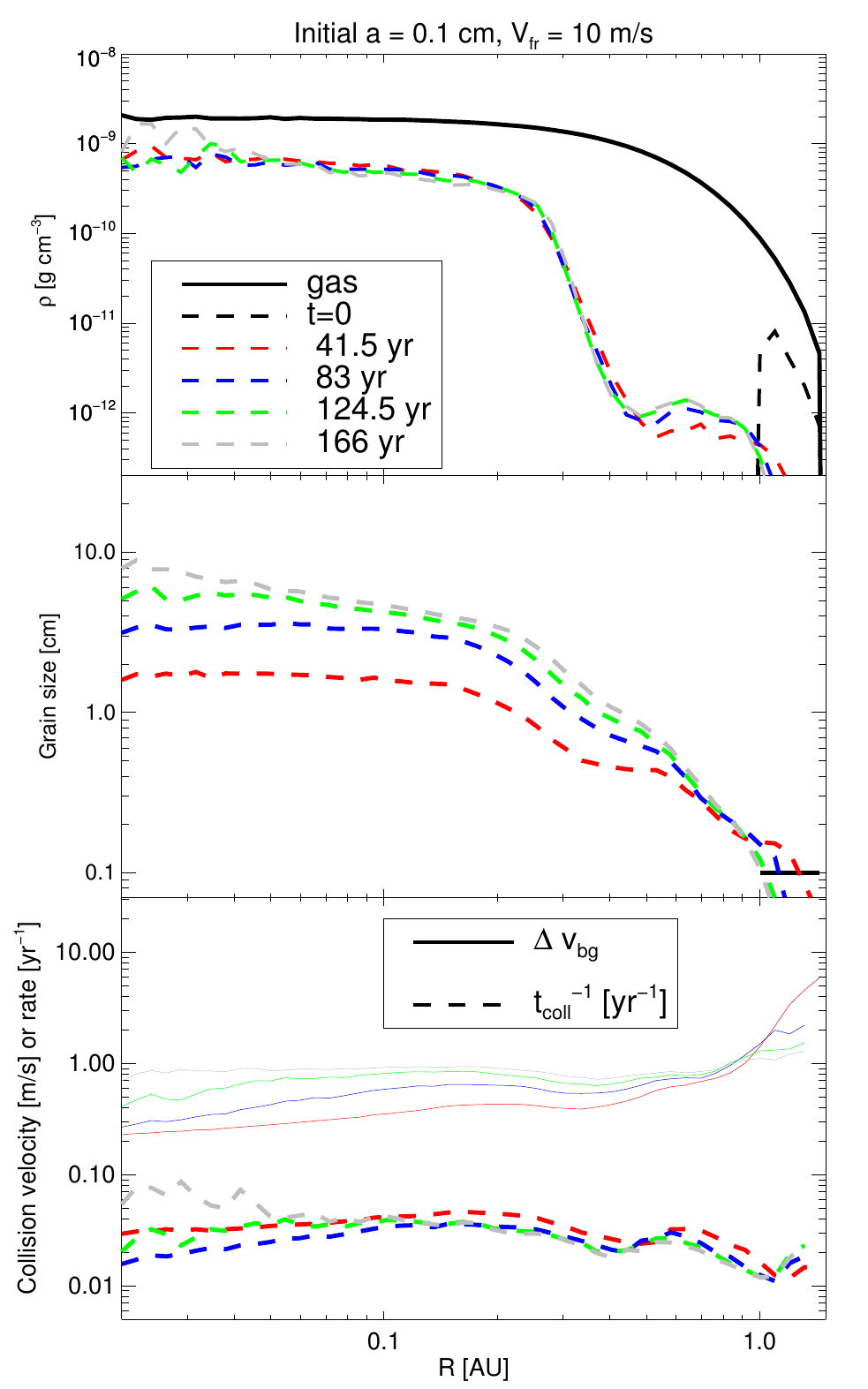}
\includegraphics[width=0.45\textwidth]{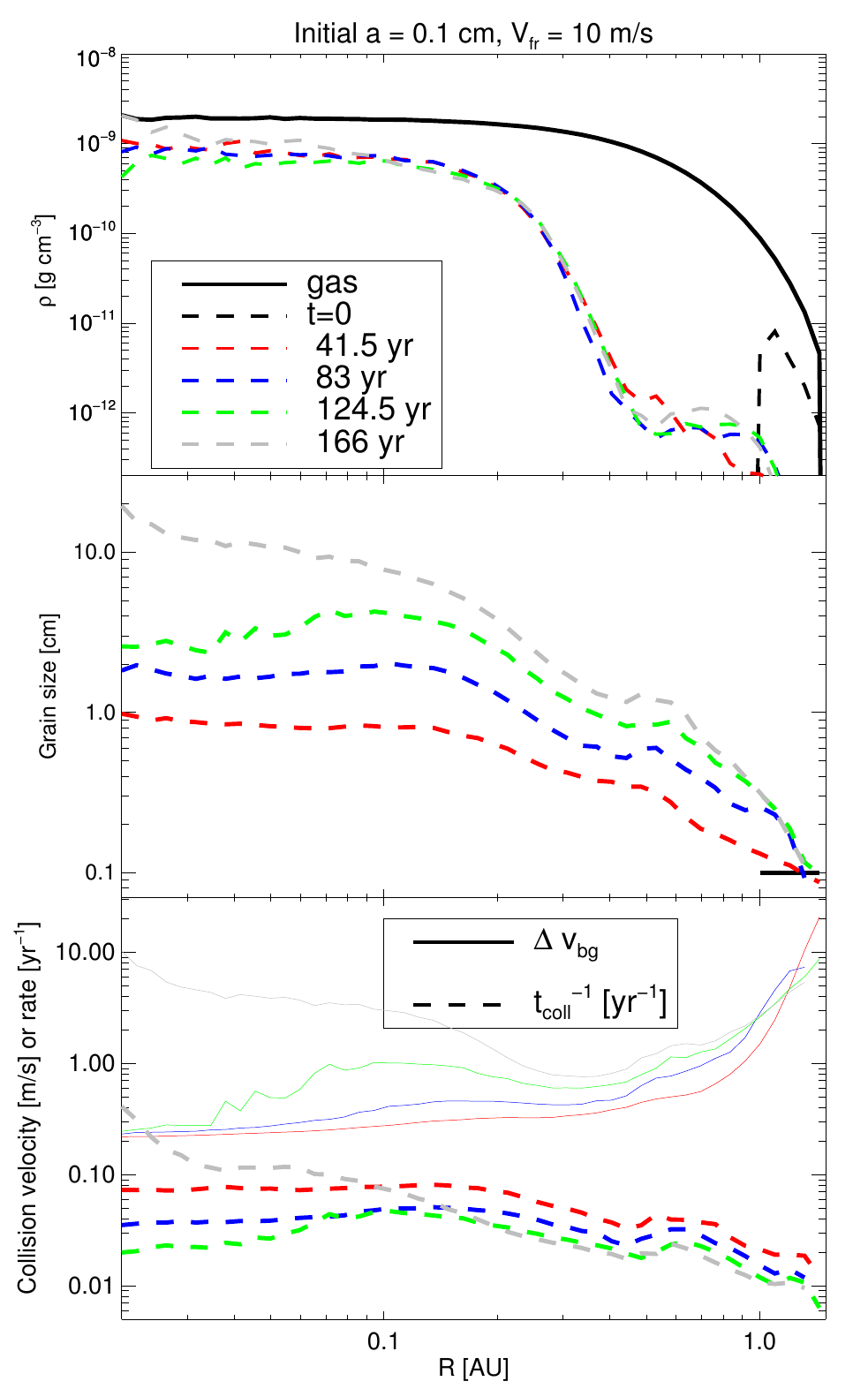}
\caption{Same as \ref{fig:Frag_a100}, but for smaller initial grain sizes for simulations WedgeZ2a01V1 and WedgeZ2a01V10, left and right panels, respectively. Note that in this case the grain size increases with time rather than decreases due to grain growth.}
\label{fig:Frag_a01}
\end{figure*}

Fig. \ref{fig:Frag_a01} shows simulations WedgeZ2a01V1 and WedgeZ2a01V10, for which the initial $a= 0.1$~cm. Only at the very edge of the clump pebble collisions appear to be fragmenting, and only for the $v_{\rm fr} = 1$~m/s run (left panels). Everywhere else inside the clump grain collisions lead to grain growth. The collision velocity is in fact dominated at early times by the Brownian motion. Grain sizes in these two runs hence increase with time, and are also larger in the clump centre than on the outskirts, in direct contrast to the results of the large initial grain cases.

Summarising, the overall effects of grain-grain collisions is to reduce the differences between the small initial and the large initial grain simulations, as grains evolve towards sizes that satisfy the local collisional equilibrium. This implies that massive cores may well be made rapidly, e.g., within hundreds of years, inside of the gas clumps even if pebbles supplied from the protoplanetary disc are small, $a\lesssim 0.1$~cm.

\subsection{Grain vaporisation: Fuzzy Cores}\label{sec:vap}

Pebbles may be vaporised if temperature of gas surrounding pebbles is high enough. After vaporisation, further sedimentation of pebble material into the core may be possible via droplets  \citep[e.g.,][]{CameronEtal82,BrouwersEtal18}, but here for simplicity we assume that pebbles transition from the solid into the gaseous form directly, so that, once they are vaporised, they are tightly coupled to the gas and no longer move with respect to it. 

The rate of pebble vaporisation, expressed in terms of the grain size reduction rate, $(da/dt)_{\rm vap}$, is calculated as in \S 2.5 of \cite{Nayakshin14b}, and is based on earlier work by \cite{PodolakEtal88,HS08}. 
%The vaporisation rate of pebbles is a strong function of temperature. We consider two pebble compositions as two opposite extremes. Water ice is a very volatile material, vaporising at about 150 K; Silicates (rocks) is a refractory material that is vaporised at $T \gtrsim 1500$~K.
In this section we consider icy pebbles. To isolate and differentiate the effects of grain vaporisation from grain fragmentations, $v_{\rm fr}$ in this section is set to infinity. The grain vaporisation rate $(da/dt)_{\rm vap}$ is then added to the rate of pebble size change due to sticking collisions (eq. \ref{dadt_switch}). We do not allow pebble size to drop to less than $a_{\rm min} = 0.1$~cm in this section. 
%At such a size pebbles are co-moving with the surrounding gas (since the clumps are quite dense), and hence $a=0.1$~cm is numerically same as $a = 0$ on the time scales of interest.

The initial conditions for the runs presented here are identical to those presented in \S \ref{sec:exp_growth} except that the clumps are shrunk or expanded in a homologous way to obtain clumps with different values for the initial central temperature, which we vary here between $T_{\rm c} = 100$ K and 400 K. Pebbles have initial size $a=10$~cm and are located in a wedge-like disc; pebble metallicity is set at $Z=0.02$. The simulations names are WedgeZ2a10Tc100 to WedgeZ2a10Tc400.

\begin{figure}
\includegraphics[width=0.9\columnwidth]{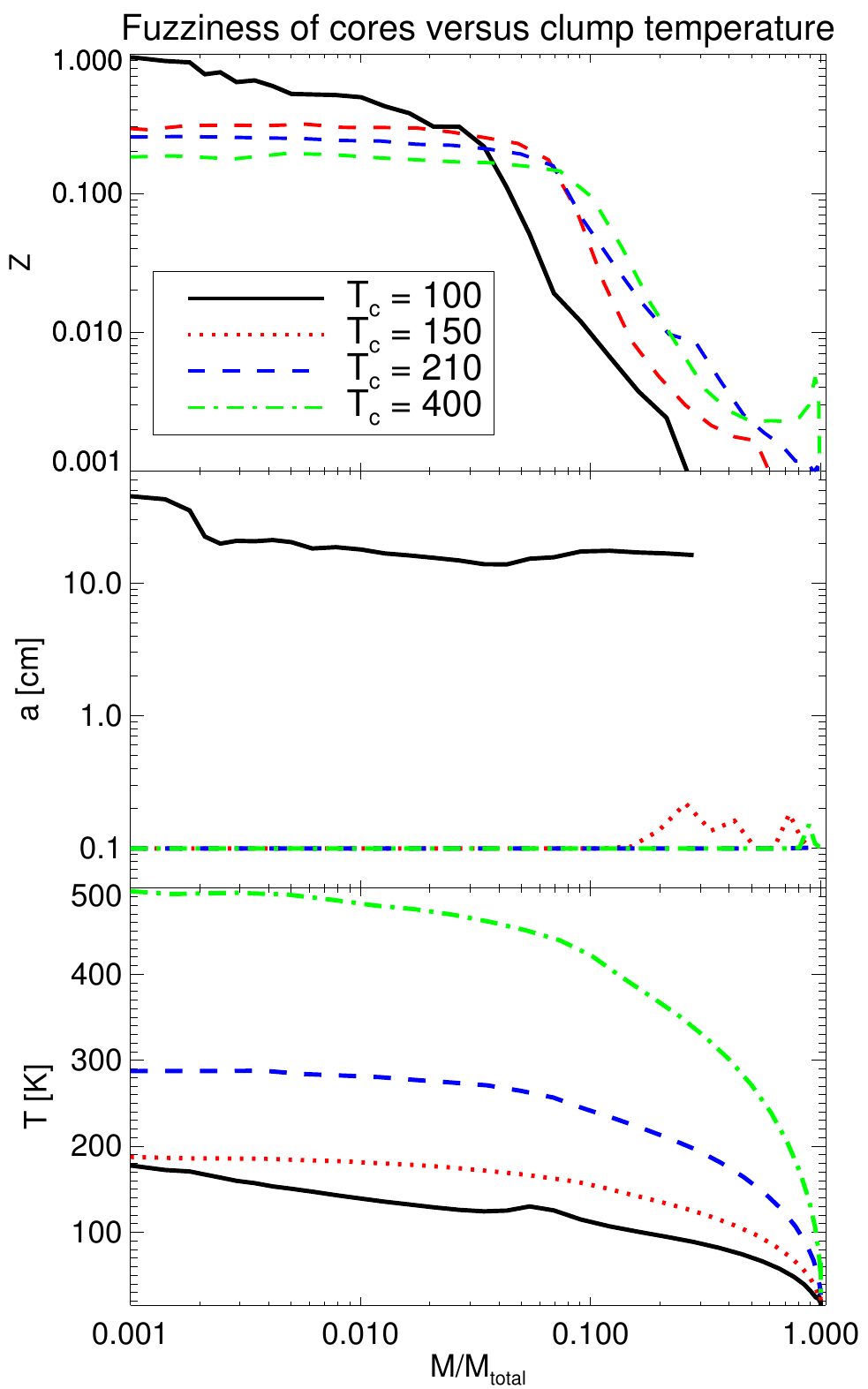}
\caption{Properties of gas clumps with four initial central temperatures loaded with water ice grains for simulations WedgeZ2a10Tc100 to WedgeZ2a10Tc400. {\bf Top:} Local metal abundance as a function of enclosed mass. {\bf Middle:} Pebble grain size. {\bf Bottom:} Gas temperature. Grains are vaporised in the regions hotter than $\sim 180$~K. A solid core is assembled only in the $T_{\rm c} = 100$~K clump, all other cases result in "fuzzy" cores.}
\label{fig:evap}
\end{figure}

Fig. \ref{fig:evap} shows the internal structure of the gas clumps averaged on concentric shells at time $t=80$~yr. The bottom panel shows the gas temperature profiles. All of the clumps actually heat up somewhat. For example, the clump with the initial $T_{\rm c} = 400$~K  heats up to $T_{\rm c} \approx 500$~K. The increase in the central temperature of the clump is due to the additional weight of pebbles which causes the clump to contract somewhat (see \S \ref{sec:loading}). The middle panel in fig. \ref{fig:evap} presents pebble size profiles. For the three hotter clumps, pebbles are at the minimum size imposed, $a_{\rm min} = 0.1$~cm, except in the oute rcool  regions. For the cooler clump, pebbles remain large all the way to the core. There is some pebble growth in this case.

The upper panel of fig. \ref{fig:evap} plots the local pebble abundance, defined on concentric shells, as $Z \equiv M_{\rm Z}/(M_{\rm Z} + M_{\rm gas})$, where $M_{\rm Z}$ and $M_{\rm gas}$ are the masses of the pebbles and gas, respectively, in the shell. The coolest of the four clumps shows formation of a solid core whereas the three hotter clumps do not. In those other cases, the core is best described as a diffuse or a fuzzy one. Note that pebble abundance in the centre of these three clumps is enhanced by at least an order of magnitude compared with the clump-average of 0.02 given our initial condition. For higher values of pebble mass deposited onto the clumps, the metal abundance in the centre would be yet higher, perhaps approaching $Z \sim 1$. As realistic clumps are likely to be hotter than the coolest of the clumps considered here \citep[since cool clumps contract and become much hotter quite rapidly, see fig. 1 in][]{Nayakshin15a}, we conclude that water is not a likely constituent of cores for the gravitational instability model \citep[as concluded by previous authors, e.g.][]{HS08,HelledEtal08}.

In this section we investigated water ice grains only but silicate grains can also make fuzzy/diffuse cores in cases when the central temperature of the clump is sufficiently high, e.g., $T \gtrsim 1500$~K. Nevertheless, such situation is less likely since gas clumps hotter than $\sim 2,000$~K collapse by H$_2$ dissociation \citep{Bodenheimer74}, so that the window of opportunity for a fuzzy Fe/silicate core formation is narrower than for water. This model thus predicts that solid cores made inside gas clumps are likely to be composed of silicates and Fe.%, as again concluded by previous authors \citep[e.g.,][]{NayakshinFletcher15}.

\section{Metal loading and dark collapse}\label{sec:loading}

So far the relative abundance of metals added to the clump was moderate, $Z \lesssim$ a few \%.  \cite{Nayakshin15a} showed that evolution of gas clumps is very sensitive to adding extra mass via pebble accretion. Modeling the gas clump as a polytropic sphere with adiabatic index $\gamma = 1 + 1/n$, and assuming a uniform composition for the clump,  an analytical solution for the central temperature, $T_{\rm c}(Z)$, was found:
\begin{equation}
T_{\rm c} =  T_0
\left[{1-Z_0\over 1-Z}\right]^{6\over 3-n}\;,
\label{tc1}
\end{equation}
where $Z_0$ and $T_0$ are the  initial metallicity and central temperature of the clump.

%This may be expected if convection is vigorous and the pebbles are small, well coupled to the gas, and are added to the clump slowly. Note that this is not the case in the simulations presented so far in this paper, where pebbles are loaded instantaneously in an outer layer. We have seen that RT instabilities, although leading to a rapid transport of pebbles in, did not lead to a uniform composition. A continuous slow deposition of pebbles, on time scales much longer than our runs here, may be better at recreating the conditions envisaged in \cite{Nayakshin15a} (although we shall not find this below).

Clump contraction under the weight of pebbles may allow the clump to collapse on reaching the central temperature of $\sim 2,000$~K \citep{Bodenheimer74} more rapidly than possible by radiative cooling of the clump. Since this mode of collapse requires no radiative losses, it may be called "dark collapse" to distinguish from the radiation-driven collapse. 

%Dark collapse may help the clump to escape tidal disruption if it is migrating inward rapidly, hence changing the planet formation outcome and producing a positive rather than negative correlation between gas giant planet survival and the host star metallicity \citep{Nayakshin15b}.

%\subsection{Idealised test}\label{sec:ideal_test}

\begin{figure}
\includegraphics[width=0.4\textwidth]{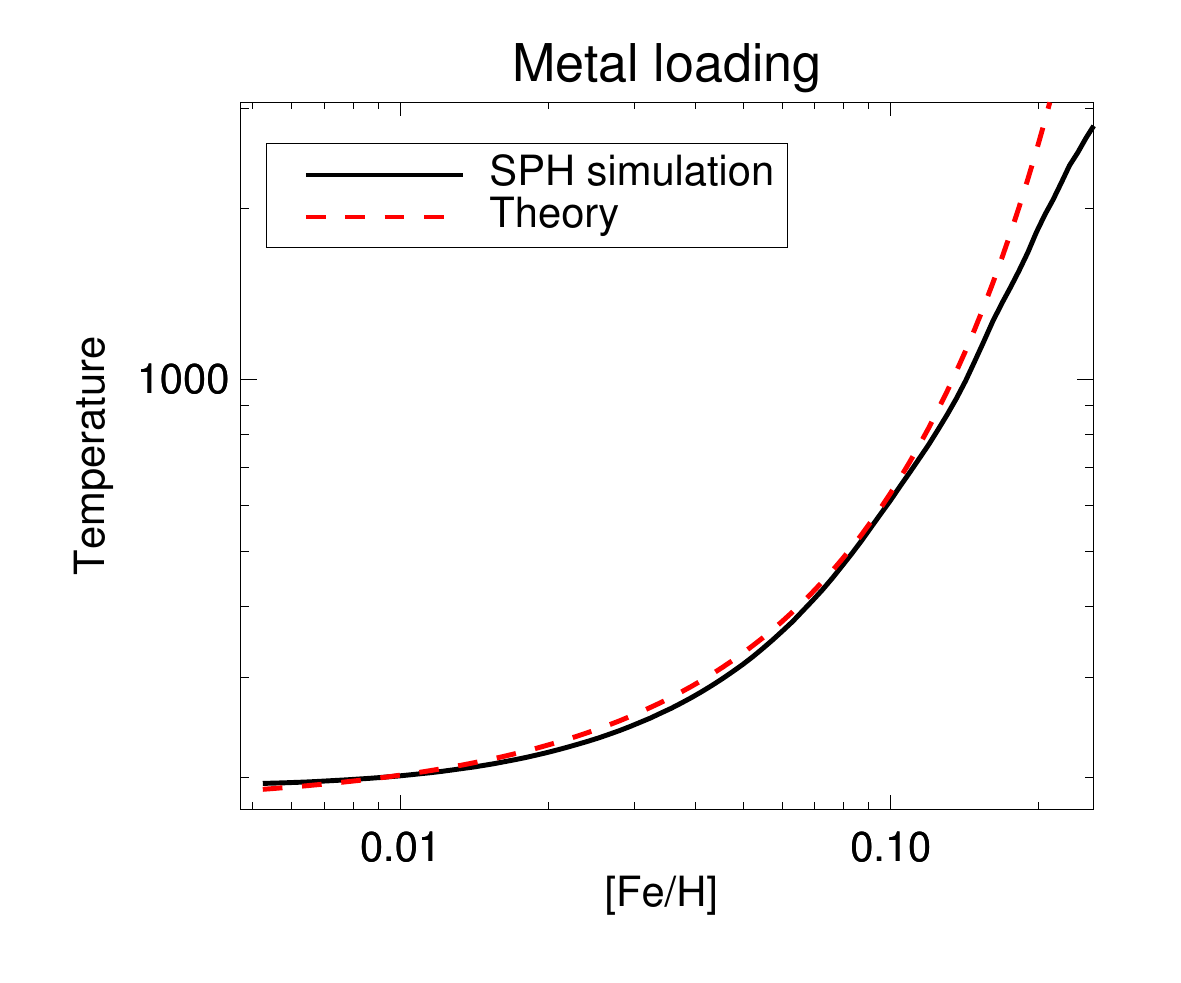}
 \caption{Simulation UniZload that tests polytropic sphere contraction under a uniform metal loading (\S \ref{sec:ideal_test}).}
   \label{fig:test1}
 \end{figure}

In deriving eq. \ref{tc1} it is assumed that pebbles entering the clump are spread around the clump by convection {\em uniformly}. Our 3D simulations can go beyond this simplifying assumption but first we present simulation UniZload that recreates the assumption of a uniform pebble abundance. 
%As there are no instabilities or any non-spherical structures in this simulation, a modest number of SPH particles, $N_{\rm sph}=10^5$ is employed.
For this test only, a polytropic gas clump with $\gamma=5/3$ and initial $T_{\rm 0} = 200$~K is used. 
%The initial mass of the clump is $3\mj$. 
To ensure a uniform dust abundance throughout the clump, dust particles are introduced at $t=0$ by copying the locations of all SPH particles. We set $a=0.01$~cm, ensuring that pebbles are very closely coupled to the SPH particles. The initial pebble particle weight is such that clump metallicity is $Z_0 = 0.005$. Instead of adding new dust particles to increase pebble abundance in the cloud we simply increase the mass of the existing pebble particles exponentially, $m_{\rm d} = m_{\rm d0} \exp(t/t_Z)$, where $t_Z = 16$ years. 
%This particular choice for the metal loading timescale $t_Z$ is not important as long as it is much longer than the dynamical time of the clump so that its contraction is quasi-hydrostatic. 
Figure \ref{fig:test1} compares the analytical solution given by eq. \ref{tc1} with the SPH simulation. The numerical solution starts to deviate from the theory slightly at the highest temperatures due to the finite gravitational softening of $h_{\rm soft} = 0.04$~AU for the pebble particles employed for this simulation.
%, which under-estimates their contribution to the gravitational potential at the highest temperatures.

\subsection{Realistic 3D simulation of dark collapse}\label{sec:dark}

%\subsubsection{Setup}\label{sec:Zload_set}

%Simulation SpN8a1T900

Simulation DarkCollapse starts with a gas clump with an initial $T_{\rm c} = 900$~K. The SPH particle number is $N= 8\times 10^5$.
%presented in this section explores dark collapse ideas in 3D.  For now we wish to avoid additional effects caused with formation of a massive core \citep[note that][also assumed that no massive core forms inside the clump]{Nayakshin15a}, we choose a gas clump with a high initial central temperature of $T_{\rm c} = 900$~K.  
%The initial condition for this clump is obtained by a homologous contraction by factor of 3 of the initial condition of our fiducial gas clump with $T_{\rm c} = 300$~K. 
%Due to this contraction, the clump is much denser, and the dynamical timescale is shorter, $t_{\rm dyn} \sim (G\rho_{\rm c})^{-1/2} = 0.46$~yr.
Our simplified (fixed $\gamma = 7/5$) equation of state (EOS) becomes grossly inaccurate above $T=$2,000~K as it does not take into account H$_2$ molecule dissociation. Calculations with a more detailed EOS show that the gas clump collapses when the central temperature exceeds $2000 - 2500$~K \citep{BodenheimerEtal80,HB11,Nayakshin15a}. The collapse is approximately isothermal until the central density reaches $\sim 10^{-3}$ g/cm$^3$ \citep[e.g., see fig. 2 in][]{Bodenheimer74}. Once most of H$_2$ is dissociated, the central temperature in the clump rises to above $\sim 10^{4}$ K and the gas density increases further. In the temperature region $2,000 < T < 10,000$~K an effective value of $\gamma$ is as low as $\gamma \approx 1.1$. Modeling H$_2$ dissociation in detail is beyond the scope of our paper. Instead, we set the maximum temperature that the gas can reach in our simulations to $T_{\rm max} =2,000$ K. This leads to collapse of gas clumps when they cool to $T_{\rm max}$.

We introduce a sink particle if gas density exceeds $\rho = 10^{-6}$ g/cm$^{3}$. The sink particle is allowed to accrete SPH gas and pebble particles if they are bound to it gravitationally and are within accretion radius $r_{\rm a} = 6\times 10^{-3}$~AU. For all the simulations explored, the sink particle creation criterion is triggered only when the gas clump in the centre was indeed collapsing with a large negative velocity.

No pebbles are present in the clump at $t=0$. They are added to the clump at a constant rate. At every time step, each SPH particle  has probability $(1-\exp(-\Delta t/t_{\rm b}))$  of "giving birth" to a dust particle, where $\Delta t$ is the time step for the particle and $t_{\rm b} = 80$~yr. The initial dust particle position is offset from the position of the parent gas particle by 0.4 AU (the initial extent of the clump) in the direction from the clump centre to the SPH particle location. Thus, pebbles materialise at a constant rate  in a spherical shell just outside of the gas clump. 
%Pebble injection procedure employed here is isotropic, so that any anisotropies developing later on are a result of the dust-RT instability.

Note that $t_{\rm b}$ is much shorter than expected time scales on which significant pebble mass can be accreted in realistic protoplanetary discs, which are $\sim O(10^3)$ yr \citep{HN18}. However, $t_{\rm b}$ is $ \sim 200$ times longer than the clump dynamical time, meaning that pebbles are added slowly (adiabatically). We rerun the simulation with $t_{\rm b}$ equal to 20 and 40 years, respectively, and obtained results virtually identical to those presented here: clump evolution, presented as $t/t_{\rm b}$, is independent of $t_{\rm b}$ as long as $t_{\rm b} \gg t_{\rm dyn}$. This implies that the results of this simulation should also apply to gas clumps with $t_{\rm b}$ as long as $10^3-10^4$ yrs. 
%Another limitation on $t_{\rm b}$, from the top this time, is that $t_{\rm b}$ must be significantly shorter than the clump radiative cooling time, which is somewhat shorter than $10^5$~yr for $M_{\rm p} = 3\mj$.

We consider pebbles made of rocks, set their initial size to $a=1$~cm, mass to 0.1 SPH particle mass, and allow for pebble vaporisation but not collisional growth or fragmentation. Furthermore, as pebbles are vaporised, we include the latent heat of grain vaporisation, $E_{\rm vap0}$, in the energy balance. For rocks, the specific value for $E_{\rm vap0}\approx 10^{11}$ erg/g \citep[see table I in][]{PodolakEtal88}.
%, which is of order the mean specific kinetic energy of H$_2$ molecules at $T=1500$~K, implying that vaporisation of silicates saps a significant amount of thermal energy from the gas. %Vaporisation of rocks is just one example of dust/molecular latent heats of materials other than H/He that may help to "cool" the gas \citep[e.g.,][]{HoriIkoma11,BrouwersEtal18} and accelerate gas clump collapse.
To follow energy transfer from the surrounding gas to pebbles as they are vaporised, we define a function $E_{\rm vap}(a)$,
\begin{equation}
E_{\rm vap}(a) = E_{\rm vap0} \; \frac{a-a_{\rm min}}{a_{\rm max} - a_{\rm min}}\;,
\label{Evap}
\end{equation}
where $a_{\rm max} = 1$~cm, the initial grain size, and $a_{\rm min} = 0.1$~cm. 
%The latter value has no bearings on the results as pebbles of such small size are very tightly coupled to gas in this high density clump. 
Eq. \ref{Evap} stipulates that grains "use up" their alloted latent heat of vaporisation completely as their size drops from $a_{\rm max}$ to $a_{\rm min}$. The change in $E_{\rm vap}(a)$ for a dust particle in a given time step is subtracted from the internal energy of the SPH neighbours of the particle, using the SPH kernel averaging, ensuring energy conservation.

%\subsubsection{Results}\label{sec:Zload_results}

%We first describe the results of the simulation qualitatively (cf. animations at TBD). 
An animation of the simulation is available in the online supplementary material. Pebbles born on the outskirts of the clump initially sink in rapidly, but then stall where the stopping time is long. As more pebbles arrive, RT fingers develop and carry the pebbles in. Since new pebbles are loaded onto the clump continuously, hundreds of RT fingers develop, one after the other. These fingers build up a dust-enriched central region. The gas clump contracts in reaction to the extra weight in pebbles, and the central region becomes hot enough to vaporise the dust. The dust particles are vaporised in the clump centre, preventing solid core formation.  The clump becomes denser and hotter with time, $T_{\rm c}$ exceeds $T=2,000$~K, at which point the central region becomes isothermal. The centre of the gas clump collapses, bringing the rest of the clump down with it as well. This results in formation of a hot-start gas giant with  bulk metallicity of about $Z=0.1$.

\begin{figure*}
\includegraphics[width=0.45\textwidth]{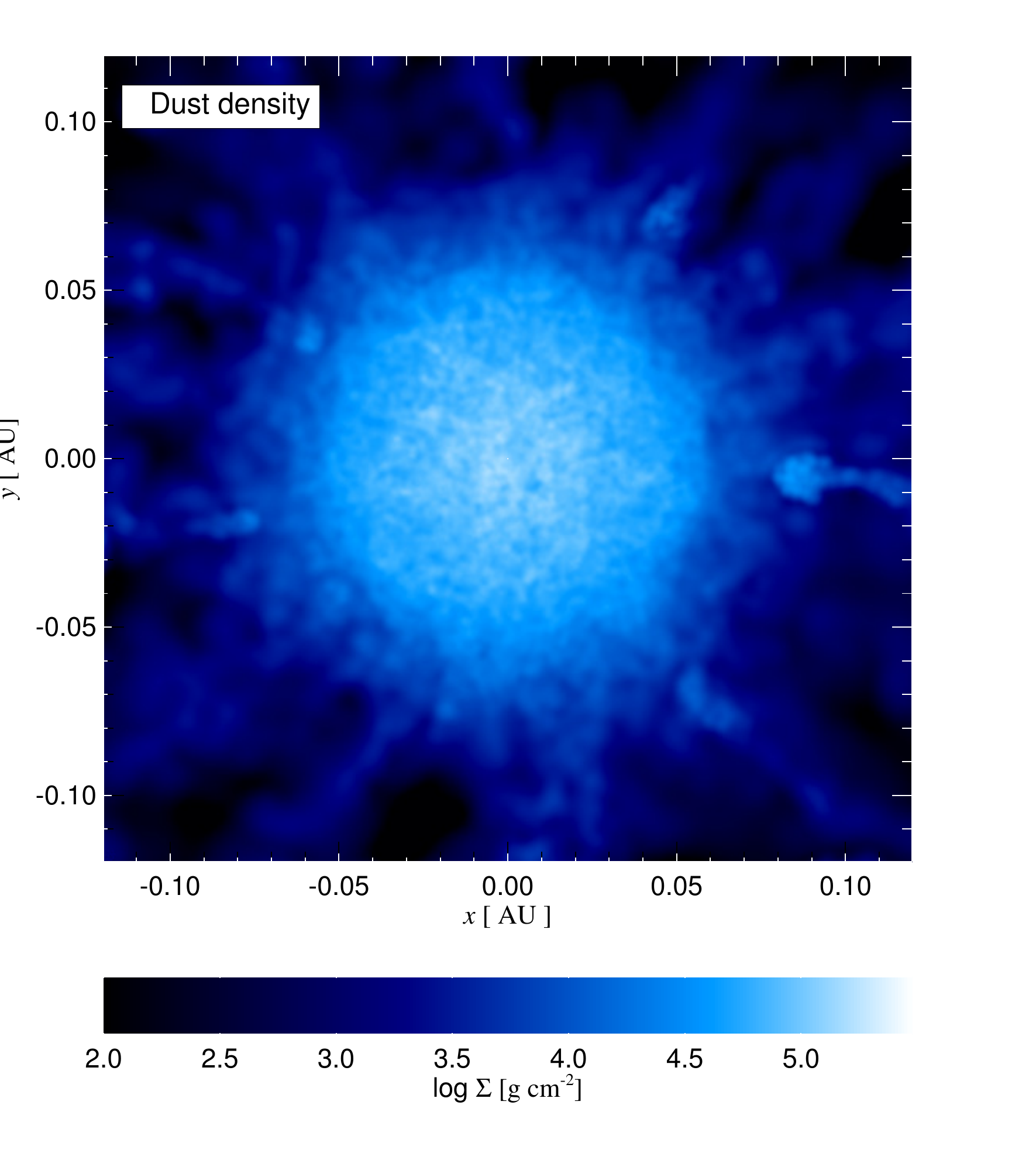}
\includegraphics[width=0.45\textwidth]{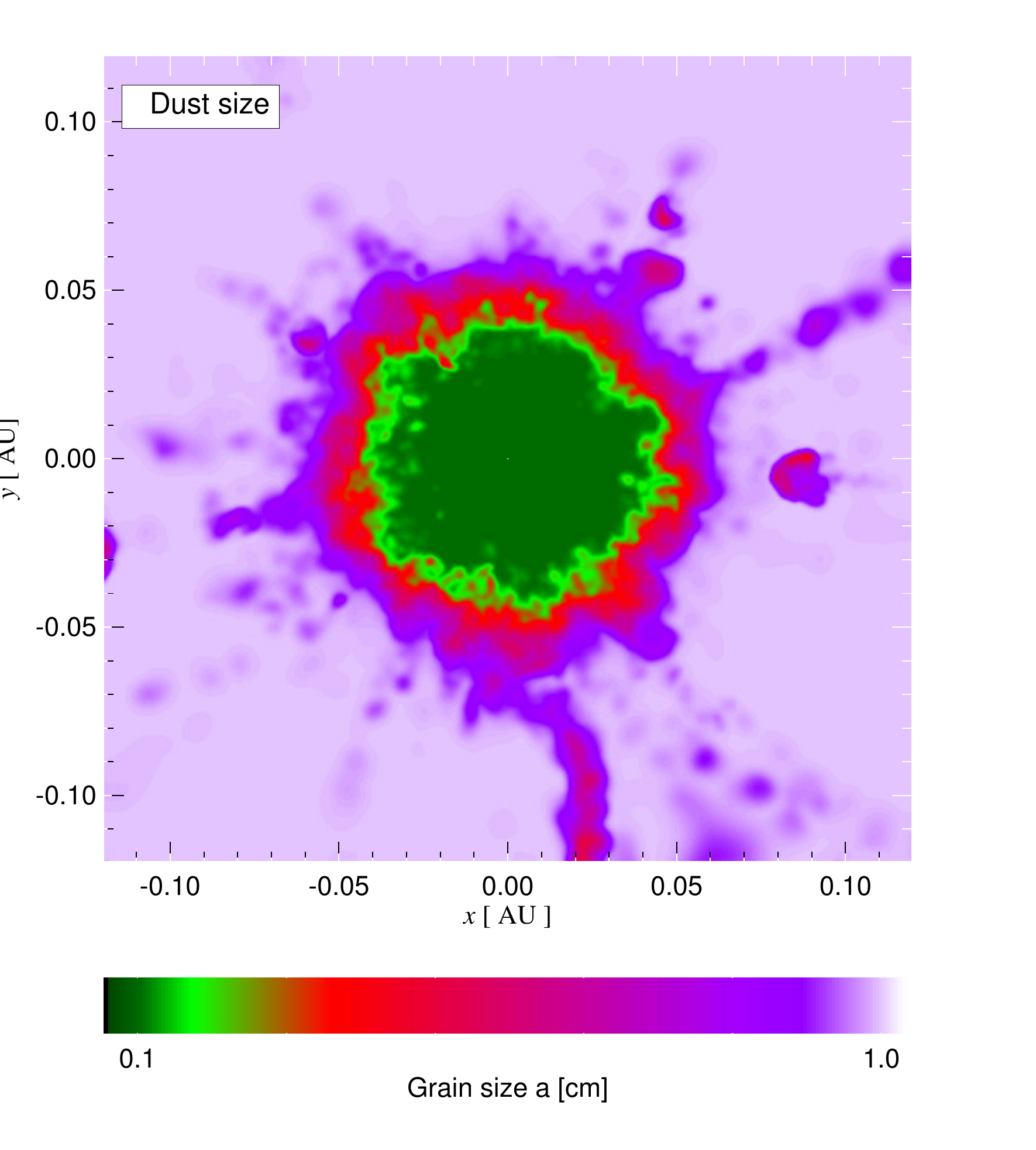}
\includegraphics[width=0.45\textwidth]{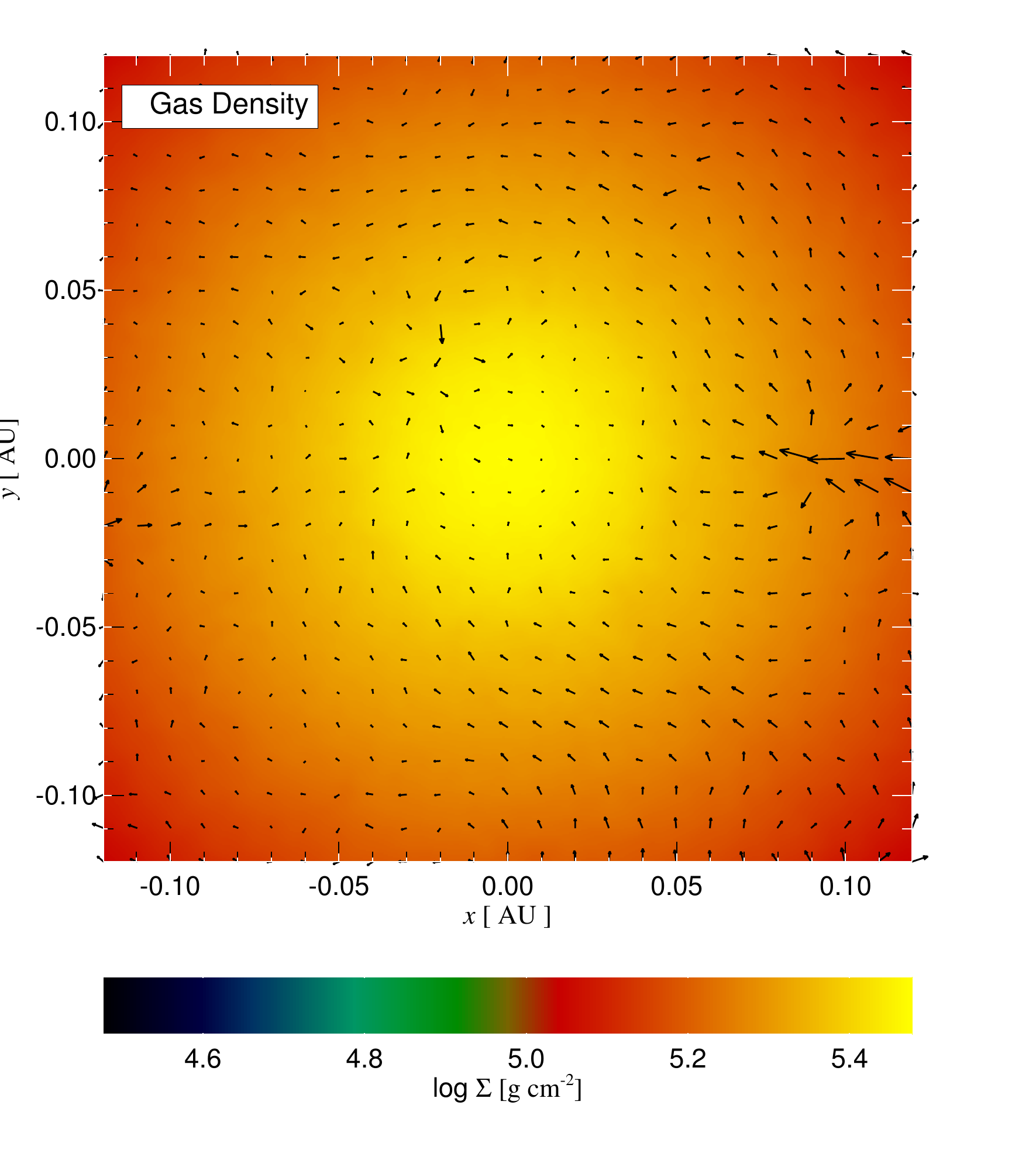}
\includegraphics[width=0.45\textwidth]{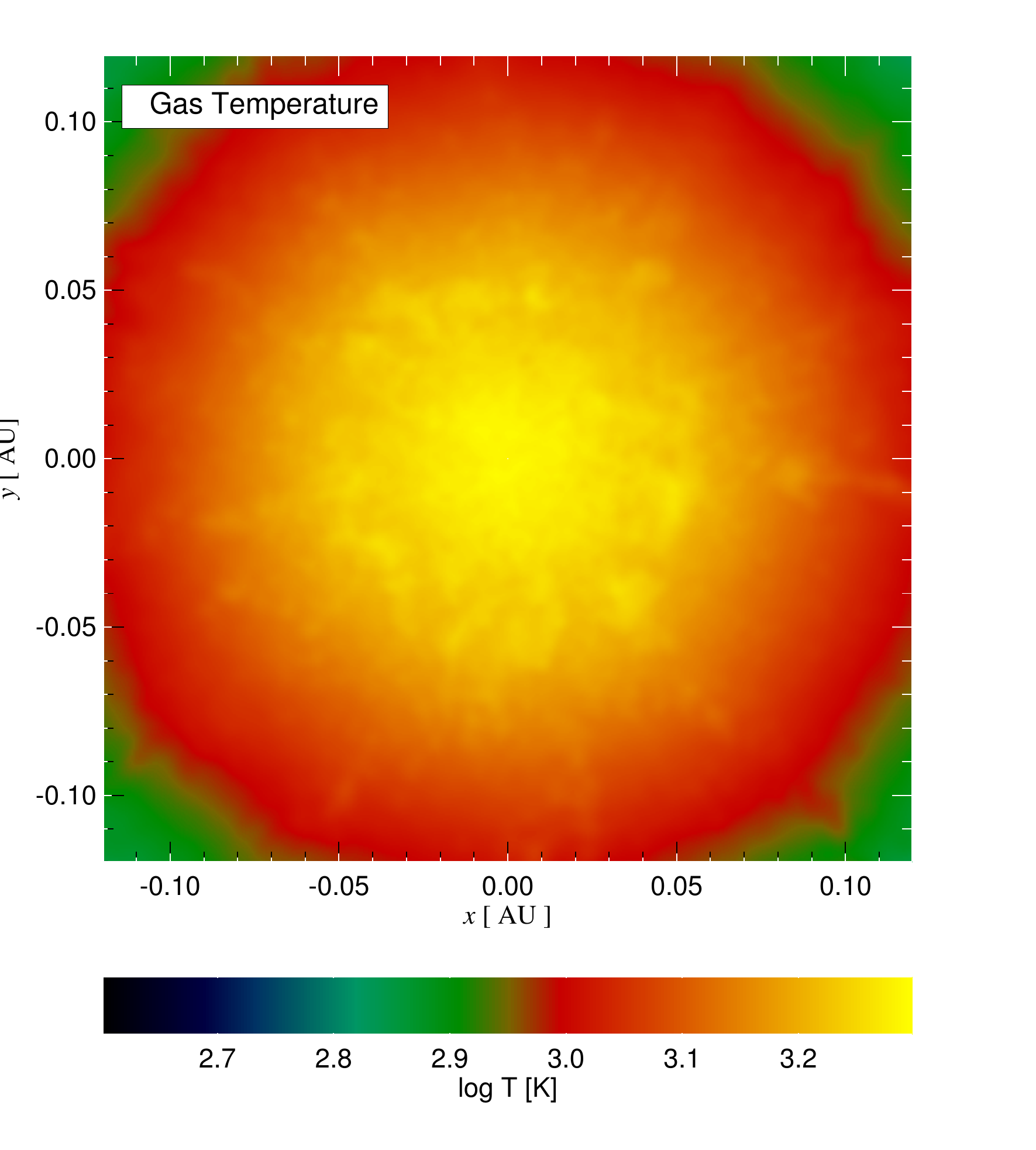}
\caption{Thin slice projections of gas and dust properties for simulation DarkCollapse at time $t=79.7$~yr, three years before it collapses into a hot start gas giant with a fuzzy core.}
\label{fig:Dark_Collapse_Map}
\end{figure*}

Fig. \ref{fig:Dark_Collapse_Map} shows thin slices of the pebble density and size (top row), gas density and temperature (bottom row) in the
 central region at $t=79.7$ yr, several years before collapse. Comparing the gas and the dust projected densities in the left panels of fig. \ref{fig:Dark_Collapse_Map}, we conclude that the dust is almost as abundant as gas in the clump centre by mass. Dust distribution is far more inhomogeneous compared with that of the gas. Individual RT fingers can be seen; most of these filaments are remnants/tails of the fingers that fell in earlier. The gas temperature map shows that the temperature distribution is not spherically symmetric, showing lumpy structure probably related  to individual RT fingers. For example, the RT finger with a mushroom like head seen at $(x,y) \approx (0.08, 0)$~AU has a corresponding local peak in gas temperature and also a local minimum in the dust particle size. This is caused by the heating due to gas-dust aerodynamical friction.

\begin{figure*}
\includegraphics[width=0.45\textwidth]{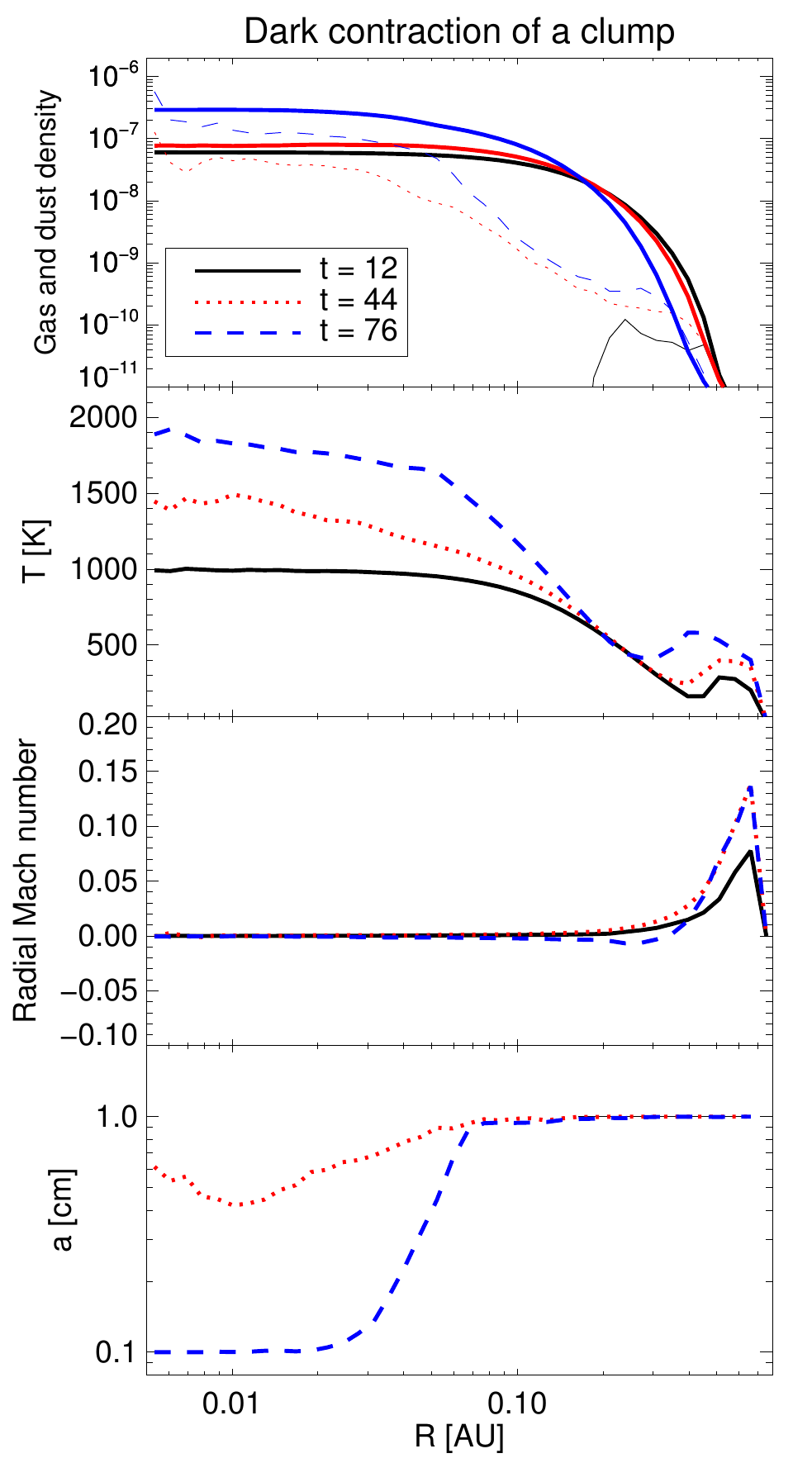}
\includegraphics[width=0.45\textwidth]{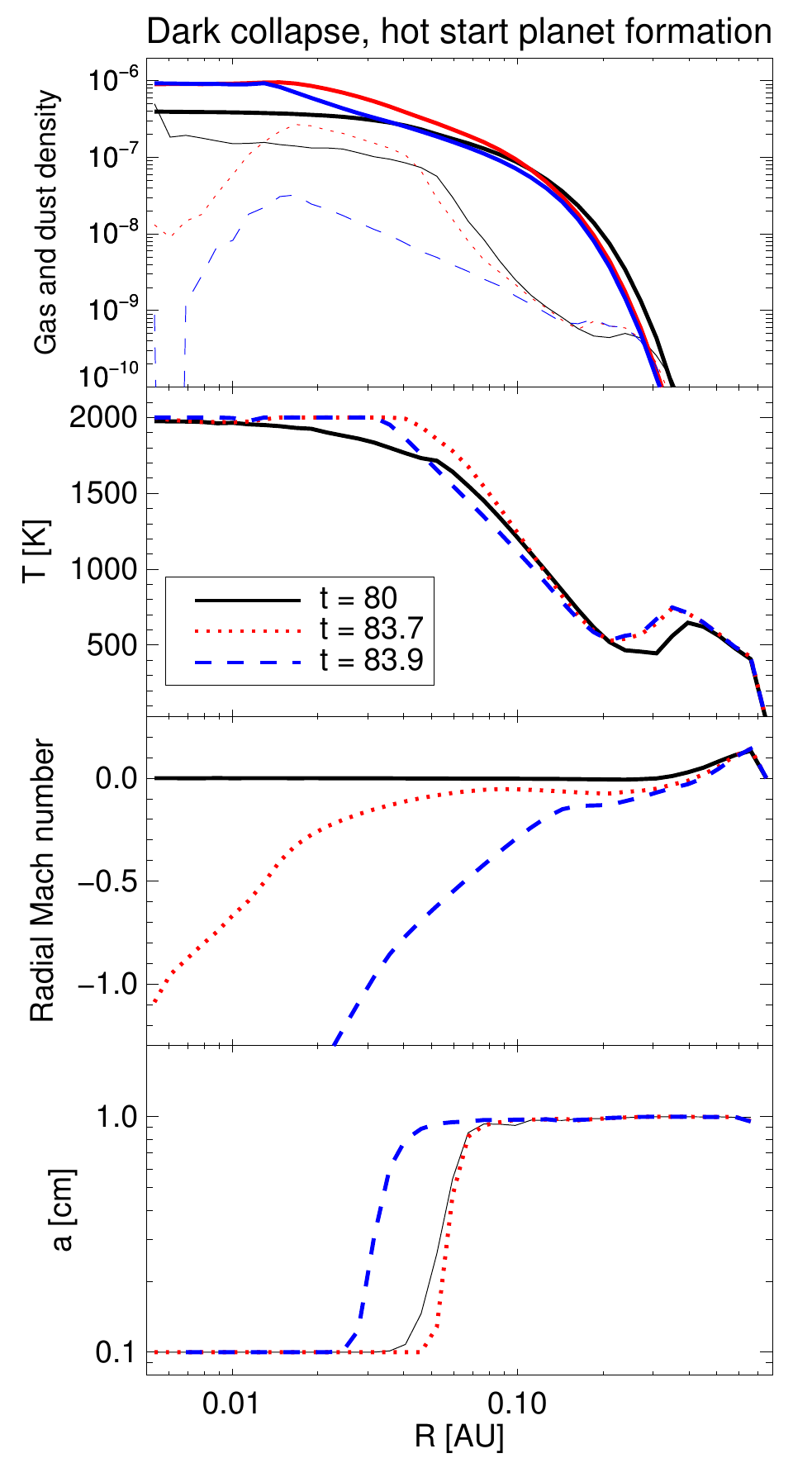}
\caption{Gas and dust properties averaged on concentric shell for simulation DarkCollapse {\bf Left panels:} Contraction phase during which pebbles are loaded onto the clump from outside. {\bf Right panels:} Same quantities but shown at times just before and during hydrodynamical collapse of the clump due to H$_2$ molecule dissociation.}
\label{fig:Dark_Collapse}
\end{figure*}

Fig. \ref{fig:Dark_Collapse} shows evolution of the gas clump profile. The left panels show a relatively slow hydrostatic phase. The radial Mach number plot demonstrates that this phase is very nearly hydrostatic.  An outward expansion of the outer gas layers is driven by the aerodynamical heating of these layers as pebbles sediment through those regions. 
%As explained earlier in \S \ref{sec:spherical}, pebbles go through the outer low density layers not via collective mode but rather due to individual particle motion through the gas, which results in gas heating. Since these layers contain a small fraction of the gas clump mass and yet the mass of pebbles loaded onto the clump is significant ($\sim 10$\% of the clump mass by the end of the simulation), the gravitational potential energy released by the pebbles as they go through the outer layers heats the gas significantly, up to temperatures of hundreds of K. 
The very central region of the clump is hot enough by $t= 76$~yr to vaporise all the grains there, so that they reach the minimum allowed size, $a_{\rm min}$. The right panels of fig. \ref{fig:Dark_Collapse} show  same quantities but very near and during the collapse phase. The temperature reached $T=2000$~K at $t=80$~yrs but the collapse is not immediate as $v_{\rm rad}$ remains very nearly zero at that time. This is because the isothermal region is yet too small in terms of enclosed mass. However, the region eventually becomes massive enough to collapse under its own self-gravity (although the weight of the outer clump layers certainly helps). By $t= 83.7$~yr the sink particle is present in the centre of the gas clump. The gas infall velocity in the centre becomes super-sonic and the collapse cannot be reversed. The clump collapses dynamically in a small fraction of a year.
%\newpage

%\subsubsection{Comparison with analytic theory}\label{sec:DK_implications}

%\cite{Nayakshin15a} worked out an idealized spherically symmetric theory for 
evolution of a polytropic gas clump loaded with pebbles due to accretion from the parent disc. A uniform composition and no solid core formation was assumed. The theory predicts (eq. \ref{tc1}) that gas clumps that increase their bulk metallicity due to pebble accretion contract and heat up rapidly, collapsing when accreting $\sim 5-20$\% of their mass in pebbles, depending on the initial central temperature. 

%In contrast to the classical radiative cooling contraction and collapse of gas clumps \citep{Bodenheimer74}, this scenario requires no 

The 3D simulation presented here confirms that dark collapse of gas clumps is possible in realistic 3D simulations. However, the composition of the clump is non uniform: pebbles are concentrated in the clump centre (fig. \ref{fig:Dark_Collapse}). There is some outward gas motion on the outer clump edge due to aerodynamical friction between the gas and the pebbles, making collapse more difficult. This may be the reason why the clump collapses at a significantly higher metal abundance than expected from eq. \ref{tc1}. According to eq. \ref{tc1}, $T_{\rm c}$ should reach 2,000~K at $Z =0.065$. In the simulation, $T_{\rm c} \approx 2,000$~K at $Z=0.091$, and the clump actually collapses at $Z = 0.096$.

\section{Discussion and Conclusions}

%The instability presented here may be also related to the "drafting instability" recently found by \cite{LambrechtsEtal16} in the context of vertical grain settling in protoplanetary discs. The authors also predicted existence of this instability in the envelopes of growing gas giant planets in the context of the Core Accretion model for planet formation. Here we re-discovered their results for planets formed via gravitational disc instability. An idealised model developed by  \cite{LambrechtsEtal16} predicts that the instability growth rate is independent of grain size (their equation 12), which is also the case 

%Simulations presented here show that gas clumps born by gravitational instability of protoplanetary discs are promising environs for making planets with very diverse properties. 

Simulations presented here show that clumps that accrete pebbles from their parent discs can make high-Z cores more rapidly than assumed based on earlier closed-box 1D models of the clumps \citep[e.g.,][]{HS08,BoleyEtal10,Nayakshin10c,ForganRice13b}. In general, pebbles loaded onto the clump sediment through the outer envelope rapidly in the test particle regime, and then stall in higher density regions. The dust Rayleigh-Taylor instability then develops, transporting them in in a matter of tens to hundreds of years. Small grains grow and large grains fragment in the metal enriched central part of the clump, also on time scales of tens to hundreds of years. A few cm or larger sized pebbles then get locked into solid cores. The outcome of these processes depends on pebble composition, how hot the centre of the clump is, and how long it has to live before its tidal disruption. Although these external factors are not modeled in this paper, previous 1D models of planet formation by gravitational instability that {included both dust physics and pebble accretion} \citep[e.g.,][]{NayakshinFletcher15} compare with many observational facts favorably \citep[for details see \S 9 in][]{Nayakshin_Review}. 

3D simulations presented here however show that cores can be made even more rapidly due to dust-RT instability, and that even small pebbles loaded into the clump tend to concentrate into the clump central regions rather than be spread around the clump uniformly \citep[as was assumed in][]{Nayakshin15a}. As a result, we found that gravitational instability clumps may form gas giants with fuzzy cores if the central regions of the clump are hotter than $\sim 1500$~K. This may be relevant to the recent {\em Juno} satellite Jupiter's gravity measurements that indicate that its core may be fuzzy rather than solid \citep{WahlEtal17}.

Formation of cores inside the gaseous clumps formed by gravitational instability is a promising and probably required mechanism to explain planets found in circumstances unfavorable to formation by Core Accretion \citep{PollackEtal96}. For example, the suspected $\sim$ Neptune mass planets in the $\sim 1$~Myr-old young disc of HL Tau \citep{BroganEtal15,DipierroEtal15} should have formed after $\sim 10^8$ years in the classical Core Accretion scenario \citep[e.g.,][]{KB15}. These low mass planets could not form by the pure gas disc fragmentation as such objects are at least $\sim 1 \mj$ in mass \citep{BoleyEtal10}. 

Another promising application of the theory is close circum-binary planets, where binary kicks lead to very violent planetesimal-splitting collisions \citep{LinesEtal14}. In the context of gravitational instability, these planets could have formed inside the protective envelope of the self-gravitating gas clump, initially at large separation from the binary centre. When such a clump migrates close enough to be disrupted, its ready-made-core or planet could be safely deposited onto a much smaller orbit.

Our simulations however do not include radiative cooling of the clumps and feedback from growing massive cores \citep{Nayakshin16a}. These effects may dictate the resulting metallicity correlations of objects made by gravitational instability, from planetary debris and sub-Neptune planets \citep{FletcherNayakshin16a} to massive planets and brown dwarfs \citep{NayakshinFletcher15}. 3D simulations including these effects shall be reported elsewhere.

Finally, note that the instability presented here is probably related to the "drafting instability" recently found by \cite{LambrechtsEtal16} in the context of vertical grain settling in protoplanetary discs. The authors also predicted existence of this instability in the envelopes of growing gas giant planets in the context of the Core Accretion model for planet formation. Our results therefore echo their funding for planets formed via gravitational disc instability. 

\section{Acknowledgements}
%The author thanks Anders Johansen and Chris Ormel for stimulating discussions and useful references. 

Support is acknowledged from STFC grants ST/K001000/1 and ST/N504117/1, as well as the ALICE High Performance Computing Facility at the University of Leicester, and the STFC DiRAC HPC Facility (grant ST/H00856X/1 and ST/K000373/1). DiRAC is part of the National E-Infrastructure. 

%An idealised model developed by  \cite{LambrechtsEtal16} predicts that the instability growth rate is independent of grain size (their equation 12), which is also the case 

%\caption{Distribution of grain growth time scales ({\bf Top panels}) and velocity dispersion ({\bf Bottom panels}) for pebbles in the simulations with different grain sizes as shown in the legend. Note that grain growth time scales are quite short for most of pebbles for all the panels. However, the dispersion velocity increases as grain size increases, and the largest grains are shattered by collisions, leading to fragmentation instead of growth.}

\bibliographystyle{mnras}
\bibliography{nayakshin} % if your bibtex file is called example.bib

%%%%%%%%%%%%%%%%%%%%%%%%%%%%%%%%%%%%%%%%%%%%%%%%%%

%%%%%%%%%%%%%%%%% APPENDICES %%%%%%%%%%%%%%%%%%%%%

% Don't change these lines
\bsp	% typesetting comment
\label{lastpage}
\end{document}